\documentclass[aps,pra,twocolumn,superscriptaddress,floatfix]{revtex4-1}
\usepackage[table,svgnames]{xcolor}
\definecolor{darkred}{rgb}{0.6, 0, 0}
\definecolor{darkgreen}{rgb}{0, 0.5, 0}
\usepackage[colorlinks = true, urlcolor = teal, linkcolor = NavyBlue, citecolor = violet, bookmarks = false]{hyperref}
\usepackage{amsfonts,amsmath,amssymb,mathrsfs}
\usepackage{color, colortbl}
\usepackage{graphicx}
\usepackage{dsfont}
\usepackage[normalem]{ulem}

\makeatletter
\def\@ssect@ltx#1#2#3#4#5#6[#7]#8{%
  \def\H@svsec{\phantomsection}%
  \@tempskipa #5\relax
  \@ifdim{\@tempskipa>\z@}{%
    \begingroup
      \interlinepenalty \@M
      #6{%
       \@ifundefined{@hangfroms@#1}{\@hang@froms}{\csname @hangfroms@#1\endcsname}%
       {\hskip#3\relax\H@svsec}{#8}%
      }%
      \@@par
    \endgroup
    \@ifundefined{#1smark}{\@gobble}{\csname #1smark\endcsname}{#7}%
  }{%
    \def\@svsechd{%
      #6{%
       \@ifundefined{@runin@tos@#1}{\@runin@tos}{\csname @runin@tos@#1\endcsname}%
       {\hskip#3\relax\H@svsec}{#8}%
      }%
      \@ifundefined{#1smark}{\@gobble}{\csname #1smark\endcsname}{#7}%
      \addcontentsline{toc}{#1}{\protect\numberline{}#8}%
    }%
  }%
  \@xsect{#5}%
}%
\makeatother

\usepackage{braket}
\usepackage{slashed}
\usepackage{mathtools}
\usepackage{hyperref}
\usepackage{multibib}
\usepackage{enumerate}
\usepackage{booktabs}
\usepackage{youngtab}

\usepackage{pdfpages} 
\makeatletter
\AtBeginDocument{\let\LS@rot\@undefined}
\makeatother

\usepackage{tikz}
\usetikzlibrary{arrows,topaths,shapes.geometric,calc,decorations.markings,decorations.pathmorphing,shadows}
\usepackage{tikz-3dplot} 

\def\bra#1{\mathinner{\langle{#1}|}}
\def\ket#1{\mathinner{|{#1}\rangle}}
\newcommand{\ii}{{\rm i}}
\newcommand{\dd}{{\rm d}}

\def\ol#1{\bar{#1}}

\newcommand{\calT}{\mathcal{T}}
\newcommand{\calY}{\mathcal{Y}}
\newcommand{\scrT}{\mathscr{T}}
\newcommand{\scrY}{\mathscr{Y}}

\newcommand{\rH}{\mathrm{H}}
\newcommand{\rE}{\mathrm{E}}
\newcommand{\rX}{\mathrm{X}}
\newcommand{\rQ}{\mathrm{Q}}

\newcommand{\U}{\mathrm{U}}
\newcommand{\rO}{\mathrm{O}}
\newcommand{\SU}{\mathrm{SU}}
\newcommand{\SO}{\mathrm{SO}}
\newcommand{\Sp}{\mathrm{Sp}}
\newcommand{\USp}{\mathrm{USp}}

\usepackage{pdfpages} 
\makeatletter
\AtBeginDocument{\let\LS@rot\@undefined}
\makeatother

\begin{document}

\title{Superuniversality of superdiffusion}

\author{Enej Ilievski}
\affiliation{Faculty for Mathematics and Physics, University of Ljubljana, Jadranska ulica 19, 1000 Ljubljana, Slovenia}
\author{Jacopo De Nardis}
\affiliation{Department of Physics and Astronomy, University of Ghent, Krijgslaan 281, 9000 Gent, Belgium}
\author{Sarang Gopalakrishnan}
\affiliation{Department of Physics and Astronomy, CUNY College of Staten Island, Staten Island, NY 10314;  Physics Program and Initiative for the Theoretical Sciences, The Graduate Center, CUNY, New York, NY 10016, USA }
\author{Romain Vasseur}
\affiliation{Department of Physics, University of Massachusetts, Amherst, Massachusetts 01003, USA}
\author{Brayden Ware}
\affiliation{Department of Physics, University of Massachusetts, Amherst, Massachusetts 01003, USA}

\begin{abstract}

Anomalous finite-temperature transport has recently been observed in numerical studies of various integrable models in one dimension; 
these models share the feature of being invariant under a continuous non-abelian global symmetry.
This work offers a comprehensive group-theoretic account of this elusive phenomenon.
For an integrable quantum model invariant under a global non-abelian simple Lie group $G$, we find that finite-temperature transport of 
Noether charges associated with symmetry $G$ in thermal states that are invariant under $G$ is universally superdiffusive and 
characterized by dynamical exponent $z = 3/2$. This conclusion holds regardless of the Lie algebra symmetry, local degrees of freedom 
(on-site representations), Lorentz invariance, or particular realization of microscopic interactions: we accordingly dub it as 
\emph{superuniversal}. The anomalous transport behavior is attributed to long-lived giant quasiparticles dressed by thermal 
fluctuations. We provide an algebraic viewpoint on the corresponding dressing transformation and elucidate formal connections to fusion 
identities amongst the quantum-group characters.  We identify giant quasiparticles with nonlinear soliton modes of classical field 
theories that describe low-energy excitations above ferromagnetic vacua. Our analysis of these field theories also provides a complete 
classification of the low-energy (i.e., Goldstone-mode) spectra of quantum isotropic ferromagnetic chains.

\end{abstract}

\maketitle

\tableofcontents

\section{Introduction}

A complete characterization and classification of dynamical properties of isolated interacting quantum many-body systems remains one of 
the central unsettled problems in statistical mechanics, whether classical or quantum.
Especially in one-dimensional systems, a range of exotic dynamical phenomena have been demonstrated, both theoretically and experimentally.
Two prominent examples are integrable and many-body localized systems~\cite{doi:10.1146/annurev-conmatphys-031214-014726, VM_review, RevModPhys.91.021001}, which feature extensively many conserved quantities and therefore can persist in nonthermal ``generalized Gibbs states'' that are measurably different from the standard thermal ensemble~\cite{Deutsch91,Srednicki94,Rigol07,ETH_review, VR_review,QLreview}. 
Because these extensive conservation laws lead to nonstandard equilibrium states, and because hydrodynamics begins with an assumption 
of local thermal equilibrium, it follows that hydrodynamics is also modified for integrable systems. Thus, instead of normal diffusion, 
non-disordered integrable systems typically exhibit ballistic transport with
finite Drude weights \cite{CZP95,Bertini_transport_review}, whereas in localized models transport is entirely absent~\cite{VM_review}.

Integrable systems feature coherent quasiparticle excitations 
with infinite lifetime \cite{ZZ78,ZZ79} that propagate through the system in a ballistic manner while scattering elastically off one 
another. The same picture remains valid in thermal ensembles at finite temperature, where one can think of quasiparticles being 
`dressed' due to interactions with a macroscopic thermal environment \cite{YY1969}. Thermal fluctuations are responsible for screening, 
and thus conserved charges carried by quasiparticles can be appreciably different from the bare values. This effect is captured by the 
versatile framework of generalized hydrodynamics (GHD) \cite{PhysRevX.6.041065,PhysRevLett.117.207201}. Among other results, GHD has 
led to the explicit characterization of ballistic transport \cite{IN_Drude,DS17,IN17,Bulchandani18} and analytic treatments of various 
other nonequilibrium protocols \cite{Bulchandani17,PhysRevLett.120.045301,CDDY19,Bastianello_noise,Bastianello19}.
Remarkably, despite the ballistic motion of individual excitations, certain integrable models do \emph{not} 
necessarily exhibit ballistic transport of macroscopic scales, but instead display normal diffusion or even \emph{anomalous} diffusion;
this is the case for a distinguished subset of conserved quantities linked with internal degrees of freedom, whereas
other conserved quantities (such as energy) undergo ballistic transport~\footnote{Notice that even when a quantity exhibits ballistic 
transport, the sub-leading corrections to ballistic dynamics (seen, e.g., in the low-frequency conductivity) can be
anomalous~\cite{GVW19,Agrawal20}.}.

\medskip

This work is dedicated to anomalous transport in integrable models. This unexpected phenomenon was first found
in Ref.~\cite{MarkoKPZ} in the Heisenberg spin-$1/2$ chain, which can be regarded as
an archetypal example of a quantum many-body interacting system.
Nowadays there is numerical evidence \cite{Ljubotina19} that the dynamical exponent and asymptotic scaling 
profiles of dynamical structure factors belong to the Kardar-Parisi-Zhang (KPZ) universality class \cite{KPZ}.
Specifically, the diagonal dynamical spin correlations $C^{j j}(x,t)\equiv \langle S^{j}(x,t) S^{j}(0,0) \rangle$
among the spin components $S^{j}$ (for all $j \in \{{\rm x,y,z}\}$), evaluated in thermal equilibrium \emph{at half filling},
have been found to comply with the scaling form
\begin{equation}
C^{j j}(x,t) \simeq \frac{C^{j j}}{(\lambda_{\rm KPZ}\,t)^{2/3}}\,f_{\rm KPZ}\left[\frac{x}{(\lambda_{\rm KPZ}\,t)^{2/3}}\right],
\label{eqn:KPZ_scaling_functions}
\end{equation}
characterized by scaling function $f_{\rm KPZ}$ (tabulated in \cite{Prahofer2004}), KPZ nonlinearity coupling parameter $\lambda$,
and normalized by diagonal static charge susceptibilities $C^{j j}=\int \dd x\,C^{j j}(x,t)$.
Kardar--Parisi--Zhang physics is a widespread phenomenon in \emph{stochastic} growth processes and dynamical interfaces
\cite{SS10,CD11,Corwin2012,Takeuchi18}. Its occurrence in \emph{deterministic} Hamiltonian dynamical systems is therefore
an interesting curiosity. The microscopic mechanism responsible for superdiffusion is not yet understood in full detail.
The fact that spin (or charge) superdiffusion of the KPZ type has also been numerically observed in a number of other quantum chains, 
such as the higher-spin $\SU(2)$ integrable chains and the $\SO(5)$-symmetric spin ladder \cite{DupontMoore19},
the Fermi--Hubbard model \cite{Fava20}, and even in the classical Landau-Lifshitz equation \cite{Bojan,PhysRevE.100.042116,KP20}
(and its higher-rank analogues that exhibit symmetry of non-abelian unitary groups \cite{MatrixModels}), give a strong indication that 
there is a general principle behind superdiffusion in integrable systems with non-abelian symmetries that awaits to be uncovered.

Generalized hydrodynamics (GHD) has already provided a number of invaluable theoretical insights into this question.
Specifically for the case of the quantum Heisenberg spin chains, both the dynamical exponent $z=3/2$ \cite{GV19} and the KPZ coupling 
constant $\lambda_{\rm KPZ}$ \cite{NGIV20} (though not the scaling function) have been inferred with aid of a heuristic extension of
the GHD framework, using full advantage of the exact knowledge of the Bethe ansatz quasiparticles.
Recently, a phenomenological explanation for the observed KPZ phenomenon has been given in \cite{Vir20} which invokes the notion
of hydrodynamic `soft modes' coupled to an effective noisy environment. The current understanding is that such soft models
are a manifestation of the so-called giant quasiparticles in the long-wavelength regime, whose emergent dynamics is governed
by a classical action \cite{NGIV20}. This appears to suggest that that the observed superdiffusive spin dynamics in 
\emph{classical} rotationally symmetric (e.g. $\SO(3)$-invariant) spin chains has the same microscopic origin as that of
quantum spin chains.
In spite of this theoretical progress however, a fundamental question nonetheless remains unsettled:
\begin{quote}
\emph{In integrable Hamiltonian dynamical systems, what are the necessary and sufficient conditions for
the superdiffusive charge dynamics characterized by anomalous dynamical exponent $z=3/2$?}
\end{quote}
We dedicate this paper to answering this question on general grounds, both in the realm of integrable lattice models and integrable 
quantum field theories.

Before delving into mostly technical aspects, we would like to offer a broader perspective on the problem at hand.
By invoking the universal GHD formulae for the spin diffusion constants, it is evident from the outset that divergent charge diffusion 
constants are only possible in systems with infinitely many quasiparticle species. This necessary condition is quite generally 
fulfilled in integrable lattice models, including in particular the models with non-abelian symmetries that will be our primary focus.
Given that such models \emph{do} have infinitely many quasiparticle species, however, there are two scenarios that
seem particularly plausible:
\begin{enumerate}[(I)]
\item All integrable Hamiltonians that are symmetric under a non-abelian Lie group $G$ display universal superdiffusive dynamics
of the Noether charges in $G$-invariant (i.e. unpolarized) Gibbs states,
\item integrable models accommodate for a wider range of dynamical exponents, depending possibly
on the rank of type of Lie algebras and their representations assigned to local degrees of freedom.
\end{enumerate}

There are several recent studies, e.g. \cite{Ilievski18,DupontMoore19,Fava20}, which speak in favor of (I).
Another strong piece of evidence comes from a recent study \cite{MatrixModels} of \emph{classical} integrable matrix models, providing
integrable space-time discretizations of higher-rank analogues of $\SU(n)$-symmetric Landau--Lifshitz field theories on complex 
projective spaces and Grassmannian manifolds.
In Ref.~\cite{MatrixModels} the authors provide clear numerical evidence for the KPZ scaling profiles independently of rank $r=n-1$
and further conjectured that the phenomenon occurs across all classical non-relativistic $G$-invariant integrable field theories with 
hermitian symmetric spaces as their target manifolds. Indeed, the class of models considered in Ref.~\cite{MatrixModels} govern
low-energy spectra of integrable $\SU(n)$-invariant Lai--Sutherland ferromagnetic spin chains~\cite{Lai1974,Sutherland1975} that
are included as a part of this study. On this basis, it is reasonable to anticipate that coherent semiclassical modes analogous to the 
aforementioned giant quasiparticles to emerge as a general feature of integrable quantum chains. Their microscopic description is 
nevertheless not known at this moment.

Nonetheless, option (II) cannot be a priori rejected either. We note that within the phenomenological framework of nonlinear 
fluctuating hydrodynamics (NLFHD) \cite{Kulkarni15,Das19}, an infinite family of anomalous dynamical exponents can 
arise~\cite{Popkov15}. Now, NLFHD does not directly apply to integrable systems, and there are good reasons to doubt that it can 
capture superdiffusion in the integrable case~\cite{Vir20}: notably, in NLFHD, superdiffusion arises as a correction to ballistic 
transport, rather than as the leading dynamical behavior.
Regardless, the existing numerical evidence on dynamical exponents in integrable systems is also ambiguous: for instance, a study of
the integrable $\SU(4)$ spin chain found a \emph{distinct} exponent $z \approx 5/3$ \cite{MarkoSU4}, suggesting
the possibility that different non-abelian symmetries may after all realize distinct universality classes of anomalous transport.
In addition to its fundamental interest, this question is experimentally relevant because interacting quantum lattice systems
possessing $\SU(N)$ symmetry can be implemented using ultracold alkaline-earth atoms~\cite{Gorshkov:2010aa,PhysRevLett.103.135301}.
In the context of solid state physics, several strongly coupled ladder compounds also realize (approximately) the $\SU(4)$-symmetric 
ladder in the presence of fields~\cite{Dagotto618,Batchelor_2003,doi:10.1080/00018730701265383}.


\subsection*{Summary}

In this work we outline a systematic theoretical analysis of anomalous charge transport, specializing to the class of
integrable quantum ``spin'' chains symmetric under global non-abelian simple Lie groups. In order to provide
a universal algebraic description we shall heavily rely on representation theory of quantum groups and the associated fusion relations.
For clarity, we here first summarize our main findings.

Our central results is that an anomalous algebraic dynamical exponent $z = 3/2$ associated with transport of the Noether charges
is a common feature of integrable Hamiltonian systems invariant under a general non-abelian Lie group $G$, provided the equilibrium 
state does not break the global symmetry by the presence of finite chemical potentials. This statement holds independently of the type 
of simple Lie algebra and on unitary representations associated with local Hilbert spaces (in quantum chains) or local degrees
of freedom (in integrable QFTs): the only requirement is that the charge $Q$ whose correlation functions we study must transform 
nontrivially under $G$ (unlike, e.g., the energy). We thereby establish \emph{superuniversality} of superdiffusive charge transport.

Our analysis incorporates the following complementary approaches:
\begin{enumerate}[(i)]
\item We carry out a scaling analysis of the universal Nested Bethe Ansatz dressing equations, concluding that
the spectrum of giant quasiparticles (governing the semiclassical long-wavelength dynamics of the charge density in highly 
excited eigenstates) exhibit the same type of asymptotic scaling relations irrespectively of the non-abelian symmetry algebra.
This implies that the kinetic theory argument outlined in Ref.~\cite{GV19} carries through in general. To solidify this conclusion, we 
derive an explicit closed-form solution of the dressing equations for the case of higher-rank unitary groups $\SU(N)$ (including the
general dependence on the $\U(1)$ chemical potentials coupling to the Cartan charges).

\item Secondly, we verify our predictions through tensor-network based numerical simulations by computing dynamical charge correlations 
functions for a number of representative cases, including unitary, orthogonal and symplectic Lie groups
(shown in Figs. \ref{fig:all_tebd} and \ref{fig:sun_profilecollapse});
evidently, all the cases we have studied yield the exponent $z = 3/2$.

\item Lastly, we elucidate the physical nature of the giant quasiparticles.
These are none other than semiclassically quantized classical soliton modes, which from the spin chain viewpoint correspond to 
macroscopically large coherent states made out of interacting, quadratically dispersing (i.e., ``magnon-like''), Goldstone modes above 
a ferromagnetic vacuum. At first glance it may seem counter-intuitive that long-wavelength modes matter to high-temperature 
physics. Here integrability comes into play, ensuring that these large bound-state excitations remain well-defined quasiparticles even 
in thermal states; however, their (bare) properties do get dressed by thermal fluctuations. To corroborate this, we establish a
one-to-one correspondence between the Bethe ansatz quasiparticles and the spectrum of Goldstone modes.
This correspondence is subtle: there are many more distinct Goldstone modes for a given symmetry-breaking pattern than there are magnon 
species (flavors) in the Bethe ansatz spectrum. Counting the Goldstone modes correctly therefore requires the notion of composite 
quasiparticles called `stacks'. To our knowledge, these multiflavored stacks have not been explicitly classified thus far.
\end{enumerate}

While we mostly focus on integrable quantum chains with ferromagnetic exchange, we also
shortly discuss in Sec.~\ref{sec:QFTs} how KPZ superuniversality emerges even in Lorenz-invariant integrable quantum 
field theories possessing internal isotropic degrees of freedom which take values in compact simple Lie groups $G$ or coset spaces 
thereof. Such integrable QFTs are characterized by non-diagonal scattering, signifying that their elementary quasiparticle excitations 
can exchange isotropic degrees of freedom upon elastic collisions. The outcome of that are dynamically produced massless 
pseudoparticles that are responsible for charge transport at finite temperature. These so-called `auxiliary quasiparticles' are 
interacting magnon waves and bound states thereof, which formally resemble magnons of the quantum ferromagnetic spin
chains \cite{ZZ92}. Divergent charge diffusion constants are thus once again attributed to the presence of interacting giant magnons.

\paragraph*{Outline.}
The paper is structured as follows. In Sec.~\ref{sec:transport} we succinctly review the formalism of generalized hydrodynamics, 
provide closed formulae for the charge diffusion constants, and proceed by summarizing the general physical 
picture and scaling arguments that lead to superuniversality.
In Sec.~\ref{sec:numerics} we discuss our numerical results on transport.
The remainder of the paper consists of a longer technical section \ref{sec:theory} with all the background information,
divided into various subsection devoted to various theoretical aspects, including the details of the Nested Bethe Ansatz 
diagonalization technique and the algebraic structure of the quasiparticle spectra for a family of quantum spin chains.
In Sec.~\ref{sec:semiclassical} we elaborate on the semiclassical limit of integrable models and discuss related concepts.
In Sec.~\ref{sec:conclusion} we summarize our results and propose areas for future exploration.

\section{Overview of methods and results}
\label{sec:transport}

We aim to characterize charge dynamics in a Hamiltonian dynamics invariant under a non-abelian continuous symmetry group $G$.
We begin our presentation by first outlining the general setting and introducing the key quantities of the linear transport theory.
We shall provide compact spectral representations for the charge diffusion constant in terms of quasiparticle spectra,
following largely previous works on the subject \cite{DeNardis2018,Gopalakrishnan18,DeNardis_SciPost,GV19,NMKI19}.

We specialize to simple Lie groups $G$ of rank $r$, generated by Lie algebra $\mathfrak{g}$.
Owing to the Noether theorem, the system possess local conserved (Noether) charges,
\begin{equation}
Q^{(\sigma)}=\int \dd x\, q^{(\sigma)}(x),
\label{eqn:Noether_charge}
\end{equation}
with local densities $q^{(\sigma)}(x)$, one per each hermitian generator ${\rm X}^{\sigma}$ of Lie algebra $\mathfrak{g}$.
For simplicity of notation we shall not make explicit distinction between lattice and continuum models, i.e. in lattice model spatial 
integration in Eq.~\eqref{eqn:Noether_charge} should be understood as a discrete summation.

Our first objective is to obtain closed-form expressions for transport coefficients in thermal equilibrium. With no loss of generality
we can specialize exclusively to grand-canonical Gibbs ensembles (at inverse temperature $\beta$), corresponding to density matrices
of the form
\begin{equation}
\varrho_{\beta,{\bf h}} = \frac{1}{\mathcal{Z}_{\beta,\boldsymbol{h}}} \exp{\left[-\beta\,H+\sum_{i=1}^{r}h_{i}Q^{(i)}\right]},
\label{eqn:gc_ensemble}
\end{equation}
with normalization (partition function) $\mathcal{Z}_{\beta,\boldsymbol{h}}\equiv {\rm Tr}\,\varrho_{\beta,{\bf h}}$.
Parameters ${\bf h}\equiv \{h_{i}\in \mathbb{R}\}$ are the $\U(1)$ chemical potentials which have been assigned to a maximal set
of commuting (Cartan) charges $Q^{(i)}$ with $i \in \mathcal{I}_{r}\equiv \{1,2,\ldots,r\}$.

The presence of finite chemical potentials ${\bf h}$ allows us to study transport at generic values of the charge densities.
Their values are of profound importance for the transport phenomena. By adopting generic non-vanishing chemical potentials.
When all $h_{i}\neq 0$, the ensembles \eqref{eqn:gc_ensemble} explicitly violate the global symmetry of $G$, and one is left
with a residual symmetry of the maximal  abelian subgroup $T\equiv \U(1)^{r}\subset G$ (called the maximal torus).
The complete set of ${\rm dim}(\mathfrak{g})$ Noether charges can be accordingly separated into two sectors:
the `longitudinal' charges $Q^{(i)}$ assigned to the `unbroken' (Cartan) 
generators and the non-abelian set of `transversal' charges $Q^{(\sigma)}$ for $\sigma \notin \mathcal{I}_{r}$,
satisfying $[Q^{(i)},Q^{(\sigma \notin \mathcal{I})}]\neq 0$, associated to the `broken' generators.
In this work we are exclusively interested in emergent anomalous charge transport that arises in the limit $h_{i}\to 0$ where the 
distinction between longitudinal and transversal correlators disappears (since $G$-invariance of $\varrho_{\beta,{\bf h}}$ gets 
restored); it will be thus sufficient to focus exclusively to the longitudinal sector (see e.g. \cite{undular} for remarks about
the nature of transversal correlators).

\subsection{Generalized hydrodynamics}
\label{sec:GHD}

We now formulate the transport theory for integrable systems, known as generalized hydrodynamics 
\cite{PhysRevX.6.041065,PhysRevLett.117.207201} (cf. \cite{Doyon_notes} for an overview).
As our starting point, we begin by an \emph{infinite} tower of \emph{local}
(including quasilocal \cite{IMP15,QLreview}) conservation laws
\begin{equation}
\partial_{t}q_{k}(x,t) + \partial_{x}j_{k}(x,t) = 0.
\label{eqn:local_conservations_laws}
\end{equation}
For the time being $k \in \mathbb{N}$ is just a formal discrete mode index. Although the complete set of (quasi)local charges can
be constructed explicitly with aid of the Algebraic Bethe Ansatz techniques \cite{IMP15},
recently adapted in to construct the associated currents \cite{BPP20,Pozsgay_algebraic},
this step can be in practice circumvented as long as one operates at the level of thermal averages.

The GHD formalism provide an explicit prescription to describe large spatio-temporal variations of thermally averaged conservation
laws state by Eqs.~\eqref{eqn:local_conservations_laws}. The key ingredient is to express expectation values of the current 
densities as functionals of the charge averages. There are various meaningful choices for the thermodynamic state functions one can consider. A particularly useful one is to use quasiparticle rapidity densities $\rho_{A}(\theta)$, where the `type' label $A$ runs
over the entire model-specific (thermodynamic) quasiparticle content, and $\theta$ is the corresponding rapidity variable parametrizing
their bare momenta, that is $p=p(\theta)$. An infinite collection of functions $\{\rho_{A}(\theta)\}$ uniquely specifies
a \emph{macrostate}, representing unbiased microcanonical ensembles of locally indistinguishable thermodynamic eigenstates.
By exploiting a one-to-one correspondence between the expectation  values of the (quasi)local conservation laws and quasiparticle 
content in an equilibrium macrostate \cite{PhysRevLett.115.157201,StringCharge}, it proves useful to recast 
Eq.~\eqref{eqn:local_conservations_laws} as a continuity equation for the quasiparticle densities
\begin{equation}
\partial_{t}\rho_{A}(\theta;x,t) + \partial_{x}j_{A}(\theta;x,t) = 0.
\end{equation}
The quasiparticle current densities take a simple factorized form at the Euler scale~\cite{PhysRevX.6.041065,PhysRevLett.117.207201}
\begin{equation}
j_{A}(\theta) = v^{\rm eff}_{A}(\theta)\rho_{A}(\theta).
\end{equation}
Here $v^{\rm eff}_{A}$ are state-dependent effective group velocities which determined from the dressed quasiparticles dispersions
\begin{equation}
v^{\rm eff}_{A}(\theta) = \frac{\partial_{\theta}\varepsilon_{A}(\theta)}{\partial_{\theta}p_{A}(\theta)}.
\label{eqn:effective_velocities}
\end{equation}

For a simple Lie group $G$ of rank $r$, elementary quasiparticle excitations (defined with respect to a reference ferromagnetic order 
parameter) come in exactly $r$ different types. We shall call these `flavors'. In addition to that, quasiparticles participate in 
formation of bound states, a quasiparticle that carries $s$ quanta of flavor $a$ is accordingly assigned a pair of quantum numbers 
$A=(a,s)$. All of the models we consider have infinitely many species of bound states, i.e., $s$ takes values in an
infinite countable set (typically ranging from $1$ to $\infty$).

Expanding above a reference equilibrium densities $\rho_{A}(\theta)$, that is
$\rho_{A}(\theta;x,t)=\rho_{A}(\theta)+\delta \rho_{A}(\theta;x,t)$, the hydrodynamic evolution of density (or charge) fluctuations
$\delta \rho_{A}(\theta;x,t)$ on Euler scale is encoded in the linear propagator (flux Jacobian)
$\mathbf{A}=\partial \mathbf{j}/\partial \boldsymbol{\rho}$, reading
\begin{equation}
\partial_{t} \delta \boldsymbol{\rho}(x,t) + \partial_{x}(\mathbf{A} \delta \boldsymbol{\rho}) = 0,
\label{eqn:GHD_fluctuations}
\end{equation}
where (for compactness of presentation) we have employed the tensor notation by flattening the quasiparticle and rapidity labels. 
Equation \eqref{eqn:GHD_fluctuations} can be diagonalized by performing a basis transformation
\begin{equation}
\delta \boldsymbol{\phi} = \boldsymbol{\Omega}[{\bf n}]\,\delta \boldsymbol{\rho},
\end{equation}
depending (non-linearly) on Fermi functions ${\bf n}$ of the underlying equilibrium state, where $\delta \boldsymbol{\phi}$
are the \emph{normal modes} of GHD (defined uniquely modulo normalization of static covariances).
In the basis of normal modes, the propagator $\mathbf{A}$ acts diagonally, with eigenvalues given by the
effective velocities \eqref{eqn:effective_velocities}, that is
\begin{equation}
\boldsymbol{\Omega}\,\mathbf{A}\,\boldsymbol{\Omega}^{-1} = \mathbf{v}^{\rm eff}.
\end{equation}
The physical interpretation of $\boldsymbol{\Omega}[{\bf n}]$ becomes transparent in the formalism of the Thermodynamic Bethe 
Ansatz, where it plays the role of a dressing operator for conserved quantities,
\begin{equation}
\mathbf{q}^{\rm dr} = \boldsymbol{\Omega}[{\bf n}]\,\mathbf{q}.
\end{equation}

\subsection{Diffusion constants}
\label{sec:diffusion_constants}

We shall now introduce the linear transport coefficients. Here we define them based on the asymptotic behavior
of dynamical structure factors (or dynamical charge susceptibilities).
For our intents it will be sufficient to only consider the Cartan charges and introduce the 
associated $r$-dimensional matrix of \emph{dynamical} susceptibilities
\begin{equation}
C^{i j}(x,t)\equiv \langle q^{(i)}(x,t)q^{(j)}(0,0) \rangle,
\end{equation}
where bracket $\langle \bullet \rangle$ designates the \emph{connected} correlator evaluated in a grand-canonical Gibbs ensemble given 
by Eq.~\eqref{eqn:gc_ensemble}. Static susceptibilities are accordingly given by the time-invariant sum rule
$C^{i j}=\int \dd x\,C^{i j}(x,t)$.

For generic values of the Cartan chemical potentials $h_{i}$, the variance of the structure factor in integrable models experiences 
ballistic spreading,
\begin{equation}
\int \dd x\,x^{2}\,C^{i j}(x,t) \simeq \mathcal{D}^{i j}\,t^{2/z},
\label{eqn:ballistic_scaling}
\end{equation}
signalled by the ballistic dynamical exponent $z=1$ and charge Drude weights $\mathcal{D}^{i j}$ \cite{CZP95,Zotos99,Ilievski12}.
The charge Drude weights admit the following mode resolution \cite{DS17,IN17}
\begin{equation}
\mathcal{D}^{i j} = \sum_{A}\!\int \dd \theta \rho_{A}(\theta)\ol{n}_{A}(\theta)
\big(v^{\rm eff}_{A}(\theta)\big)^{2}q^{(i){\rm dr}}_{A}q^{(j){\rm dr}}_{A}.
\label{eqn:charge_Drude}
\end{equation}
Here $\ol{n}_{A}(\theta)\equiv 1-n_{A}(\theta)$ denote thermal Fermi occupation functions associated with vacancies (holes).
All the thermodynamic quantities in the integrand of Eq.~\eqref{eqn:charge_Drude} depend on temperature and chemical
potentials.

Drude weights do not provide complete information about the late-time relaxation of $C^{i j}(x,t)$.
To deduce the asymptotic behavior of $C^{i j}(x,t)$ on a \emph{sub-ballistic} scale, we adopt the kinetic theory framework of
Refs.~\cite{Gopalakrishnan18,GV19}. One normally envisions a thermodynamic system divided up into large fluid cells of size $\ell$, 
with each cell being approximately in local thermal equilibrium. In a hydrodynamic description,
both the dressed charges $q^{(i){\rm dr}}_{A}$ and local chemical potentials $h_{i}$ get promoted to dynamical quantities which, in a 
macroscopic macroscopic fluid cell of length $\ell$ exhibit thermal fluctuations of the order $\mathcal{O}(\ell^{-1/2})$, which 
will, in turn, lead to diffusion.

In generic local equilibrium states, diffusion is a subleading correction to the ballistic transport characterized by the Drude weight. 
However, in \emph{unpolarized} thermal ensembles in systems with non-abelian symmetries, the charge Drude weight vanishes and the 
leading transport behavior is sub-ballistic. Charge Drude weights are proportional to \emph{dressed} charges $q^{(i){\rm dr}}_{A}$ 
carried by quasiparticles, cf. Eq.~\eqref{eqn:charge_Drude}. The latter are quite different from their bare (quantized) values 
$q^{(i)}_{A}$ and depend non-trivially on chemical potentials of the background equilibrium state, including
the $\U(1)$ chemical potentials $h_{i}$. In unpolarized thermal states that exhibit full invariance under $G$, the dressed 
quasiparticle charges vanish simply by symmetry under $G$ (see, e.g., Refs.~\cite{IN_Drude,IN17,Vir20}): one can see this as the 
screening of the quasiparticle charge by the thermal environment. Therefore, the charge Drude weight~\eqref{eqn:charge_Drude} vanishes.

The leading response therefore occurs at the diffusive scale, where one treats the chemical potentials as dynamically fluctuating 
quantities, with fluctuations that are suppressed by the hydrodynamic scale, ${\cal O}(\ell^{-1/2})$. For sufficiently small $h_i$ the 
quasiparticles carry dressed charges linearly proportional to $h_i$, i.e., they behave paramagnetically.
Fluctuations of chemical potentials induce fluctuations of thermally dressed charges in accordance with
\begin{equation}
q^{(i){\rm dr}}_{A}q^{(j){\rm dr}}_{A}
= \frac{1}{2}\sum_{k,l}\frac{\partial^{2}(q^{(i){\rm dr}}_{A}q^{(j){\rm dr}}_{A}\big)}
{\partial h_{k}\partial h_{l}}\Big|_{{\bf h}\to {\bf 0}}h_{k}h_{l} + \ldots.
\end{equation}
Notice that that chemical potentials can be simply related to densities of the Cartan charges via
\begin{equation}
h_{k}h_{l}=\sum_{k,l}\big(C^{-1}\big)^{k i}\big(C^{-1}\big)^{l j} q^{(i)}q^{(j)}.
\end{equation}
Here $C^{ij} = \partial q^{(i)}/\partial h_{j}=(\partial^{2}/\partial h_{i}\partial h_{j})\log \mathcal{Z}$ are static charge 
susceptibilities, and it will be helpful below to express them in terms of a mode expansion~\cite{DS17,IN17} analogous
to Eq.~\eqref{eqn:charge_Drude}:
\begin{equation}
C^{ij} = \sum_A \int d\theta \rho_A(\theta) \ol{n}_A(\theta) q_A^{(i)\mathrm{dr}} q_A^{(j)\mathrm{dr}}.
\label{cij}
\end{equation}
The latter also determine the magnitude of charge fluctuations, $\langle q^{(i)}q^{(j)} \rangle = C^{i j}/\ell$.
By combining these two results, thermal fluctuations of dressed charges (or dressed susceptibilities) carried by individual 
quasiparticle modes can be expressed in the form
\begin{equation}
\langle q^{(i){\rm dr}}_{A}q^{(j){\rm dr}}_{A} \rangle_{{\bf h} \to {\bf 0}} = \frac{\Upsilon^{i j}_{A}}{\ell},
\label{eqn:dressed_fluctuations}
\end{equation}
where
\begin{equation}
\Upsilon^{i j}_{A} = \frac{1}{2}\sum_{k,l}\big(C^{-1}\big)^{k l}
\frac{\partial^{2}(q^{(i){\rm dr}}_{A}q^{(j){\rm dr}}_{A})}{\partial h_{k} \partial h_{l}}\Big|_{{\bf h}\to {\bf 0}}
\!\! = C^{i j}\Upsilon_{A}.
\end{equation}
Physically, $\Upsilon^{i j}_{A}$ can be interpreted as an effective paramagnetic moment, assumes non-trivial dependence on both 
quasiparticle quantum numbers $A=(a,s)$.

In diffusive dynamics, the variance of the dynamical structure factors $C^{ij}$ which experiences linear growth at late times,
\begin{equation} \label{eqConductivity}
\int \dd x\,C^{i j}(x,t) \simeq 2\, \sigma^{i j}\,t,
\end{equation}
characterized by charge conductivity matrix
\begin{equation}
\sigma = D\,C,
\end{equation}
with static susceptibility matrix $C$ and charge diffusion matrix $D$ \footnote{Note that the factor of 2 in Eq.~\eqref{eqConductivity} follows from the familiar scaling of the variance $\delta x^2 = 2 D t$ of a diffusing particle in one dimension.}.
A full expression for the conductivity (Onsager) matrix $\sigma$ has been derived in \cite{DeNardis2018,DeNardis_SciPost} 
using the form-factors approach, and its diagonal part in~\cite{Gopalakrishnan18} using a kinetic 
theory formulation. Here we provide a compact expression for $D$ in models of higher-rank symmetry (restricted to the Cartan sector),
specializing to the limit of vanishing chemical potentials, ${\bf h}\to {\bf 0}$.
To this end, we substitute the fluctuation relation \eqref{eqn:dressed_fluctuations} into
Eq.~\eqref{eqn:ballistic_scaling} with Eq.~\eqref{eqn:charge_Drude} on a characteristic scale $\ell$ set by quasiparticles'
effective velocities, $\ell = |v^{\rm eff}_{A}(\theta)|t$~\cite{GV19}, yielding the following spectral resolution
of the conductivity matrix
\begin{equation}
\sigma^{i j}({\bf h}={\bf 0})
= \frac{1}{2} \sum_{A}\!\int\! \dd \theta\,\rho_{A}(\theta)\ol{n}_{A}(\theta)|v^{\rm eff}_{A}(\theta)|\Upsilon^{ij}_{A}.
\label{eqn:diffusion_constants}
\end{equation}
The diffusion matrix is thus proportional to the identity, corresponding to a single value of charge diffusion constant
\begin{equation}
\boxed{
D=\frac{1}{2} \sum_{A}\int \dd \theta\,\rho_{A}(\theta)\ol{n}_{A}(\theta)|v^{\rm eff}_{A}(\theta)|\Upsilon_{A}.
}
\end{equation}

Quantities $\Upsilon^{i j}_{A}$ can be easily computed explicitly at infinite temperature in various integrable spin chains.
For instance, in the ${\rm A}_{n-1}\equiv \SU(n)$ ferromagnetic integrable chains (with onsite degrees of freedom
in the fundamental irreducible representation $\mathcal{V}_{\square}$ of dimension ${\rm dim}\mathcal{V}_{\square} = n+1$),
static susceptibility matrix $C_{{\rm A}_{n}}$ evaluated at ${\bf h}\to {\bf 0}$,
written in the non-orthonormal basis $\rH^{i}$ of Cartan subalgebra $\mathfrak{t}$
(with Killing metric ${\rm Tr}(\rH^{i}\rH^{j}) = (\kappa_{{\rm A}_{n}})^{ij}$, cf. appendix \ref{app:Lie} for details and conventions), 
reads $C_{{\rm A}_{n}}=\kappa_{{\rm A}_{n}}/(n+1)$, and with aid of prescription \eqref{eqn:dressed_charges_from_logY} we find 
$\Upsilon^{{\rm A}_{1}}_{s}=\frac{4}{9} (s+1)^{4}$, $\Upsilon^{{\rm A}_{2}}_{a,s} = \tfrac{1}{16}(1 + s)^{2} (2 + s)^{2}$,
while for $n \geq 4$ it depends on the flavor label $a\in \{1,2,\ldots,n-1\}$. Its precise form will not matter for our purposes, but 
note the large-$s$ scaling $\Upsilon_{A} \sim s^{4}$.

By adopting quasiparticle densities as variational objects (see appendix \ref{app:magic} for the derivation), and comparing
the result to Eq.~\eqref{eqn:diffusion_constants}, one finds the following relation between the dressed charge fluctuations and the 
dressed differential scattering phases $K^{\rm dr}_{AA^{\prime}}$,
\begin{equation}
\langle q^{(i){\rm dr}}_{a,s}q^{(i){\rm dr}}_{a,s} \rangle_{{\bf h}\to {\bf 0}}
= \frac{C^{i i}}{\ell}\left[\sum_{a'} q^{(i)}_{a'} \! \lim_{s^{\prime} \to \infty}s^{\prime} K^{\rm dr}_{as, a's'}\right]^2,
\end{equation}
which rewards us with the following non-trivial identity
\begin{equation}
\Upsilon_{a,s} = \left[\sum_{a'} q^{(i)}_{a'} \! \lim_{s^{\prime}\to \infty} s^{\prime} K^{\rm dr}_{as,a's'}\right]^{2}.
\label{eqn:MagicFormula}
\end{equation}
We remark that this expression can thought of as a generalization of the so-called `magic formula' of Ref.~\cite{NMKI19}, which was 
originally introduced as a way to relate the different expressions of the diffusion constant of the XXZ spin chain with $\Delta >1$ 
obtained from form factors~\cite{DeNardis_SciPost}, and from a kinetic theory argument~\cite{GV19} similar to the one we used above.

\subsection{Emergence of superdiffusion}
\label{sec:superdiffusion}

From the considerations above, one might expect to find normal diffusion with $z = 2$, with diffusion
constant given by Eq.~\eqref{eqn:diffusion_constants}. Explicit computation shows, however,
that this result diverges as $h_{i}\to 0$ \cite{Ilievski18, GV19}. This divergence has been catalogued for various specific cases;
here we will explain why it holds in general. To this end we define the regularized diffusion constant by truncating the spectral sum~\eqref{eqn:diffusion_constants} at some string index $s_*$,
\begin{equation}
D_{*} = \sum_{a = 1}^r \sum_{s = 1}^{s_{*}} D_{a,s},
\label{regdiff}
\end{equation}
where $D_{a,s} \equiv D_A$ corresponds to each summand in Eq.~\eqref{eqn:diffusion_constants}. As a finite sum, Eq.~\eqref{regdiff}
is manifestly finite. 
If $D_{*}$ converges as $s_{*} \to \infty$ then one has regular diffusion. Suppose instead (i) that $D_{*}$ diverges as
$D_{*}\sim (s_{*})^{\upsilon}$. This already signals superdiffusive transport. We further posit (ii) that the time-dependent diffusion constant $D(t) \sim t^\alpha$ with $0 < \alpha \leq 1$ and (iii) that the dressed velocities of quasiparticles fall off at large $s$ as
$v^{\rm eff}_{s}\sim s^{-\nu}$ with exponent $\nu > 0$. We now derive scaling relations among these exponents. We note first that the 
distance over which a spin packet has spread in time $t$ is defined by $x_{\rm a}(t) \sim \sqrt{D(t) t} \sim t^{(1+\alpha)/2}$,
thus the dynamical exponent $z \equiv 2/(1+\alpha)$. At time $t$ we can divide quasiparticles into ``light'' quasiparticles for which 
$v_s t \sim t/s^\nu > x_{\rm a}(t)$ and ``heavy'' quasiparticles for which $v_s t < x_{\rm a}(t)$. The natural crossover scale at time 
$t$ is thus set by $t/s_*^\nu \sim t^{(1+\alpha)/2}$, so $s_* \sim t^{(1 - \alpha)/2\nu}$. Combining this with assumption~(i) we 
conclude that
\begin{equation}
D(t) \sim t^\alpha \sim t^{\upsilon(1-\alpha)/2\nu} \quad \Longrightarrow \quad z = \frac{\upsilon+2\nu}{\upsilon+\nu}.
\end{equation}

In the isotropic Heisenberg chain, analyzed in ref.~\cite{GV19}, one has specifically $\upsilon = \nu = 1$ and hence $z=3/2$.
This type of reasoning appears to suggest that more generally a larger set of anomalous dynamical exponents could be realized
in integrable models by appropriate tuning of $\nu$ and $\upsilon$. However, $\nu$ and $\upsilon$ are \emph{not} truly independent
and unrestricted parameters in integrable models, since $\upsilon$ depends on the scaling behavior of Eq.~\eqref{regdiff}, which in 
turn depends on $\nu$ through Eq.~\eqref{eqn:diffusion_constants}.

This imposes quite stringent and general constraints on $\upsilon$, as follows. Notice that as ${\bf h} \rightarrow {\bf 0}$ the static 
charge susceptibilities $C^{ij}$, given by Eq.~\eqref{cij}, must approach a constant value. At finite ${\bf h}$, the occupation factor 
for a quasiparticle specie $a$ of type $s$ (a bound state of $s$ magnons) is suppressed with a factor of $\exp(-s\,h_{a})$,
and thus quasiparticles with $s\,h_{a} \gg 1$ do not contribute to the susceptibility at chemical potential $h_a$.
The label `$a$' plays no role in the following argument, and so we will drop it. Writing uniformly $h_{a}\to h$,
truncating the sum~\eqref{cij} at $s \sim 1/h$, and using the paramagnetic behavior of the dressed
charges, we have $C^{ij} \sim h^2 \sum_{a=1}^{r}\sum_{s< 1/h} C^{ij}_{a,s}$. Requiring that the $C^{ij}$ have a nonvanishing limit
as $h \to 0$, we see that the sum over $s$ has to scale as $\sim s^2$, so the summand $C^{ij}_{a,s}$ must grow linearly in $s$.
The summands in Eqs.~\eqref{cij} and~\eqref{eqn:diffusion_constants} are the same, except for a factor of
$|v_s| \sim s^{-\nu}$. Therefore, the diffusion constant must scale as $D_* \sim \sum_{s \leq s^*} s^{1-\nu} \sim s^{2 - \nu}$, 
yielding the scaling relation
\begin{equation}
\upsilon + \nu = 2.
\end{equation}
Finally, the analytic structure of Bethe ansatz solutions implies that exponent $\nu \geq 0$ is integer-valued. The three remaining 
possibilities are ballistic transport ($\nu = 0$), which is ruled out by the vanishing Drude weight discussed above;
KPZ scaling ($\nu = 1$), which we will argue is generic; and diffusion with possible logarithmic corrections ($\nu \geq 2$), which 
occurs in certain fine-tuned models~\cite{NMKI20} where the single-magnon dispersion is fine-tuned to scale as $\omega \sim k^3$
(or slower) at long wavelengths $k\to 0$.

In what follows we will establish $z = 3/2$ within a universal algebraic description of the thermodynamic dressing 
equations and link them to the underlying symmetries structures and representation theory of quantum groups (Yangians).
We find, remarkably, that the Fermi functions assume universal algebraic scaling at large-$s$,
\begin{equation}
\boxed{
n_{a,s} \sim \frac{1}{s^{2}}
},
\end{equation}
which comes out as a direct corollary of fusion identities amongst the quantum character associated to
the Yangian symmetry~\cite{FRT1988,Drinfeld1988,Drinfeld1990,Bernard93,MacKay05}.
Similarly, we find that the total state densities and the dressed velocities of giant magnons (when multiplied
by regular function and integrate over the rapidity domain, cf. Sec.~\ref{sec:NewSection2}) decay as
\begin{equation}
\boxed{\rho^{\rm tot}_{a,s}  = \frac{\rho_{a,s}}{n_{a,s}} \sim \frac{1}{s}} \qquad {\rm and}\qquad
\boxed{v^{\rm eff}_{a,s} \sim \frac{1}{s}},
\end{equation}
respectively, for all flavors $a=1,\ldots,r$.
Most remarkably, these large-$s$ scaling properties hold irrespectively of the local on-site degrees of freedom (i.e.
finite-dimensional irreducible unitary representations of $\mathfrak{g}$); they can be understood as kinematic constraints stemming 
from the underlying quantum symmetry algebra. Finally, notice that $\Upsilon_{A} \sim s^4$ in conjunction with the above scaling 
relations implies $\upsilon=1$. The upshot is thus that the anomalous fractional algebraic dynamical exponent $z=3/2$ is deeply rooted
in the fusion relations for the quantum characters.

\section{Numerical analysis}
\label{sec:numerics}

We verify our predictions for the superdiffusive charge transport on a number of representative instances of integrable quantum spin 
chains invariant under simple Lie groups from the classical series. Specifically, we consider homogeneous spin-chain Hamiltonians
in terms of fundamental degrees of freedom given by Eqs.~\eqref{eqn:Hamiltonian}. Exceptional algebras are not included in our
analysis. We moreover leave out the $B$-series as the fundamental integrable $B_{2}\equiv \SO(5)$ chain has been already been studied 
numerically in a recent paper \cite{DupontMoore19}.

We employ numerical tensor-network based computations of the dynamical correlation function $C^{ii}(x,t)$
(as defined in Eq.~\eqref{eqn:KPZ_scaling_functions}) in the canonical Gibbs equilibrium at infinite temperature.
Owing to the non-abelian symmetry $G$ of both the Hamiltonian $H$ and infinite-temperature Gibbs density matrix, all the components
$i=1,2,\ldots,r$ are identical and we thus suppress the redundant index $i$.
We performed the time-evolving block-decimation (TEBD) algorithm in the Heisenberg 
picture~\cite{PhysRevLett.91.147902,PhysRevLett.93.207204,PhysRevLett.108.227206}, evolving the local charge density operator 
$q(0,t)$ in a matrix product operator (MPO) representation.

\paragraph*{Details.}
A fourth-order Trotter decomposition is used to propagate the operator 
forward in time-steps of size $\delta t$, followed by truncations at each bond keeping the largest $\chi$ Schmidt states.
The simulations are accelerated by taking advantage of time-reversal symmetry and translation symmetry to compute the
full correlator $C(x, t) = \braket{q(x, t/2) q(0,-t/2)}$ using a single MPO evolution; additionally, the TEBD implements the maximal 
abelian subgroup of the on-site symmetry group for efficiency. The computations are checked for convergence in $\delta t$ and
bond dimension $\chi$ (up to $\chi = 1024$).

Our TEBD scheme uses a Trotter step size of $\delta t = 0.4$. The operators are truncated initially with a constant discarded weight 
$\varepsilon=10^{-8}$, allowing the bond dimension $\chi$ to grow until it reaches a threshold $\chi_{\text{max}}=512$. Subsequent 
truncations keep only $\chi_{\text{max}}$ states. The simulations are checked for convergence in the step size $\delta t$, the 
truncation error $\varepsilon$, and the threshold bond dimension $\chi_{\text{max}}$. Additionally, the results shown here include 
rescaling of the correlations at each time to explicitly enforce the charge sum rule $\int \dd x\,C(x, t) = C(0, 0)$, which 
improves the convergence significantly.

\begin{figure}[ht]
    \vspace{0.5cm}
    \includegraphics[width=\linewidth]{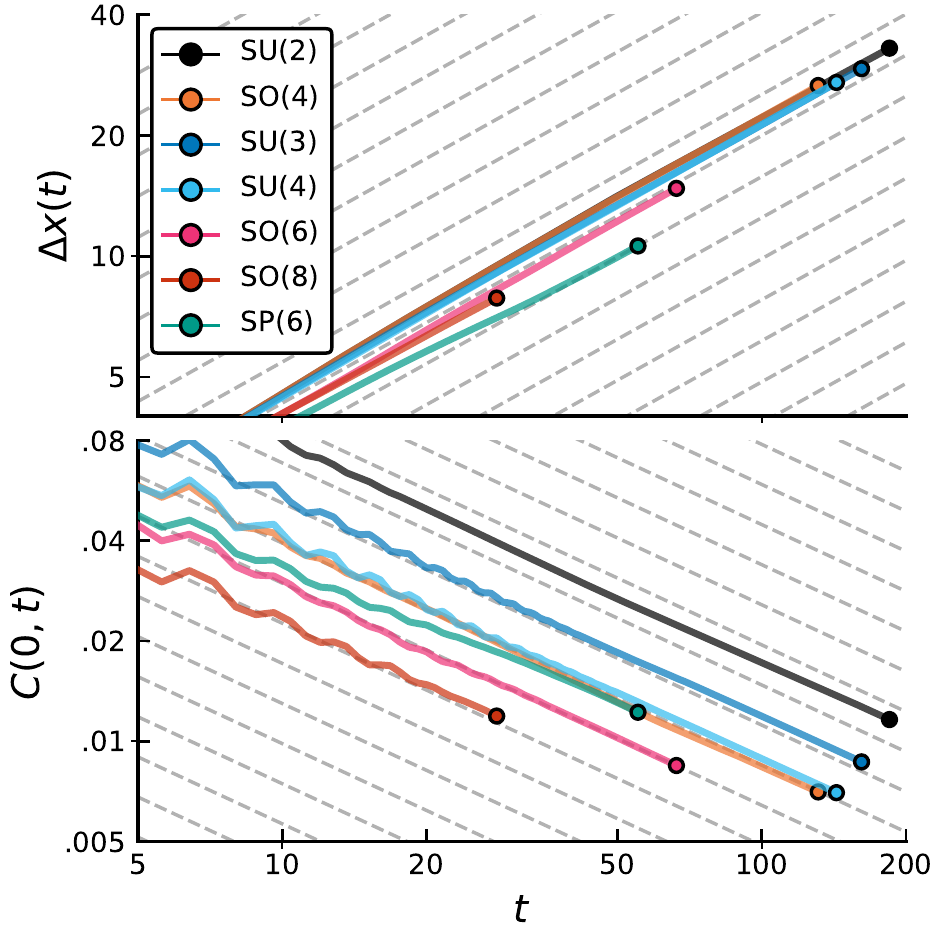}
    \caption{\label{fig:TEBD} The dynamical charge correlation functions computed with TEBD, showing asymptotic scaling
	with dynamical exponent $z=3/2$ for each of the integrable chains, measured by (top) the width $\Delta x(t)$ of the 		
	expanding charge profiles and (bottom) the return probability $C(0,t)$. Background dashed lines show
	$\Delta x \propto t^{2/3}$ and $C(0,t) \propto t^{-2/3}$.}
	\label{fig:all_tebd}
\end{figure}

\begin{figure}[ht]
    \vspace{0.5cm}
    \includegraphics[width=\linewidth]{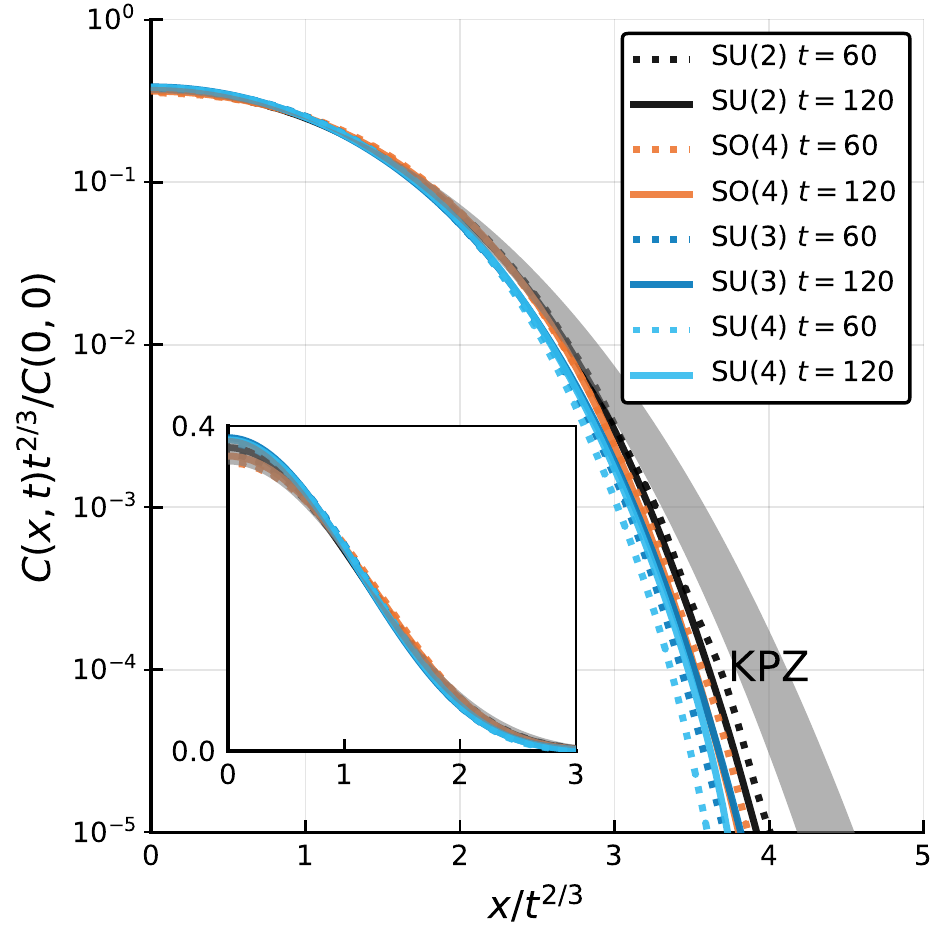}
    \caption{\label{fig:TEBDcollapse} The asymptotic charge correlator profiles rescaled using the dynamical exponent $z=3/2$ collapse. The highlighted region shows the KPZ scaling function for a range of widths fit to these profiles --- the fit fails, as the tails of the scaling function appear to fall off faster than in KPZ scaling. The inset shows the same plot with a linear axis.}
    \label{fig:sun_profilecollapse}
\end{figure}

\paragraph*{Results.}
Our main results are succinctly summarized in Fig.~\ref{fig:TEBD} where we display the dynamical width of the charge profiles
\begin{equation}
[\Delta x(t)]^{2} = \frac{\int \dd x\,x^{2} C(x, t)}{\int \dd x\, C(x, t)}.
\end{equation}
We find clear signature of an asymptotic power-law growth $ \Delta x(t) \propto t^{1/z}$, and the return probability shows
power-law decay $C(0, t) \propto t^{-1/z}$, with numerically estimated dynamical exponent $z = 3/2$ with great accuracy.
The scaling collapse in Fig.~\ref{fig:TEBDcollapse} shows the correlators obey the asymptotic scaling form
$C(x,t) \simeq t^{-1/z} f_{\rm sc}(x\,t^{-1/z})$. Comparing $f_{\rm sc}$ to the KPZ scaling function $f_{\rm KPZ}$ given by
Eq.~\eqref{eqn:KPZ_scaling_functions} we find some discernible deviations in the tails (including the basic $\SU(2)$ case studied
previously in \cite{Ljubotina19}). On accessible timescales, we are unable to infer whether this discrepancy persists at late times or 
is it merely due to transient effects.

\section{Theoretical background}
\label{sec:theory}

In the rest of the paper we provide the necessary technical details using the language and framework of the Algebraic Bethe Ansatz.
Our aim is mostly to give a concise description the quasiparticle content and a group-theoretical formulation of the TBA equations.
This will require us to briefly revisit the key notions from the theory quantum integrability.
Despite a vast body of work on the subject, we are not aware of a self-contained exposition of the Nested Bethe Ansatz (NBA)
techniques and its various mathematical underpinnings; even some of the fundamental results appear to be dispersed across
several specialized articles. We would thus like to use this opportunity to partly fill this gap and offer a comprehensive 
exposition aiming at physicists with interest in the field of many-body statistical physics out of equilibrium.

Our analysis crucial relies on several basic notions from the representation theory of quantum groups known in the literature
as \emph{Yangians}. Specifically, we shall rely on the knowledge of functional identities amongst their `quantum characters'.
The study of Yangian modules has been initiated by Kirillov and Resthetikhin in their seminal papers \cite{KR86,Kirillov89,KR90}.
Many key developments in this field are summarized in the review article by Kuniba et al. \cite{KNS11} which will be extensively used.

Concerning physics applications, it is worthwhile mentioning that quantum character formulae proved invaluable for
deriving the dilogaritm identities for the conformal central charges (that can be extracted from the low-temperature scaling limit
for a wide range of integrable quantum chains), see e.g. Refs.~\cite{Kirillo89,KN92,KNS93,KNS11}. We are nonetheless not aware
of any applications in the nonequilibrium physics department. Readers not interested in the details of the Bethe ansatz solutions to 
these models can feel free to skip this more technical section, although some of the scaling properties we derive in 
subsections~\ref{sec:NewSection1} and~\ref{sec:NewSection2} are central to our analysis;
the results derived in these two subsections are, to our knowledge, new.

For the most part we shall concentrate our analysis to quantum lattice models with isotropic interactions, adopting
finite-dimensional irreducible representations of non-abelian simple Lie algebras $\mathfrak{g}$ as
local degrees of freedom. These models can be viewed as higher-rank variants of ferromagnetic quantum chains that possess manifest 
global continuous symmetry of a Lie group $G$.

We shall make extensive use of the group-theoretical language. For a brief reminder on simple Lie algebras we refer
the reader to appendix \ref{app:Lie}, where we also fix some of our notations.
The complete classification of simple Lie algebras is a landmark achievement due to Cartan.
The list comprises four infinite families
\begin{align}
A_{r} &\equiv \mathfrak{su}(r+1,\mathbb{C}),&\qquad B_{r} &\equiv \mathfrak{so}(2r+1,\mathbb{C}),\nonumber \\
C_{r} &\equiv \mathfrak{sp}(2r,\mathbb{C}),&\qquad D_{r} &\equiv \mathfrak{so}(2r,\mathbb{C}),
\end{align}
commonly referred to as the \emph{classical series}, alongside five exceptional algebras
$\mathfrak{g}_{2}$, $\mathfrak{f}_{4}$, $\mathfrak{e}_{6}$, $\mathfrak{e}_{7}$, $\mathfrak{e}_{8}$.
While these exceptional cases can treated in parallel to the classical algebras, they will be of secondary interest for us.
We also remind that there are several exceptional isomorphisms amongst the low-rank algebras: $A_{1}\cong B_{1}\cong C_{1}$,
$D_{2}\cong A_{1} \oplus A_{1}$ and $A_{3}\cong D_{3}$.

\paragraph*{Irreducible representations.}

Finite-dimensional irreducible representations $\mathcal{V}_{\Lambda}$ of $\mathfrak{g}$ are specified by $r$ non-negative integer 
labels $m_{a}$, called \emph{Dynkin labels}. They are coordinates in the `$\omega$-basis' of the fundamental weight $\omega_{a}$.
We employ notation $\Lambda=[m_{1},m_{2},\ldots,m_{r}]$, or sometimes write simply $(\bf{n})$, referring to the dimension of 
$\mathcal{V}_{\Lambda}$. Writing $m_{i}$ inside the corresponding node of the Dynkin diagram we obtain a one-to-one
correspondence between the enumerated Dynkin diagrams and the finite-dimensional irreducible representations of
$\mathcal{V}_{\Lambda}$, as shown in Fig.~\ref{fig:Dynkin_labelled}. More more information we refer the reader
to appendix \ref{app:Lie}.

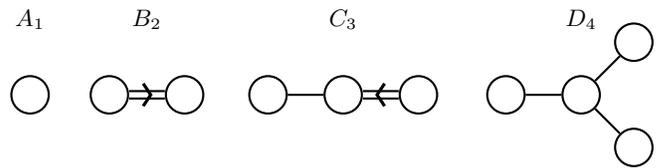
\begin{figure}[t!]
\begin{tikzpicture}
\begin{scope}
\node at (0,1) {$A_{1}$};
\node[circle,draw,thick, minimum size=0pt, inner sep = 5pt] (a) at (0,0) {};
\end{scope}
\begin{scope}[xshift=30pt]
\node at (0.5,1) {$B_{2}$};
\node[circle,draw,thick, minimum size=0pt, inner sep = 5pt] (a) at (0,0) {};
\node[circle,draw,thick, minimum size=0pt, inner sep = 5pt] (b) at (1,0) {};
\draw[-,thick,double distance = 2pt,postaction = {decorate}] (a) -- (b);
\draw[-,very thick] (0.45,0.15) -- (0.55,0) -- (0.45,-0.15);
\end{scope}
\begin{scope}[xshift=90pt]
\node at (1,1) {$C_{3}$};
\node[circle,draw,thick, minimum size=0pt, inner sep = 5pt] (a) at (0,0) {};
\node[circle,draw,thick, minimum size=0pt, inner sep = 5pt] (b) at (1,0) {};
\node[circle,draw,thick, minimum size=0pt, inner sep = 5pt] (c) at (2,0) {};
\draw[-,thick] (a) -- (b);
\draw[-,thick,double distance = 2pt] (b) -- (c);
\draw[-,very thick] (1.55,0.15) -- (1.45,0) -- (1.55,-0.15);
\end{scope}
\begin{scope}[xshift=180pt]
\node at (1,1) {$D_{4}$};
\node[circle,draw,thick, minimum size=0pt, inner sep = 5pt] (a) at (0,0) {};
\node[circle,draw,thick, minimum size=0pt, inner sep = 5pt] (b) at (1,0) {};
\node[circle,draw,thick, minimum size=0pt, inner sep = 5pt] (c) at (1+0.7,0.7) {};
\node[circle,draw,thick, minimum size=0pt, inner sep = 5pt] (d) at (1+0.7,-0.7) {};
\draw[-,thick] (a) -- (b);
\draw[-,thick] (b) -- (c);
\draw[-,thick] (b) -- (d);
\end{scope}
\end{tikzpicture}
\caption{Several representative examples of Dynkin diagrams, displaying the simplest one from each of the classical series.
Node $a$ and $b\neq a$ share a bond if $\mathcal{K}_{ab}\neq 0$. The number of bonds between adjacent nodes equals
$-\mathcal{K}_{ab}$, corresponding to different angles between simple roots. For non-simply-laced algebras, the arrow
is pointing towards the short root.}
\label{fig:Dynkin_empty}
\end{figure}

To every irreducible representation of the highest weight we can also bijectively associate a Young diagram (partition).
The latter consists of $r$ rows, with each $\ell_{a}$ boxes in each row such that moving from top to bottom the number of boxes never 
increases. The upper row contains $\ell_{1}=\sum_{i=1}^{r}m_{i}$ boxes, while in every subsequent row the number is lowered by 
successively subtracting Dynkin labels, namely $\ell_{a}=\sum_{i=1}^{a}m_{i}$. This means that the $a$th fundamental weight 
$\omega_{a}$ has a single non-zero Dynkin label $1$ in the $a$th node. The associated diagram is a single column with $a$ boxes.
It is important to stress that rectangular irreducible representations play a distinguished role as the constitute a closed set
of fusion relations. In the context of quantum integrable models solvable by the Bethe Ansatz, rectangular diagrams indeed
correspond to distinct quasiparticles species that can be excited in the thermodynamic eigenstates.

\subsection{Integrable lattice ferromagnets}
\label{sec:models}

We now proceed with the algebraic foundations of the Bethe Ansatz. We shall describe two complementary principles which
underlie quantum integrability. From a formal perspective, integrability arises from solutions to the quantum
\emph{Yang--Baxter equation} (YBE) \cite{Korepin_book}
\begin{equation}
\mathbf{R}_{12}\mathbf{R}_{13}\mathbf{R}_{23} = \mathbf{R}_{23}\mathbf{R}_{13}\mathbf{R}_{12}.
\label{eqn:YBE}
\end{equation}
Here matrices $\mathbf{R}_{ij}=\mathbf{R}_{ij}(u_{i},u_{j})$ act non-trivially on two copies of vector spaces $\mathcal{V}_{i}$ and 
$\mathcal{V}_{j}$ (irreducible $\mathfrak{g}$-modules of Lie algebra $\mathfrak{g}$) and depend additionally on a pair complex 
parameters $u_{i,j} \in \mathbb{C}$. Geometrically speaking, equation \eqref{eqn:YBE} therefore establishes the equivalence of two
{\it a priori} distinct intertwining protocols on a three-fold space $\mathcal{V}_{1}\otimes \mathcal{V}_{2}\otimes \mathcal{V}_{3}$. 
In algebraic terms, $\mathbf{R}$-matrices provide `structure constants' for the defining relations of quasi-triangular Hopf algebras,
widely known in the literature as \emph{quantum groups} \cite{FRT1988,Drinfeld1988,Drinfeld1990,Kassel_book}.

For every simple Lie algebra $G$, the YBE admits a class of $G$-invariant $R$-matrices $\mathbf{R}_{ij}(u_{i},u_{j})$ acting on
$\mathcal{V}_{i}\otimes \mathcal{V}_{j}$, which are \emph{rational} functions of the difference parameter $u=u_{i}-u_{j}$.
The associated quantum algebras are known as \emph{Yangians} (see e.g. Refs.~\cite{FRT1988,Drinfeld1988,Kassel_book}).

Equation \eqref{eqn:YBE} has a suggestive physical meaning when viewed as an equivalence relation for a factorizable
three-body scattering processes, signifying that two apriori different three-body scattering events decompose in terms of sequential 
two-body scatterings in two equivalent ways. This powerful symmetry principle is intimately tied to presence of coherent long-lived 
interacting quasi-particles that undergo completely elastic scattering \cite{ZZ79}. In this respect, the variables $u_{i}$ parametrize
quasi-particle momenta.

\subsection{Commuting transfer matrices}
\label{sec:TM}

Integrability most prominently manifests itself through the existence of infinitely many commuting operators conserved under
time evolution. Let $\mathcal{H}=\mathcal{V}^{\otimes L}_{\Lambda_{\rm p}}$ denote the Hilbert space of a homogeneous 
quantum chain of length $L$, with an irreducible onsite representation $\mathcal{V}_{\Lambda_{\rm p}}$.
The \emph{row} transfer matrices $\mathbf{T}_{\Lambda_{\rm a}}(u) = {\rm Tr}_{\Lambda_{\rm a}}{\bf M}_{\Lambda_{\rm a}}(u)$
on $\mathcal{H}$, i.e. auxiliary traces of monodromies
\begin{equation}
{\bf M}_{\Lambda_{\rm a}}(u) = {\bf R}_{\Lambda_{\rm a},L}(u)\cdots {\bf R}_{\Lambda_{\rm a},2}(u){\bf R}_{\Lambda_{\rm a},1}(u),
\label{eqn:monodromy}
\end{equation}
provide the generating operators for an infinite family commuting charges.
Above we have used a standard notation ${\bf R}_{\Lambda_{\rm a},i}$ denote the embedded ${\bf R}$-matrices which operate non-trivially 
only on the auxiliary space $\mathcal{V}_{\Lambda_{\rm a}}$ and physical lattice site $i$.
Mutual commutativity for $u,u^{\prime}\in \mathbb{C}$ and for any two arbitrary finite-dimensional $\mathfrak{g}$-modules of the 
highest weight type $\Lambda_{\rm a}$ and $\Lambda^{\prime}_{\rm a}$, namely
\begin{equation}
\left[\mathbf{T}_{\Lambda_{\rm a}}(u),\mathbf{T}_{\Lambda^{\prime}_{\rm a}}(u^{\prime})\right] = 0,
\label{eqn:commutativity}
\end{equation}
follows as a direct corollary of (leaving dependence on $u$ implicit)
\begin{equation}
{\bf R}_{\Lambda_{\rm a},\Lambda^{\prime}_{{\rm a}}}{\bf M}_{\Lambda_{\rm a}}{\bf M}_{\Lambda^{\prime}_{{\rm a}}}
= {\bf M}_{\Lambda^{\prime}_{{\rm a}}}{\bf M}_{\Lambda_{\rm a}}{\bf R}_{\Lambda_{\rm a},\Lambda^{\prime}_{{\rm a}}}.
\end{equation}
The latter lifts the local property \eqref{eqn:YBE} to the entire many-body Hilbert space $\mathcal{H}$.

The simplest ${\bf R}$-matrices operates on two copies of fundamental $\mathfrak{g}$-modules corresponding to the one-box tableaux, 
$\mathcal{V}_{\omega_{1}}\equiv \mathcal{V}_{\square}$, and we accordingly denote them by ${\bf R}_{\square,\square}(u)$.
Other ${\bf R}$-matrices, acting in higher-dimensional (physical or auxiliary) representations,
can be systematically constructed via an appropriate fusion procedure. This procedure is described in detail in
e.g. \cite{KLWZ97,Zabrodin98}. Amongst ${\bf T}_{\Lambda_{\rm a}}(u)$, we find an infinite subset of transfer matrices
${\bf T}_{a,s}(u)$ with \emph{rectangular} auxiliary representations of the highest weight
$\Lambda_{s\,\omega_{a}}\equiv \Lambda_{a,s}$, with with Dynkin labels $\Lambda_{a,s}\equiv [s,s,\ldots,s]$
(depicted by $a\times s$ Young diagrams). From an algebraic perspective, they are distinguished by the property that they constitute
a closed set of functional fusion identities in the form of Hirota relations. On the other hand, their representation labels $(a,s)$,
with $1\leq a\leq r$ and $s\in \mathbb{N}$, bijectively enumerate the types of quasi-particle excitations in the their thermodynamic 
eigenstates.

\subsubsection*{Quantum ${\bf R}$-matrices}
\label{sec:Rmatrices}

In this section we outline the construction of commuting transfer matrices for homogeneous quantum chains with degrees of
freedom in the fundamental representation. We shall thus only operate with the fundamental ${\bf R}$-matrix
${\bf R}_{\square,\square}(u)$. For compactness of notation, we subsequently denote simply by ${\bf R}(u)$.

We begin by $\SU(n)$-symmetric ${\bf R}$-matrices which have originally appeared
in papers of Lai \cite{Lai1974} and Sutherland \cite{Sutherland1975}.
The fundamental ones, acting on the product of $\mathcal{V}_{\square}\cong \mathbb{C}^{n}$, read simply
\begin{equation}
{\bf R}(u) = u\mathbf{1} + \boldsymbol{\Pi},
\label{eqn:unitary_R}
\end{equation}
where
\begin{equation}
\boldsymbol{\Pi} = \sum_{i,j=1}^{n}{\bf E}^{ij}\otimes {\bf E}^{ji},\qquad
\boldsymbol{\Pi}^{2}=\mathbf{1},
\end{equation}
denote permutation operators acting in $\mathbb{C}^{n}\otimes \mathbb{C}^{n}$,
with unit (Weyl) matrices $({\bf E}^{ij})_{kl}=\delta_{i k}\delta_{j l}$.

The orthogonal and symplecic cousins of \eqref{eqn:unitary_R} have been given by Resthetikhin \cite{Kolya91}.
They contain an extra non-invertible element
$\boldsymbol{\Xi}=\boldsymbol{\Pi}^{\rm T_{1}}=\boldsymbol{\Pi}^{\rm T_{2}}$,
\begin{equation}
\boldsymbol{\Xi} = \sum_{i,j=1}^{n}{\bf E}^{ij}\otimes {\bf E}^{n+1-i,n+1-j},
\end{equation}
that satisfies
\begin{equation}
\boldsymbol{\Xi}^{2} = n\,\boldsymbol{\Xi},\qquad
\big[\boldsymbol{\Pi},\boldsymbol{\Xi}\big]=0,\qquad
\boldsymbol{\Pi}\,\boldsymbol{\Xi}=\boldsymbol{\Xi}.
\end{equation}
In particular, the $\rO(n)$-invariant ${\bf R}$-matrix acting on two fundamental (vector) representations has the structure
\begin{equation}
{\bf R}(u) = u\left(u + \frac{n-2}{2}\right)\mathbf{1}
+ \left(u+\frac{n-2}{n}\right)\boldsymbol{\Pi} - u\,\boldsymbol{\Xi}.
\end{equation}

Finally, in the case of symplectic groups $\Sp(2r)$ ($r\geq 2$), the $\Sp(2r)$-symmetric fundamental ${\bf R}$-matrix acting
in the two-fold product space $\mathbb{C}^{2r}\otimes \mathbb{C}^{2r}$ reads
\begin{equation}
{\bf R}(u) = u\left(u + k+1\right)\mathbf{1}
+ (u+k+1)\boldsymbol{\Pi} - u\,\widetilde{\boldsymbol{\Xi}},
\end{equation}
with a block-form matrix
\begin{equation}
\widetilde{\boldsymbol{\Xi}} = \begin{pmatrix}
0 & 0 & 0 & 0 \\
0 & \boldsymbol{\Xi} & -\boldsymbol{\Xi} & 0 \\
0 & -\boldsymbol{\Xi} & \boldsymbol{\Xi} & 0 \\
0 & 0 & 0 & 0
\end{pmatrix},
\end{equation}
satisfying
\begin{equation}
\big[\boldsymbol{\Pi},\widetilde{\boldsymbol{\Xi}}\big]=0,\qquad
\boldsymbol{\Pi}\,\widetilde{\boldsymbol{\Xi}}=-\widetilde{\boldsymbol{\Xi}}.
\end{equation}
All the above ${\bf R}$-matrices fulfil the unitary condition ${\bf R}(u){\bf R}(-u)=\gamma(u)\mathbf{1}$,
for some appropriate scalar function $\gamma(u)$.

\subsubsection*{Spin chain Hamiltonians}
\label{sec:Hamiltonians}

An infinite tower of mutually commuting Hamiltonians with $k$-site local densities $\mathbf{h}^{(k)}$ can be produced by 
differentiation of the logarithm of $\mathbf{T}_{\square}(u)$, evaluated at a special point $u_{\star}=0$
(where ${\bf R}(0)\simeq \boldsymbol{\Pi}$). The $(k-1)$-st Hamiltonian in the family has a two-body density $\mathbf{h}^{(k)}$,
corresponding to some $G$-symmetric interaction involving $k$ adjacent fundamental degrees of freedom
(with $\Lambda_{\rm p}=\square$),
\begin{equation}
\mathbf{H}^{(k+1)} = \frac{\dd^{k}}{\dd u^{k}}\log \mathbf{T}_{\square}(u)\Big|_{u = 0} = \sum_{j=1}^{L}\mathbf{h}^{(k+1)}_{j,j+1},
\label{eqn:Hamiltonian}
\end{equation}
for $k\geq 1$. We subsequently consider the Hamiltonians $\mathbf{H}\equiv \mathbf{H}^{(2)}$ with nearest neighbor interactions 
$\mathbf{h}\equiv \mathbf{h}^{(2)}$. Depending on $\mathfrak{g}$, the Hamiltonian densities $\mathbf{h}$ assume the following form
\begin{align}
\mathbf{h}_{\mathfrak{su}(n)} &\simeq \boldsymbol{\Pi},\\
\mathbf{h}_{\mathfrak{so}(n)} &\simeq \boldsymbol{\Pi }- \frac{2}{n-2}\boldsymbol{\Xi},\\
\mathbf{h}_{\mathfrak{sp}(2n)} &\simeq \boldsymbol{\Pi} - \frac{1}{n+1}\widetilde{\boldsymbol{\Xi}}.
\end{align}

\subsection{Nested Bethe Ansatz}
\label{sec:nested}

\paragraph*{Completely elastic scattering.}

The algebraic principles of quantum integrability can be neatly translated into the scattering theory language.
The many-body ${\bf S}$-matrix, albeit being non-trivial, fulfils extremely stringent constraints owing to infinitely many local 
conservation laws; they ensure only the usual total conservation of particle momenta, but also that the sets of incoming and outgoing 
momenta are identical. In other words, quasi-particles every pair-wise scattering retain their momenta and consequently the entire
many-body scattering matrix completely factorizes into a sequence of two-particle scattering events.

To demonstrate the basic principle, let us consider a one-dimensional system system enclosed in a finite compact region of space of 
circumference $L$. Requiring periodicity of the wave-function yields a simple constraint
\begin{equation}
e^{\ii p_{j}L}\prod_{j\neq k}^{L}S(p_{j},p_{k}) = 1,
\label{eqn:Bethe_scalar}
\end{equation}
providing the quantization condition for quasi-particle momenta $p_{j}$. Since scattering is purely elastic and there is no
particle production or decay, we only had to take into account that upon every collision a quasi-particle experiences a phase shift 
encoded in a momentum-dependent $\U(1)$ scattering amplitude $S(p_{j},p_{k})$.
Equation \eqref{eqn:Bethe_scalar} evidently resembles that of free quasiparticles, `corrected' by the net effect of
interparticle interactions.

For the class of model under consideration, we deal with a more general situation where quasi-particles carry additional quantum 
numbers coming from isotropic degrees of freedom. Condition \eqref{eqn:Bethe_scalar} has to be accordingly promoted to a matrix 
equation,
\begin{equation}
e^{\ii p_{j}L}\prod_{j\neq k}^{L}\mathbf{S}(p_{j},p_{k})\ket{\Psi} = \ket{\Psi},
\label{eqn:Bethe_nested}
\end{equation}
where now $\mathbf{S}$ denotes a scattering matrix, and $\ket{\Psi}$ a many-body wave-function.
By diagonalizing this resulting equation, the quantization condition takes the form of a coupled scalar equations known
as the nested Bethe Ansatz (NBA) equations (cf. \cite{Takahashi72,ZZ92,Hubbard_book,BR08}).

\subsubsection*{Bethe Ansatz diagonalization}

The task at hand is to simultaneously diagonalize the set of commuting transfer matrices. In distinction to free systems,
many-body eigenstates depend on solutions to the NBA equations \eqref{eqn:Bethe_nested}. There are various routes to tackle this 
problem, and we shall briefly outline two standard alternative methods for constructing the complete set of Bethe eigenstates.

The best known approach is the celebrated algebraic Bethe Ansatz which mirrors the familiar second-quantization construction.
Given a symmetry group $G$ of rank $r$, one can identify a set of $r$ quasiparticle creation operators $\mathbf{B}_{a}$,
one for each simple root of $\mathfrak{g}$. The discrete label $a$ can be regarded as a flavor index.
The operators $\mathbf{B}_{a}$ appear as off-diagonal elements of the fundamental monodromy matrix defined
in Eq.~\eqref{eqn:monodromy}, see for instance ref.~\cite{GLS17}. Identifying $a\leftrightarrow \alpha_{a}$,
the Bethe eigenstates assume the form
\begin{equation}
\ket{\{\theta^{(a)}_{j}\}} = \sum_{\{a\}}F^{(a_{1},\ldots,a_{N})}\prod_{b=1}^{r}\prod_{k=1}^{N_{a}}\mathbf{B}_{a}(\theta^{(a)}_{k})\ket{\emptyset},
\label{eqn:nested_Bethe_state}
\end{equation}
for a suitable set of $N_{a}\in \mathbb{N}$ complex `quantum numbers' $\{\theta^{(a)}_{j}\}_{j=1}^{N_{a}}$
(one for each $a \in \mathcal{I}_{r}$) obtained as distinct solutions to Eq.~\eqref{eqn:Bethe_nested}. Coefficients
$F^{(a_{1},\ldots,a_{N})}$ indeed turn out to be the wavefunction amplitudes of an integrable quantum chain of rank $r-1$ and
length $N$, for which rapidities take the role of inhomogeneities (for further details and a pedagogical introduction the reader
is referred to \cite{Fedor16,Slavnov_NBA}). Here we add that one can formulate an alternative algebraic construction
of nested Bethe eigenstates by using a single excitation operators for a monodromy matrix \emph{twisted} by
a generic invertible element \cite{GLS17}.

The reference state $\ket{\emptyset}$ is a trivial (particle-less) `Fock vacuum' eigenstate usually called the Bethe pseudovacuum.
In the case of homogeneous ferromagnetic chains, it always corresponds to the fully polarized state (i.e. having the maximal weight
in $\mathcal{H}$) pointing along a specified direction. Beware that such a state is only unique after gauge-fixing since by virtue
of $G$-invariance Hamiltonians $H$ possess continuously degenerate ground-state manifolds.
This signifies that $\ket{\emptyset}$ indeed carries a polarization degree of freedom which takes values on a coset 
manifold $G/H$, where the subgroup $H\subset G$ is referred to as the stability group (leaves $\ket{\emptyset}$ intact, apart
from a phase). With no loss of generality, we have freedom to adopt the conventional gauge choice and set the vacuum polarization to 
$\ket{\emptyset}=\bigotimes_{L}\ket{0}$. The Algebraic Bethe Ansatz provides a procedure for explicit construction a complete of
the highest-weight Bethe eigenstates $\ket{\{\theta^{(a)}_{j}\}}$, each of which is ascribed a unique set of Bethe roots.

The structure of the eigenvalues of commuting transfer matrices can be most suggestively described in the language of
functional Bethe Ansatz, exploiting the fact that transfer matrices can be further decomposed in terms of the so-called Baxter
${\bf Q}$-operators (see e.g. \cite{shortcut,Frassek11} for further details and explicit construction).
Eigenvalues of commuting ${\bf Q}$-operators are called $\rQ$-functions \cite{shortcut,Frassek11,Frassek20}.
In the nested Bethe Ansatz, eigenvalues of commuting transfer matrices are all functionally 
dependent on $r$ Baxter $\rQ$-functions, given by \emph{polynomials} of the form
\begin{equation}
\rQ_{a}(\theta) = \prod_{k=1}^{N_{a}}(\theta-\theta^{(a)}_{k}),\qquad a \in \mathcal{I}_{r}.
\label{eqn:Q-function}
\end{equation}
Intuitively speaking, the $Q$-functions can be thought ot as many-body wave-functions in the first-quantization 
picture whose zeros (called Bethe roots) play the role of `nodes'. This is made most explicit in the so-called Sklaynin
SoV approach \cite{Sklyanin95}, where Baxter $\rQ$-functions become coefficients in the expansion over basis of (quantum) separated 
variables \cite{GLS17}.

\paragraph*{Nested Bethe equations.}
We consider homogeneous isotropic ferromagentic chains with unitary irreducible $\mathfrak{g}$-modules $\mathcal{V}_{\Lambda}$
as the local degree of freedom, specializing to rectangular representations with $\Lambda_{\rm p} = s_{\rm p}\,\omega_{a_{\rm p}}$
(here subscript ${\rm p}$ stands for `physical'). Requiring that a Bethe state Eq.~\eqref{eqn:nested_Bethe_state} is an eigenstate of 
the commuting transfer matrices, the cancellation of all the `unwanted terms' yields a set coupled equations of the form~\footnote{ To 
match the standard TBA convention, our parametrization differs from those of \cite{KNS11} by a $\ii/2$ rescaling of the complex 
spectral plane.}
\begin{equation}
\left[\frac{\theta^{(a)}_{j}+\tfrac{\ii}{2}\hat{m}_{a}}{\theta^{(a)}_{j}
-\tfrac{\ii}{2}\hat{m}_{a}}\right]^{L}
= -\prod_{b=1}^{r}\prod_{k=1}^{N_{b}}
\frac{\rQ_{b}\big(\theta^{(a)}_{j}+\tfrac{\ii}{2}(\alpha_{a},\alpha_{b})\big)}{\rQ_{b}\big(\theta^{(a)}_{j}
-\tfrac{\ii}{2}(\alpha_{a},\alpha_{b})\big)},
\label{eqn:BetheEqs}
\end{equation}
where $\hat{m}_{a}\equiv (\Lambda_{\rm p},\alpha_{a})=(s_{\rm p}/t_{a})\delta_{a,a_{\rm p}}$ and $(\cdot,\cdot)$ is
a bilinear form for on the dual space $\mathfrak{t}^{*}$ of Cartan subalgebra (for definition, cf. appendix \ref{app:Lie}).

\begin{figure}[t!]
\begin{tikzpicture}
\begin{scope}
\node at (1,1) {$A_{3}=\mathfrak{su}(4)$};
\node[circle,draw,thick, minimum size=0pt, inner sep = 3pt] (a) at (0,0) {\scriptsize $0$};
\node[circle,draw,thick, minimum size=0pt, inner sep = 3pt] (b) at (1,0) {\scriptsize $0$};
\node[circle,draw,thick, minimum size=0pt, inner sep = 3pt] (c) at (2,0) {\scriptsize $1$};
\draw[-,thick] (a) -- (b);
\draw[-,thick] (b) -- (c);
\end{scope}
\begin{scope}[xshift=90pt]
\node at (0.5,1.4) {$D_{3}=\mathfrak{so}(6)$};
\node[circle,draw,thick, minimum size=0pt, inner sep = 3pt] (a) at (0,0) {\scriptsize $0$};
\node[circle,draw,thick, minimum size=0pt, inner sep = 3pt] (b) at (0.7,0.7) {\scriptsize $1$};
\node[circle,draw,thick, minimum size=0pt, inner sep = 3pt] (c) at (0.7,-0.7) {\scriptsize $0$};
\draw[-,thick] (a) -- (b);
\draw[-,thick] (a) -- (c);
\end{scope}

\begin{scope}[xshift=140pt]
\node at (0.5,1) {$B_{2}=\mathfrak{so}(5)$};
\node[circle,draw,thick, minimum size=0pt, inner sep = 3pt] (a) at (0,0) {\scriptsize $1$};
\node[circle,draw,thick, minimum size=0pt, inner sep = 3pt] (b) at (1,0) {\scriptsize $0$};
\draw[-,thick, double distance = 2pt] (a) -- (b);
\draw[-,very thick] (0.45,0.15) -- (0.55,0) -- (0.45,-0.15);
\end{scope}
\begin{scope}[xshift=200pt]
\node at (0.5,1) {$C_{2}=\mathfrak{sp}(4)$};
\node[circle,draw,thick, minimum size=0pt, inner sep = 3pt] (a) at (0,0) {\scriptsize $0$};
\node[circle,draw,thick, minimum size=0pt, inner sep = 3pt] (b) at (1,0) {\scriptsize $1$};
\draw[-,thick, double distance = 2pt] (a) -- (b);
\draw[-,very thick] (0.55,0.15) -- (0.45,0) -- (0.55,-0.15);
\end{scope}

\end{tikzpicture}
\caption{Enumerated Dynkin diagrams, depicted for two pairs of isomorphic Lie algebras $A_{3}\cong D_{3}$ and $B_{2}\cong C_{2}$.
The fundamental $\mathfrak{su}(4)$ representation is equivalent to $\mathfrak{so}(6)$ representation $({\bf 4})=[0,1,0]$.
Similarly, the defining $\mathfrak{so}(5)$ vector representation is equivalent to $\mathfrak{sp}(4)$
representation $({\bf 5})=[0,1]$.
}
\label{fig:Dynkin_labelled}
\end{figure}
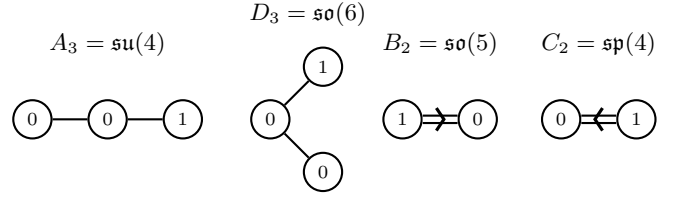

Equations \eqref{eqn:BetheEqs} are none other that the aforementioned NBA equations in disguise. To reconcile
them with Eqs.~\eqref{eqn:Bethe_nested}, the term in the square brackets on the left-hand 
side of Eqs.~\eqref{eqn:BetheEqs} needs to be interpreted as the exponential of the bare momentum,
\begin{equation}
p(\theta^{(a)}_{j}) = -\ii \log\left[\frac{\theta^{(a)}_{j}+\frac{\ii}{2}\hat{m}_{a}}{\theta^{(a)}_{j}
-\frac{\ii}{2}\hat{m}_{a}}\right],
\end{equation}
whereas the right-hand side in Eq.~\eqref{eqn:BetheEqs}, expressed abstractly in terms of the Baxter's $\rQ$-functions, should
be equated with the product of (rational) scattering amplitudes.
The physical interpretation of the resulting NBA equations is as follows:
spin waves of quasiparticles (magnons) of flavor $a+1$ propagate on a (fictitious) lattice formed by quasiparticles of
`adjacent flavor' $b$ whenever there is interaction between the two, namely $(\alpha_{a},\alpha_{b})\neq 0$.
The NBA equations \eqref{eqn:BetheEqs} are in to one-to-one correspondence with enumerated Dynkin diagrams, as exemplified
in Fig.~\ref{fig:Dynkin_labelled}.

\paragraph*{Fused transfer matrices.}
The framework of the algebraic NBA allows simultaneous diagonalization of the entire family of fused transfer matrices
${\bf T}_{a,s}$ ascribed to \emph{rectangular} irreducible auxiliary representations $\mathcal{V}_{a,s}$.
Note that an infinity family of transfer matrices ${\bf T}_{a,s}$ are not independent objects as they can all be generated out of the 
fundamental one ${\bf T}_{\square}$ via the so-called `fusion procedure' \cite{KLWZ97,Zabrodin98}.
As a consequence, the infinite family of fused row transfer matrices ${\bf T}_{a,s}(\theta)$ enclose certain functional relations,
and (due to their commutativity) likewise for their eigenvalues $T_{a,s}(\theta)$.
For example, in the $\mathfrak{su}(n)$ quantum chains, the eigenvalues satisfy a system
of coupled recurrence relations of the form
\begin{equation}
{\bf T}^{(n)+}_{a,s}{\bf T}^{(n)-}_{a,s} = {\bf T}^{(n)}_{a,s-1}{\bf T}^{(n)}_{a,s+1} + {\bf T}^{(n)}_{a-1,s}{\bf T}^{(n)}_{a+1,s},
\label{eqn:SUN_Hirota_equation}
\end{equation}
in which one can recognize the celebrated \emph{Hirota bilinear relation}. For compactness of notation we have suppressed
dependence of the spectral variable $\theta$ and employed a shorthand notation for imaginary half-unit shifts
$f^{\pm}(\theta):=f(\theta\pm k\tfrac{\ii}{2})$.
To fix a solution of an infinite hierarchy of $T$-functions it is thus sufficient to prescribe the `initial condition' in the form of 
fundamental eigenvalues $T_{a,1}(\theta)$ which, alongside the trivial boundary functions $T_{0,s}=T_{n,s}=1$, unique determine the 
remaining $T$-functions by iterative application of Eqs.~\eqref{eqn:SUN_Hirota_equation}.

It is worth mentioning that Hirota equations lie at the bedrock of classical and quantum soliton theories \cite{JimboMiwa83}.
At a formal level, one can think of its as a completely integrable classical lattice gauge theory.
The particular version of the Hirota relation has been originally found by Hirota in space-time discretization for a certain class
of nonlinear integrable partial differential equations \cite{Hirota77,Hirota_book,FaddeevVolkov}. Analogous functional relations
have been derives for other types of simple Lie algebras $\mathfrak{g}$, see e.g. \cite{KNS11} for an extensive review.
In the next section we shall outline how the Hirota-type relation reemerges as functional relation associated to certain thermodynamic
state functions.	

\paragraph*{Transfer matrix eigenvalues.}

The NBA equations \eqref{eqn:BetheEqs} are in fact implied by the fusion relations for the commuting fow fused transfer matrices
$\mathbf{T}_{a,s}(\theta)$ associated to rectangular irreducible representations. To outline the logic, we have to switch
to the functional Bethe Ansatz perspective.
To outline the basic logic, To this end, it is instructive to first shortly examine the non-nested case of
the homogeneous $\SU(2)$ spin chain with $\mathcal{V}_{\Lambda_{\rm p}}=\mathcal{V}_{\square}$.
Eigenvalues $T_{\square}(\theta)$ of ${\bf T}_{\square}(\theta)$ satisfy the Baxter's $TQ$-equation
\begin{equation}
T_{\square}(\theta) = \frac{Q^{+}_{0}(\theta)Q^{--}_{1}(\theta)+Q^{-}_{0}(\theta)Q^{++}_{1}(\theta)}{Q_{1}(\theta)},
\label{eqn:TQ_SU2}
\end{equation}
with $Q_{0}(\theta)\equiv \theta^{L}$. Taking now into account that both $T_{\square}(\theta)$ and $Q$-functions are manifestly
finite-degree polynomials in $\theta$, the roots of $Q_{1}(\theta)$ must be positioned such to ensure that all the superficial poles 
vanish. By demanding that residues on the right-hand side of Eq.~\eqref{eqn:TQ_SU2} vanish one recovers precisely the $\SU(2)$
Bethe equations.

In the of higher-rank symmetries, eigenvalues $T_{a,s}(\theta)$ can be decomposed in terms of $r$ distinct $Q$-functions.
We consider here as an example the case of unitary groups $\SU(n)$. By virtue of Eq.~\eqref{eqn:SUN_Hirota_equation},
it is sufficient to focus on the fundamental transfer matrix ${\bf T}_{\square}(\theta)$. By formally solving
Eqs.~\eqref{eqn:SUN_Hirota_equation} in a recursive manner with a chain of B\"{a}cklund transformations, the fundamental eigenvalues
$T^{(n)}_{\square}(\theta)$ can be expressed in a closed form \cite{KLWZ97,KLV16}
\begin{equation}
T^{(n)}_{\square}(\theta) = \sum_{a=1}^{n} Q_{0}(\theta)\frac{Q^{--}_{a-1}(\theta)Q^{++}_{a}(\theta)}{Q_{a-1}(\theta)Q_{a}(\theta)},
\label{eqn:sun_TQ}
\end{equation}
with boundary functions $Q_{0}(\theta)=\theta^{L}$ and $Q_{n}\equiv 0$.
The NBA equations \eqref{eqn:Bethe_nested} are implied the pole-cancellation condition,
i.e. demanding that $\mathbf{T}_{a,s}(\theta)$ are entire functions in the $\theta$-plane.

\paragraph*{Finite-volume eigenstates.}
Each highest-weight eigenstate of ${\bf T}_{\square}(\theta)$ is associated with a degree-$L$ polynomials $T_{\square}(\theta)$.
Amongst its $L$ zeros, there are $\ol{N}_{1}$ real ones (called holes). Provided that $Q_{1}(\theta)$ has only real roots, one can 
readily infer from the analytic structure of Eq.~\eqref{eqn:TQ_SU2} for large $L$ (assuming $\theta \sim \mathcal{O}(L^{0})$)
using that $Q^{\pm}_{0}Q^{\mp \mp}_{1}=T_{\square}Q_{1}$ for ${\rm Im}(\theta)\gtrless 0$. Complex zeros of $T_{\square}(\theta)$
are approximately roots of $Q_{1}(\theta)$ shifted in the imaginary direction by $\pm \ii$.
Therefore, when ${\rm deg}\,Q_{1}=N_{1}$, we can deduce the following equality $L+N_{1}=N_{1}+2N_{1}+\ol{N}_{1}$,
implying $N_{1}=(L-\ol{N}_{1})/2$. Ferromagnetic vacuum is `empty' and fully filled with $\ol{N}_{1}=L$ holes.
The opposite of it is the eigenstate with no holes, $\ol{N}_{1}=0$ and $N_{1}=L/2$, residing in the maximally filled sector.

A similar analysis applies to the higher-rank models. For example, in the $\SU(n)$-symmetric chains there are $n-1$ flavors 
of magnons and one can now deduce a set of inequalities
\begin{equation}
2N_{a}\leq N_{a-1}+N_{a+1},
\label{eqn:inequality_rule}
\end{equation}
for $a=1,\ldots,r$ (with $N_{0}\equiv L$ and $N_{r+1}\equiv 0$).
Such `triangular conditions' are equivalent to the requirement that all Bethe eigenstates 
are associated a non-negative highest-weight. The maximally saturated sector in the fundamental $\mathfrak{su}(n)$ chain thus 
corresponds to filling fractions $\nu_{a} \equiv N_{a}/L=(n-a)/n$.

\paragraph*{Complex solutions (bound states).}
The general structure of finite-volume eigenstates is in general far more complicated, mostly because Bethe roots of individual 
solutions to Eqs.~\eqref{eqn:BetheEqs} are generically complexed-valued (by reality condition, all the roots must however occur
in complex-conjugate pairs). In particular, typical large-volume eigenstates consist of compounds of complex roots
with the same real part -- known as \emph{Bethe strings}. The structure of the Bethe equations \eqref{eqn:BetheEqs} indicates that
for large $L$ (and $\theta \sim \mathcal{O}(L^{0})$), the imaginary separation of two adjacent complex roots within a string-like 
compound is precisely $\ii$, up to finite-size corrections (typically suppressed exponentially in $L$).
Since finite imaginary part in momentum signifies finite localization length, such solutions pertain to multi-magnon
bound states. For example, in the above $\SU(2)$, an $s$-string ($s\geq 2$) solution induced a complex-conjugate pair of zeros in the 
corresponding transfer matrix eigenvalue with ${\rm Im}(\theta)=\pm(s-1)\ii$.

An analogous analysis can be performed for the higher-rank models with nested spectra (see e.g. \cite{Volin12} for a general
discussion for the case of unitary superalgebras). Presently, the large-volume Bethe eigenstates in $G$-invariant ferromagnetic
chains can be partitioned in terms of Bethe strings of different flavors $a \in \mathcal{I}_{r}$.
Specifically, an $s$-string is made of $s$ constituent magnons with complex rapidities
\begin{equation}
\Big\{\theta^{(a,s)}_{k}+\frac{\ii}{2}(s+1-2j)\Big\}_{1\leq j\leq s},
\label{eqn:Bethe_string}
\end{equation}
modulo finite-size deviations which are negligible in the $L\to \infty$ limit. Notice the exact one-to-one correspondence
between the Bethe strings (bound states) and unitary irreducible rectangular representations of $\mathfrak{g}$.

For large $L$, each $Q$-function $Q_{a}(\theta)$ associated with highest-weight Bethe eigenstate can be thus partitioned into
strings, $Q_{a}(\theta)=\prod_{a=1}^{r}\prod_{s=1}^{\infty}(\theta - \theta^{(a,s)}_{k})$, neglecting vanishing finite-size 
corrections. The total number of $s$-strings of flavor $a$, denoted by $n_{a,s}$, can be computed with an explicit combinatorial 
formula (given by Eq.~\eqref{eqn:multiplicity_factor} in appendix \ref{app:multiplicities}), counting all admissible rearrangements 
of `particles' and `holes'.
We mention in passing that despite such a formula establishes `combinatorial completeness' of solutions to the NBA 
equations, it is known the certain legal but irregular solutions involving `collapsed strings' do occur finite-volume
spectrum \cite{Vladimirov84,Isler93}. Such exceptional solutions nonetheless amount to a thermodynamically vanishing fraction and hence 
are not detrimental to the string hypothesis and the validity of the TBA approach (completeness of the finite-volume NBA equations has 
been recently proven using an alternative Wronskian formulation in Ref.\cite{Chernyak20}).

\paragraph*{Irreducible multiplets.}
Owing to $G$-invariance of the considered Hamiltonians, eigenvectors organize into irreducible subspaces
(multiplets) of $\mathfrak{g}$. The total number of distinct highest-weight eigenstates equals the number of inequivalent irreducible 
multiplets in the decomposition of the $L$-site Hilbert space $\mathcal{H}\cong \mathcal{V}^{\otimes L}_{\Lambda_{\rm p}}$.
The outlined NBA construction only gives access to the highest weight Bethe eigenstates, whereas the descendant eigenstates
have to be generated by successively acting with the global lowering generators on every highest-weight Bethe state.
In effect, eigenstates within a given multiplet are all characterized by the same set of Bethe roots and consequently yield the 
same value of (quasi)local conserved charges deriving from ${\bf T}_{a,s}(\theta)$ -- they are thus only distinguished by the valued 
of the Cartan $\U(1)$ charges.

Commuting row transfer matrices ${\bf T}_{a,s}(\theta)$ acting on the tensor product Hilbert spaces are not irreducible objects
but instead decompose non-trivially into finite-dimensional irreducible representations of Yangians $Y(\mathfrak{g})$
(commonly known as the Kirillov--Reshetikhin modules \cite{Kirillov89,KR90}, or $\mathcal{W}$-modules, for short).
To be specific, a homogeneous transfer matrix acting in an $L$-fold tensor product of local Hilbert spaces
$\mathcal{V}_{\Lambda_{\rm p}}$ corresponds to a reducible $\mathcal{W}$-module $\mathcal{W}^{\otimes L}_{\Lambda_{\rm p}}$.
Multiplicities of irreducible components are precisely in agreement with the number of distinct highest-weight Bethe eigenstates.
However, when viewed as $\mathfrak{g}$-modules, irreducible $\mathcal{W}$-modules generically reducible in a non-trivial manner
(with the exception of $A_{n}$-modules which stay irreducible).
Further details on reducibility of Kirillov--Reshetikhin modules are provided in appendix \ref{app:modules}.

\subsection{Thermodynamic Bethe Ansatz}

Having described the structure of finite-volume eigenstate, we proceed to formulate the thermodynamic description
of a nested Bethe Ansatz system. This will be achieved in the standard manner by employing the universal functional integral approach 
introduced originally by Yang and Yang \cite{YY1969}, adapting it to accommodate for integrable quantum chains invariant under Lie 
symmetry group $G$.

\paragraph*{Thermodynamic limit.}
To infer thermodynamic properties, one has to consider the large-volume limit and retain only eigenstates with an extensive number
of excitations, $N_{a}\sim \mathcal{O}(L)\to \infty$, while keeping all the filling fractions $N_{a}/L$ fixed.
In doing so, one can afford to omit certain details about individual eigenstates which considerably simplifies analytic treatment.
In a generic thermodynamic eigenstate, a typical separation of nearby rapidities (or string centers $\theta^{(A)}_{k}$, in case of 
bound states) is of the order $\mathcal{O}(1/L)$, which makes it natural to introduce an infinite set rapidity densities,
\begin{equation}
\rho_{A}(\theta) = \lim_{L\to \infty}\frac{1}{L}\frac{1}{\theta^{(A)}_{k+1}-\theta^{(A)}_{k}},
\label{eqn:rho_def}
\end{equation}
one for each thermodynamic quasiparticle specie $A=(a,s)$. Quantities $\rho_{A}(\theta)\dd \theta$ representing
the total number of quasi-particles of type $A$ residing in an infinitesimal interval $[\theta,\theta+\dd \theta]$ per unit length.
Similarly, one introduces the total densities of available states $\rho^{\rm tot}_{A}(\theta)$,
encoding densities of the available modes (with mode numbers ${\rm n}_{i}$) in the spectral plane,
\begin{equation}
\rho^{\rm tot}_{A}(\theta) = \lim_{L\to \infty}\frac{1}{L}\frac{{\rm n}^{(A)}_{k+1}-{\rm n}^{(A)}_{k}}{\theta^{(A)}_{k+1}-\theta^{(A)}_{k}},
\label{eqn:rhotot_def}
\end{equation}
corresponding to taking the logarithm of the thermodynamic limit of Eq.~\eqref{eqn:BetheEqs}.
In contrast with free theories, where the latter takes a constant value of $1/2\pi$,
the net contribution of inter-particle interactions induces a non-trivial dependence on a state.
It is also convenient to define the densities of holes (i.e. unoccupied modes)
$\ol{\rho}_{A}(\theta)\equiv \rho^{\rm tot}_{A}(\theta)-\rho_{A}(\theta)$.

\paragraph*{Bethe--Yang equations.}
Owing to interparticle interactions, quasiparticle densities $\rho_{A}(\theta)$ and total densities of available states
$\rho^{\rm tot}_{A}(\theta)$ are not independent of one another. The relation between the two can be inferred from the thermodynamic 
limit of the Bethe equations \eqref{eqn:BetheEqs}, yielding a coupled system of integral (\emph{Bethe--Yang})
equations \cite{YY1969,Takahashi1971,Gaudin1971}
\begin{equation}
\rho^{\rm tot}_{A} \equiv \rho_{A} + \ol{\rho}_{A} = \frac{p^{\prime}_{A}}{2\pi} - K_{AB}\star \rho_{B}.
\label{eqn:BetheYang}
\end{equation}
expressing $\rho^{\rm tot}_{A}$ as a functional of $\rho_{A}$. Here and subsequently, $\star$ is an abbreviation for the convolution 
type integrals, which (for dummy functions $f_{A}$, $g_{A}$ and $F_{AB}$) read
\begin{align}
F_{AB}\star g_{B} &\equiv \sum_{B} \int_{\mathbb{R}} \dd \theta' F_{AB}(\theta-\theta')g_{B}(\theta'),\\
f_{A}\star g_{A} &\equiv \sum_{A} \int_{\mathbb{R}} \dd \theta' f_{A}(\theta)g_{A}(\theta-\theta').
\end{align}
The two-body kernels in Eq.~\eqref{eqn:BetheYang} are given by the logarithmic derivative of
the scattering amplitudes $S_{AB}(\theta)$,
\begin{equation}
K_{AB}\equiv \frac{1}{2\pi \ii}\partial_{\theta}\log S_{AB}(\theta),
\end{equation}
attributed to a two-body scattering event between quasiparticles of types $A=(a,s)$ and $B=(a^{\prime},s^{\prime})$.
The complete set of scattering amplitudes $S_{AB}$ can be obtained from the elementary magnon amplitudes via fusion;
this amounts to merge a finite `string' of scattering amplitudes of the form
$S_{s}(\theta) \equiv (\theta - s\tfrac{\ii}{2})/(\theta + s\tfrac{\ii}{2})$ that appear on the right-hand side
of Eqs.~\eqref{eqn:BetheEqs} for every given pair of string configurations of types $A$ and $B$. For instance, fused scattering
amplitudes of an $s$-string and an $s^{\prime}$-string of the same flavors are of the form
\begin{equation}
S_{(a,s),(a,s^{\prime})}(\theta) = \prod_{j=-\tfrac{s-1}{2}}^{\tfrac{s-1}{2}}\prod_{j^{\prime}=-\tfrac{s^{\prime}-1}{2}}^{\tfrac{s^{\prime}-1}{2}}S_{2(j+j^{\prime}+1)}(\theta).
\end{equation}

\paragraph*{Macrostates.}
Any given admissible set of quasiparticles densities $\rho_{A}(\theta)$ (that is a solution to coupled integral
equations \eqref{eqn:BetheYang}) is compatible with exponentially many distinct thermodynamic Bethe eigenstates.
It is important to stress that individual eigenstates are indistinguishable at the level of local observables \cite{MP14,MTPW15,PVW17}, 
which can be viewed as a manifestation of the `eigenstate thermalization principle' (in a generalized sense).
In this sense, set $\{\rho_{A}(\theta)\}$, commonly referred to as a \emph{macrostate}, can be understood as a equivalence classes of 
thermodynamic eigenstates with a finite density of excitations. In other words, momentum (or rapidity) densities
$\{\rho_{A}(\theta)\}$ provide a complete spectral resolution of a local equilibrium state.
For the present class of model, every macrostates is ascribed a Fermi--Dirac statistics, i.e. the associated
combinatorial weight has the form
\begin{equation}
\mathfrak{s}_{A}(\theta) = \rho_{A}\log\left(\frac{\ol{\rho}_{A}}{\rho_{A}}\right)
+\ol{\rho}_{A}\log\left(\frac{\rho_{A}}{\ol{\rho}_{A}}\right).
\end{equation} 
This expression coincides with the von-Neumann entropy of a generalized Gibbs ensemble associated with a
macrostate $\{\rho_{A}(\theta)\}$.

Equilibrium macrostates can alternatively be (uniquely) parametrized in terms of the expectation values of the local conservation laws
\cite{PhysRevLett.115.157201,StringCharge,FP20}. The caveat here is that when the quasiparticle spectrum involves bound states,
the standard set of local conservation laws (obtained from the series expansion of fundamental transfer matrices
${\bf T}_{\square}(\theta)$) are actually not sufficient (see Refs.~\cite{Wouters2014,MTPW15}) and it is imperative not to leave
out the \emph{quasilocal} conserved charges (generated out of fused row transfer matrices, see Refs.~\cite{IMP15,StringCharge,FP20}
and a review \cite{QLreview}). For explicit construction and complete classification of generalized Gibbs ensembles exemplified on the 
emblematic case of the Heisenberg spin-$1/2$ chain we refer the reader to Refs.~\cite{IQC17,IQ19}.

\subsubsection*{TBA equations}

We now outline how thermodynamic free energy can be computed in the framework of the Thermodynamic Bethe Ansatz.
For definiteness we now confine ourselves to the grand-canonical Gibbs states \eqref{eqn:gc_ensemble}.

Firstly, to every Cartan generator $\rH^{i} \in \mathfrak{t}$ we associate a conserved Cartan charge $Q^{(i)}=\int \dd x\,q^{(i)}(x)$,
with local density $q^{(i)}$. To fix a basis of the Cartan subalgebra, we define $\rH^{i} \in \mathfrak{t}$ via
$\rH^{i}=[\rE^{\alpha_{i}}, \rE^{-\alpha_{i}}]$ (for every simple roots $\alpha_{i}$), after adopting a particular normalization
convention for the Weyl generators, such as the common one $\kappa_{\alpha_{i},-\alpha_{i}}=1$ (which also fixes $\kappa_{ij}$).
The most general element $g_{0}\in T$ from the torus subgroup can be parametrized as
\begin{equation}
g_{0} = {\rm diag}(x_{1},x_{2},\ldots,x_{n}),\qquad 
\prod_{i=1}^{n}x_{i}=1,
\end{equation}
in terms of $r$ scalars $x_{i}\in \mathbb{R}$ being certain functions of the $\U(1)$ chemical potentials $h_{i}$.
Eigenvalues of $Q^{(i)}$ acting on the highest-weight Bethe eigenstates read
\begin{equation}
Q^{(i)}\ket{\{\theta^{(a)}_{j}\}} = \left(\sum_{a=1}^{r}N_{a}q^{(i)}_{a}\right)\ket{\{\theta^{(a)}_{j}\}},
\end{equation}
where $q^{(i)}_{a}$ are (bare) quanta of the $i$th Cartan charge carried by a quasi-particle of flavor $a$.
Taking further into account that general thermodynamic eigenstates involve bound states (Bethe strings), these further split amongst 
$N_{a,s}$ Bethe's $s$-strings (with $N_{a}=\sum_{s}s\,N_{a,s}$) carrying bare charges $q^{(i)}_{a,s}$.

The defining property of macrostate densities $\{\rho_{A}(\theta)\}$ is that they correspond to the variational extremum
(i.e. the saddle-point) of the free-energy functional for a specific choice of chemical potentials.
Employing quasiparticle densities as variational objects and eliminating the variation of hole densities,
using $\delta \ol{\rho}_{A}=-K_{AB}\star\delta \rho_{B}$ (cf. Eq.~\eqref{eqn:BetheYang}), the equilibrium partition integral
\begin{equation}
\mathcal{Z}_{\beta,{\bf h}} = \int \mathscr{D}[\{\rho_{A}(\theta)\}]e^{-L\,\mathcal{F}[\{\rho_{A}(\theta)\}]},
\label{eqn:partition_integral}
\end{equation}
is specified by the free energy functional
\begin{equation}
\mathcal{F}[\{\rho_{A}(\theta)\}] = \sum_{A}\int \dd \theta \Big(\mu_{A}(\theta)\rho_{A}(\theta)-\mathfrak{s}_{A}(\theta)\Big).
\label{eqn:free_energy_functional}
\end{equation}
Here each quasiparticle mode in the spectrum has been assigned a chemical potential $\mu_{A}(\theta)$.
Specifically for the grand-canonical Gibbs equilibrium states, these have the form
\begin{equation}
\mu_{A}(\theta) = \beta\,e_{A}(\theta)-\sum_{i=1}^{r}h_{i}q^{(i)}_{A},
\label{eqn:mu_functions}
\end{equation}
which can be easily inferred by resolving the energy density in terms quasi-particle modes,
$\lim_{L\to \infty}(E/L)=e_{A}\star \rho_{A}$, where $e_{A}(\theta)$ denote bare quasiparticle energies carrying bare
Cartan charges $q^{(i)}_{A}$. For the class of Hamiltonians \eqref{eqn:Hamiltonian}
(with the anti-ferromagnetic exchange coupling), we find
\begin{equation}
E = L - 2\pi\sum_{j=1}^{N_{1}}K_{1}(\theta^{(1)}_{j}),
\end{equation}
with
\begin{equation}
K_{1}(\theta)=\frac{1}{2\pi \ii}\partial_{\theta}\log S_{1}(\theta) = \frac{1}{2\pi}\frac{1}{\theta^{2}+1/4}.
\end{equation}

In the large $L$ limit, the partition integral \eqref{eqn:partition_integral} is dominated by the saddle-point.
By evaluating the variational minimum of Eq.~\eqref{eqn:free_energy_functional},
$(\delta/\delta \rho_{A}) \mathcal{F}[\{\rho_{A}\}]=0$,
one arrives at the following infinite system of coupled integral equations
\begin{equation}
\log \calY_{A} = \mu_{A} + K_{AB}\star \log(1+1/\calY_{B}),
\label{eqn:canonical_TBA}
\end{equation}
called the \emph{canonical TBA equations}. As customary, we have introduced here a set of new thermodynamic variables,
\begin{equation}
\calY_{A}(\theta) \equiv \frac{\ol{\rho}_{A}(\theta)}{\rho_{A}(\theta)},
\end{equation}
which we call the thermodynamic $\calY$-functions; they play the role of `Boltzmann weights',
and thus in canonical Gibbs equilibrium assume the form
\begin{equation}
\calY_{A}(\theta) \equiv e^{-\mu^{\rm dr}_{A}(\theta)},
\label{eqn:Boltzmann_weight}
\end{equation}
where we have introduced the dressed quasiparticle dispersions
\begin{equation}
\mu^{\rm dr}_{A}(\theta) = \beta\,\varepsilon_{A}(\theta) - \sum_{i=1}^{r}h_{i}q^{(i){\rm dr}}_{A},
\end{equation}
with quantities $\varepsilon_{A}(\theta)$ and $q^{(i){\rm dr}}_{A}$ corresponding to the physical energy and
the so-called dressed charges of the quasiparticle mode $A$, respectively.
Accordingly, one defined the Fermi occupation functions $n_{A}(\theta)$ as ratios
\begin{equation}
n_{A}(\theta) = \frac{\rho_{A}(\theta)}{\rho^{\rm tot}_{A}(\theta)} = \frac{1}{1+\calY_{A}(\theta)}.
\end{equation}

Microscopic details of the model under consideration enter into equations \eqref{eqn:BetheYang} and 
\eqref{eqn:canonical_TBA} implicitly through the scattering kernels $K_{AB}(\theta)$. We shall demonstrate in turn that
the underlying Yangian symmetry $Y(\mathfrak{g})$ can be explicitly exhibited by inverting the integral kernel,
namely by computing the Fredholm resolvent $\mathcal{R}_{AB}(\theta)$, defined through
\begin{equation}
(1-\mathcal{R}_{AC})\star (1+K_{CB}) = 1,
\end{equation}
where the identity operator on the right-hand side is understood as $1\equiv \delta_{AB}\delta(\theta)$.

\paragraph*{Unitary spin chains.}
For concreteness, we consider here explicitly the unitary cases $\mathfrak{g}=\mathfrak{su}(n)$.
For the basis of Weyl generators ascribed to simple roots we can take $\rE^{\alpha_{a}}=\ket{a}\bra{a+1}$,
such that $a$th Cartan element $\rH^{\alpha_{a}}$ takes the form
\begin{equation}
\rH^{\alpha_{a}}=\ket{a}\bra{a}-\ket{a+1}\bra{a+1}.
\end{equation}
With this choice, the Killing metric $\kappa$ coincides with the Cartan matrix, $\kappa \equiv \mathcal{K}_{\mathfrak{su}(n)}$.

The Fredholm resolvent has a remarkably simple structure
\begin{equation}
\mathcal{R}_{AB}(\theta) = s(\theta)I_{AB},
\end{equation}
where the $s$-kernel (defined as the solution to equation $K_{1}-s\star K_{2}=s$) reads explicitly
\begin{equation}
s(\theta) = \frac{1}{2\cosh{(\pi\,\theta)}},
\end{equation}
and
\begin{equation}
I_{AB} \equiv I_{(a,s),(a's')} = \delta_{a,a'}I^{A_{\infty}}_{s,s'} + \delta_{s,s'}I^{A_{n-1}}_{a,a'},
\end{equation}
is an adjacency tensor of the form $A_{n-1}\times A_{\infty}$, where $I^{A_{n}}$ pertain to
incidence matrices of dimension $n-1$ associated to $A_{n}$-type Dynkin diagrams whose `bulk part'
reads $I_{s,s'}=\delta_{s,s'-1}+\delta_{s,s'+1}$.

The transformed source term are given by functions
\begin{equation}
d_{A}=(1-\mathcal{R})_{AB}\star \mu_{B},
\end{equation}
which allows to recast the canonical TBA equations \eqref{eqn:canonical_TBA} in a `quasi-local' form
\begin{align}
\log \calY_{a,s} = d_{a,s} &+ s\star I_{s,s'}\log(1 + Y_{a,s'}) \nonumber \\
&+ s\star I_{a,a'}\log(1 + \calY_{a',s}).
\label{eqn:quasi-local_TBA}
\end{align}
At this step one encounters an important subtlety: operating with $(1-\mathcal{R})_{AB}$ erases information about bare charges 
$q^{(i)}_{a,s}=s\,q^{(i)}_{a}$ from the canonical source terms \eqref{eqn:mu_functions} and
consequently Eqs.~\eqref{eqn:quasi-local_TBA} do not admit a unique solution unless one supplements them with appropriate boundary 
(asymptotic) conditions by prescribing the large-$s$ asymptotics
\begin{equation}
\lim_{s\to \infty}\frac{1}{s}\log \calY_{a,s}(\theta) \equiv y^{(\infty)}_{a}.
\end{equation}
As explained in the subsequent section, asymptotic quantities $y^{(\infty)}_{a}$ depend solely on the $\U(1)$ chemical potentials 
$h_{i}$ assigned to the Cartan charges $Q^{(\alpha_{i})}$ (associated with $i$th simple root $\alpha_{i}$).

For example, in the grand-canonical Gibbs ensemble in a homogeneous $\SU(n)$ quantum chain with
onsite representations $\mathcal{V}_{s_{\rm p} \omega_{a_{\rm p}}}$, the local source terms read explicitly
\begin{equation}
d_{a,s}(\theta)\simeq \delta_{a,a_{\rm p}}\delta_{s,s_{\rm p}}\beta\,s(\theta),
\end{equation}
modulo a multiplicative factor which depends on normalization.

\subsubsection*{Universal dressing equations}

We would now like to present a more suggestive physical interpretation of the outlined TBA formalism, revealing itself
upon interpreting Eqs.~\eqref{eqn:BetheYang} and \eqref{eqn:canonical_TBA} as an non-perturbative renormalization of the 
quasiparticle's bare dispersion laws. Transforming quasiparticle's bare properties into the dressed ones effectively amounts to taking 
into account the net contribution from macroscopically many elastic two-body interactions with a thermal background.
To be more specific, to every \emph{bare} charge density $q_{A}(\theta)$ per quasi-particle mode $A=(a,s)$ (with rapidity $\theta$) we
associated the dressed counterpart $q^{\rm dr}_{A}(\theta)$ by applying the `dressing transformation'
\begin{equation}
q_{A}(\theta) \mapsto q^{\rm dr}_{A}(\theta).
\end{equation}
It is remarkable that such a dressing transformation is a particular functional of the equilibrium Fermi function which directly 
reflects the structure of the underlying symmetry algebra.

The first thing to note is that the Bethe--Yang integral equations \eqref{eqn:BetheYang} for the total densities of available states 
and the TBA integral equations \eqref{eqn:canonical_TBA} for the equilibrium free energy are indeed merely two different manifestations 
of one and the same dressing transformation: the former one in fact encodes information about the dressed momenta via
\begin{equation}
(p^{\prime}_{A})^{\rm dr} = (\delta_{AB}\delta(\theta) + K_{AB}n_{B})^{-1}\star p^{\prime}_{B},
\end{equation}
while the latter one provides the dressed quasi-particle energies.
Indeed, the total state densities precisely coincide with dressed momentum derivatives,
\begin{equation}
(p^{\prime}_{A})^{\rm dr} = 2\pi\rho^{\rm tot}_{A}.
\label{eqn:dressed_momentum}
\end{equation}
which is transparent already at the level of Eq.~\eqref{eqn:BetheEqs} (being the finite-volume counterpart of
Eq.~\eqref{eqn:dressed_momentum}). The dressing operator therefore admits the following compact universal form
\begin{equation}
\boldsymbol{\Omega}[{\bf n}] \equiv (\mathbf{1}+\mathbf{K}\,\mathbf{n})^{-1}.
\label{eqn:Omega}
\end{equation}

\paragraph*{Dressing of Cartan charges.}
The dressed values of the Cartan charges can be most easily inferred from the thermodynamic $\calY$-functions using Eq.~\eqref{eqn:Boltzmann_weight}, that is
\begin{equation}
q^{(i){\rm dr}}_{a,s}(\theta) = \frac{\partial \log\calY_{a,s}(\theta)}{\partial h_{i}}.
\label{eqn:dressed_charges_from_logY}
\end{equation}
Several remarks are in order at this point. We first wish to stress that (i) the dressing operator \eqref{eqn:Omega}
does \emph{not} in general commute with rapidity differentiation and that (ii) it should not be confused with conventional dressing
of excitations provided by the so-called shift function (or backflow), cf. \cite{Korepin_book,QF13}.

Let us remind that the presence of finite chemical potentials $h_{i}>0$ breaks $G$-invariance of a state down to $\U(1)^{\times r}$
generated by the Cartan elements. To uniquely specification elementary quasiparticles, one has to select
a basis of simple roots; there are different choices which are related by a discrete symmetry transformation of the root system called 
Weyl group. For instance, in the case of $\mathfrak{su}(n)$ the latter is just the permutation group ${\rm S}_{n}$ amongst the roots
(equivalently, nodes of the Dynkin diagram). While such a `bosonic duality' does not affect the form of the $\SU(n)$ NBA equations,
it does however change the notion of the quasiparticle vacuum (hence there are $n$ distinct ferromagnetic vacua and
$|S_{n}|=n!$ sets of nested Bethe equations in total).
Fixing a reference Bethe vacuum and the basis of simple roots of $\Delta_{+}$, one can only generate Bethe eigenstates with a
non-negative highest weight $\Lambda=L\,\Lambda_{\rm p}-\sum_{i=1}^{r}N_{i}\alpha_{i}$, which imposes certain limitation
on the filling fractions $\nu_{i}=N_{i}/L$. Specifically, the mapping between the charge densities and the corresponding filling
fractions is provided by the Killing metric $\kappa$, reading $q^{(a)} = (\Lambda_{\rm p})_{a} - \kappa_{ab}\nu_{b}$
(with $a,b \in \mathcal{I}_{r}$). The maximally saturated state corresponds to an unbiased (i.e. $G$-invariant) ensemble with
vanishing Cartan charge densities $q^{(i)}=0$, reached by switching all the chemical potentials off, $h_{i}\to 0$.
Moreover, by the requirement the densities of Cartan charge must be positive semi-definite,
$q^{(i)}({\bf h})=\partial f({\bf h})/\partial h_{i}\geq 0$,
which singles out an admissible region for chemical potentials $h_{i}$ inside the hyperoctant,
$\mathscr{W}_{\mathfrak{g}} \subset [0,\infty)^{r}$. To illustrate this on a simple case we can think of the $\SU(3)$ chain
with fundamental representation $\Lambda_{\rm p}={\bf (3)}\equiv [1,0]$. There are $r=2$ flavors of magnons, and the maximal filling 
fractions are given by $\nu^{\rm max}_{a} = (\kappa^{-1})_{ab}(\Lambda_{\rm p})_{b}$, yielding $\nu^{\rm max}_{1} = 2/3$,
$\nu^{\rm max}_{2}=1/3$. The admissible set $\mathscr{W}_{A_{2}}(h_{1},h_{2}) \subset [0,\infty)^{2}$ then represents a wedge region
$h_{2}/2 \leq h_{1}\leq 2h_{2}$. This non-trivial requirement precludes extraction of the bare values of Cartan charge $q^{(i)}_{A}$
by simply taking the $h_{1}\to 0$ limit of $q^{(i){\rm dr}}_{A}({\bf h})$. The correct asymptotic values of
bare charges of giant magnons can nevertheless be retrieved from the suppression of the Fermi factors in the large-$s$ limit,
where one finds
\begin{equation}
n_{a,s}({\bf h}) \simeq \exp{\Big(-\sum_{i=1}^{r}h_{i}q^{(i)}_{a,s}\Big)},\qquad {\bf h}\in \mathscr{W}_{\mathfrak{g}}.
\end{equation}

\paragraph*{Unitary quantum chains.}
In analogy to Eqs.~\eqref{eqn:quasi-local_TBA}, the dressing transformation for local charge densities $q_{A}(\theta)$ can be also 
be cast in a concise group-theoretic form. To this end, we consider below the simplest case of simply-laced
algebras $\mathfrak{su}(n)$. Introducing the `Baxterized' Cartan matrices,
\begin{equation}
\mathscr{K}^{(n)}_{a,b}(\theta) = \delta_{a,b} - s(\theta)I^{A_{n}}_{a,b},
\end{equation}
the dressing transformation can presented in the form
\begin{equation}
\mathscr{K}^{(\infty)}_{s,s'}\star F_{a,s'} - \mathscr{K}^{(n)}_{a,a'}\star \ol{F}_{a',s} = d_{a,s}.
\end{equation}
where for the moment $F_{a,s}$ are just some dummy functions. One can easily verify that equations \eqref{eqn:BetheYang} correspond
to the special case and can be retrieved by setting $F_{A}\to \ol{\rho}_{A}$ and $\ol{F}_{A}\to -\rho_{A}$;
likewise Eqs.~\eqref{eqn:canonical_TBA} by taking $F_{A}\to \log(1+\calY_{A})$ and $\ol{F}_{A}\to \log(1+1/\calY_{A})$.

Employing for convenience rapidity derivatives, $q^{\prime}_{A}(\theta) \equiv \partial_{\theta}q_{A}(\theta)\equiv $,
the dressing transformation can be brought into the following compact universal form \cite{Ilievski18}
\begin{equation}
\mathscr{K}^{{\rm dr}}_{s,s'}\star (q^{\prime}_{a,s'})^{\rm dr}
+ \mathscr{K}^{{\rm dr}}_{a,a'}\star (q^{\prime}_{a',s})^{\rm dr} = q^{\prime}_{a,s},
\label{eqn:universal_dressing_equations}
\end{equation}
where we have simultaneously introduced the following \emph{dressed} (i.e. state-dependent) analogues of the Baxter--Cartan matrices
\begin{align}
\mathscr{K}^{{\rm dr}}_{s,s'}[{\ol{n}_{a,s}}] &\equiv \delta_{s,s'} - s(\theta)I^{A_{\infty}}_{s,s'}\ol{n}_{a,s'},\\
\mathscr{K}^{{\rm dr}}_{a,a'}[{n_{a,s}}] &\equiv \delta_{a,a'} - s(\theta)I^{A_{n}}_{a,a'}\ol{n}_{a',s}.
\end{align}

\paragraph*{Spin-$S$ $\SU(2)$ ferromagnetic chains.}
To demonstrate usefulness of the above formulation, we shortly consider as a concrete example the structure for the case of $\SU(2)$ 
quantum chains of spin-$S$ degrees of freedom (with $2S \in \mathbb{N}$). Since $r=1$, there is a single magnon species in the 
spectrum. Let us moreover denote by $p^{(2S)}_{s}(\theta)$ bare momenta of bound states ($s$-strings) in a spin-$S$ ferromagnetic 
state, and furthermore introduce for convenience the `$G$-tensor'
\begin{equation}
G_{s,2S}(\theta) = \frac{1}{2\pi}\partial_{\theta}p^{(2S)}_{s}(\theta),
\end{equation}
obeying the matrix convolution identity $(1+K)\star G = s$. The above $F$-functions can be identified with their dressed counterparts,
that is $F^{(2S)}_{s}\equiv G^{{\rm dr}}_{s,2S}$. By applying the dressing operator,
one arrives at the system of coupled equations \cite{NMKI19}
\begin{equation}
\big(s^{-1}\delta_{s,s'}-I^{A_{\infty}}_{s,s'}\ol{n}_{s'}\big)\star F^{(2S)}_{s'} = \delta_{s,2S},
\label{eqn:SU2_higher_dressing}
\end{equation}
where the source term in Eqs.~\eqref{eqn:SU2_higher_dressing} `jumped' at the $2S$-th node owing to
\begin{equation}
(1+K)^{-1}_{s,s^{\prime}}G_{s^{\prime},2S}=\delta_{s,2S}s.
\end{equation}
To solve Eqs.~\eqref{eqn:SU2_higher_dressing}, the strategy is to first solve the homogeneous part of the recurrence relation and then 
solve the `gluing condition' at the $2S$-th node. Explicit results can be found in Ref.~\cite{NMKI19} and we do not reproduce them 
here. Let us just note that, thanks to the kernel identity $K^{\rm dr}_{s,s'}=G^{\rm dr}_{s,s'-1}+G^{\rm dr}_{s,s'+1}$, the complete 
set of solutions $F^{(2S)}_{s'}$ for all $2S\in \mathbb{N}$ allow to reconstruct the complete information about the dressed
quasiparticle scattering kernels $K_{AB}$ in the \emph{fundamental} (Heisenberg $S=1/2$) $\SU(2)$ chain \cite{NMKI19}.

\subsection{Infinite temperature limit}

Every admissible set of quasi-particle densities $\rho_{A}(\theta)$ bijectively corresponds to
thermodynamic $\calY$-functions $\calY_{A}(\theta)$. State-specific information thus reveals itself upon analyzing
the analytic structure of the $\calY$-functions. To this end $\calY_{A}(\theta)$ are analytically continued them into the complex
domain (i.e. $\theta$-plane), where they exhibit isolated zeros and poles (or even branch cuts \cite{IQ19}, in principle).
This analytic data is induced by the structure of the local source terms source terms $d_{a,s}(\theta)$, which
are, in turn, directly linked to values of chemical potentials $\mu_{A}(\theta)$ specifying a (generalized) Gibbs ensemble.

It is rather unfortunate that even in the presence of integrability, the TBA equations \eqref{eqn:canonical_TBA} do not in general 
permit non-perturbative analytic solutions. At this junction one is typically resorts to numerical evaluations and solves a truncated 
system of coupled integral equations with use of an appropriate iterative scheme. There is nevertheless an important exception
in this regard, corresponding to grand-canonical Gibbs states in the limit of infinite temperature $\beta \to 0$.
In lattice models invariant under a non-abelian compact Lie group $G$ such a state is always well defined.
Since as $\beta \to 0$ the source terms evidently disappear, $d_{A}\to 0$, the thermodynamic $\calY$-functions
$\calY_{A}(\theta)$ become accordingly flat (i.e. constant functions of rapidity $\theta$) and,
to distinguished them explicitly from $\calY$-functions, we denote them subsequently by $\scrY_{A}$.
In effect, the TBA dressing equations severely simplify and take the form of \emph{algebraic} equations which can be solved in closed 
form. The remainder of this section is devoted to studying this particular case.

\paragraph*{Unitary series.}
Let us briefly return to the basic example of the classical unitary series $\mathfrak{su}(n)$.
Using property $1\star s = \frac{1}{2}$, Eq.~\eqref{eqn:quasi-local_TBA} can be brought in the form
\begin{equation}
\scrY^{2}_{a,s} = \frac{(1+\scrY_{a,s-1})(1+\scrY_{a,s+1})}{(1+1/\scrY_{a-1,s})(1+1/\scrY_{a+1,s})}.
\label{eqn:constant_Y_system}
\end{equation}
where for clarify we have suppressed dependence on the $r=n-1$ $\U(1)$ chemical potentials.
Algebraic relations \eqref{eqn:constant_Y_system} are the celebrated $\calY$-system relation (originally the context
of certain integrable QFTs \cite{Zamolodchikov91}, see also \cite{RVT93}), presently specialized for the absence of the the spectral
parameter $\theta$.

While finite temperature and in general the thermodynamic $\calY$-functions inherit non-trivial $\theta$-dependence
through the local source terms $d_{A}(\theta)$, this nonetheless does not modify their large-$s$ asymptotics,
\begin{equation}
\lim_{|\theta|\to \infty}\calY_{a,s}(\theta) = \scrY_{a,s},
\end{equation}
which is fully determined by the set of distinguished $\U(1)$ chemical potentials $h_{i}$ (for $i \in \mathcal{I}_{r}$).
This particular property makes the task of studying the large-$s$ asymptotic scaling properties much easier, enabling
to carry out the entire analysis in the algebraic setting of the infinite-temperature grand-canonical Gibbs state
(given by Eq.~\eqref{eqn:constant_Y_system}).
For the particular case of $\mathfrak{su}(n)$ algebras, the solution $\scrY_{a,s}(\{x_{i}\})$ of Eqs.~\eqref{eqn:constant_Y_system}
can be expressed as ratios of classical rectangular characters $\chi_{a,s}(g_{0})$,
\begin{equation}
\scrY_{a,s} = \frac{\chi_{a,s-1}\chi_{a,s+1}}{\chi_{a-1,s}\chi_{a+1,s}},
\end{equation}
with $\chi_{0,s} = \chi_{n,0} = 1$, where ${\rm diag}(x_{1},x_{2},\ldots,x_{n})$ (with $\prod_{i=1}^{n}x_{i}=1$) parametrizing the
maximal torus $T$, i.e. the maximal abelian subgroup of $G=\SU(n)$. Classical $\mathfrak{su}(n)$ characters
obey the constant Hirota bilinear relation \cite{Hirota77,KLWZ97,Zabrodin98}
\begin{equation}
\chi^{2}_{a,s} = \chi_{a,s-1}\chi_{a,s+1} + \chi_{a-1,s}\chi_{a+1,s}.
\label{eqn:character_Hirota}
\end{equation}
We provide additional details, including explicit determinant expressions, in appendix \ref{app:characters}.
As a corollary, the infinite-temperature Fermi occupation functions admit a simple algebraic form
\begin{equation}
n_{a,s} = \frac{\chi_{a-1,s}\chi_{a+1,s}}{\chi^{2}_{a,s}},\qquad
\ol{n}_{a,s} = \frac{\chi_{a,s-1}\chi_{a,s+1}}{\chi^{2}_{a,s}}.
\label{eqn:algebraic_Fermi_functions}
\end{equation}

\subsubsection*{Yangian characters}

In order to further examine the formal structure of the TBA $\scrY$-functions, we make a short excursion into the theory
of representations of Yangians $Y(\mathfrak{g})$. As already outlined previously for the particular case of $\mathfrak{su}(n)$,
it is remarkable that the $\scrY$-functions $\scrY_{A}$ (and the associated Fermi functions $n_{A}$)
associated to Yangian symmetry $Y(\mathfrak{g})$ for \emph{any} simple Lie algebra $\mathfrak{g}$
can be expressed in terms of classical $\mathfrak{g}$-characters \cite{KR86}.

Thermodynamic $\mathscr{Y}$-functions $\mathscr{Y}_{a,s}$ can be expressed as certain nonlinear transformations of $\scrT$-functions
as follows
\begin{equation}
\scrY_{a,s} = \frac{\mathbb{T}_{a,s}}{\scrT_{a,s-1}\scrT_{a,s+1}}.
\end{equation}
An infinite set of $\scrT$-functions $\scrT_{a,s}$ here formally correspond to characters associated with rectangular representations 
of Yangians $Y(\mathfrak{g})$, namely
\begin{equation}
\scrT_{a,s} = \chi(\mathcal{W}_{a,s}),
\label{eqn:W_characters}
\end{equation}
satisfying a variant of the Hirota bilinear relation \cite{KNS11}
\begin{equation}
\scrT^{2}_{a,s} = \scrT_{a,s-1}\scrT_{a,s+1} + \mathbb{T}_{a,s}.
\label{eqn:general_Hirota}
\end{equation}
For compactness of notation, we keep dependence on chemical potentials $x_{i}$ of the Cartan $\U(1)$ charges implicit.
The last term in Eq.~\eqref{eqn:general_Hirota} can be formally expressed as
\begin{equation}
\mathbb{T}_{a,s} = \scrT^{2}_{a,s}\prod_{(a',s')}(\scrT_{a',s'})^{{\rm A}_{a s,a's'}},
\end{equation}
for an appropriate $\mathfrak{g}$-dependent `adjacency tensor' ${\rm A}$ which can be found in e.g. \cite{KNS11}
(there, and elsewhere in the literature, Eq.~\eqref{eqn:general_Hirota} is referred to as the `unrestricted Q-system',
which however should not be confused with functional relations amongst the Baxter $Q$-functions).
In the simplest case of simply-laced $\mathfrak{g}$, the last term in Eq.~\eqref{eqn:general_Hirota} assume
a simple form $\mathbb{T}_{a,s} = \prod_{b\sim a}\scrT_{b,s}$, with $\sim$ running over all the adjacent nodes in the Dynkin diagram.
The exact form of $\mathscr{T}$-systems functional relations \eqref{eqn:general_Hirota} are spelled out
in appendix \ref{app:character_expansions}. There we explain how $\scrT_{a,s}$ can be expanded in terms of
$\mathfrak{g}$-characters $\chi_{a,s}$ for the case of classical Lie algebras $\mathfrak{g}$, and exemplify it on
a number of simple cases.

\paragraph*{Remark.}
It is instructive to mention that functional relations \eqref{eqn:general_Hirota} are just a special case of a more general
system of functional relations for rapidity-dependent thermodynamic $\calT$-functions $\calT_{a,s}(\theta)$, namely
the $\theta$-dependent counterpart of Eq.~\eqref{eqn:general_Hirota}.
The latter go commonly under the name of  the $\calT$-system \cite{KP92} (see \cite{KLV16} for an overview), which 
are omnipresent in the literature of exactly solvable models of statistical mechanics.
Thermodynamic state functions $\calT_{a,s}(\theta)$ can be understood as a distinguished gauge-covariant parametrization of
the thermodynamic $\calY$-functions. Every vacuum solution to the $\calT$-system relations represents a distinct equilibrium 
macroscate, where state-dependence is reflected through analytic properties of $\calT$-functions in the `physical strip'
$\mathcal{P}=\{\theta \in \mathbb{C};|{\rm Im}(\theta)|<\tfrac{1}{2}\}$, with large-$\theta$ asymptotics
$\lim_{|\theta|\to \infty}\calT_{a,s}(\theta)=\scrT_{a,s}$.
From a more physical perspective, $\calT_{a,s}(\theta)$ can be identified with eigenvalues of (inhomogeneous) commuting \emph{column}
transfer matrices which associated to the `auxiliary dimension' of a 2D vertex-model realization of a GGE density matrix
(see Ref.~\cite{IQ19} for details and explicit construction). It is thus important to not confuse functions $\calT_{a,s}(\theta)$
with eigenvalues $T_{a,s}(\theta)$ of commuting row transfer matrices ${\bf T}_{a,s}(\theta)$ which act on the physical Hilbert space.

\subsection{Asymptotic analysis}
\label{sec:NewSection1}

After having place the TBA formalism firmly in place, we now return back to superdiffusive transport. We remind that the key properties 
determining charge transport are the asymptotic behavior of the TBA functions for large strings $s \to \infty$.
To this end, we subsequently examine certain formal properties of the dressing equations for the specific algebraic case of the
grand-canonical Gibbs equilibrium ensembles in the limit of infinite temperature. The subject of study are states with unbroken 
symmetry $G$, which requires to set all the $\U(1)$ chemical potentials to zero, that is $x_{i}\to 1$.

To begin with, we notice that $\scrT$-functions $\scrT_{a,s}$ are, by construction, polynomials in $s$
(for all $a \in \mathcal{I}_{r}$). This readily implies that the $\scrY$-functions $\scrY_{a,s}$ are \emph{rational} functions which 
only depend on quantum numbers $(a,s)$. By virtue of Eqs.~\eqref{eqn:algebraic_Fermi_functions}, the Fermi functions scale
as $n_{a,s}\sim 1/s^{2}$ (and hence $\scrY_{a,s}\sim s^{2}$) at large $s$. This is a direct corollary by the Hirota relation
(explicit formulae are given in appendix \ref{app:Tsystem}).
This in turn implies the asymptotic relation $\rho^{\rm tot}_{a,s} \sim s^{2}\rho_{a,s}$.
While both state densities $\rho_{a,s}(\theta)$ and $\rho^{\rm tot}_{a,s}(\theta)$ depend algebraically on $s$,
in distinction to Fermi functions they admit also non-trivial dependence on $\theta$ (even in the infinite temperature limit).

Equipped with this knowledge, we are now in a position to analyze the large-$s$ behaviour of the dressing equations.
We first inspect the basic case of the spin-$1/2$ Heisenberg chain $\SU(2)$ with a single type of magnon species.
It will be sufficient to analyze for the canonical form of the Bethe--Yang equations
\begin{equation}
\rho^{\rm tot}_{s} = K_{s}-\sum_{s^{\prime}\geq 1}K_{s,s'}n_{s'}\rho^{\rm tot}_{s'}.
\end{equation}
Two observations can be readily made: (i) for fixed value of $\theta$ $K_{s}(\theta)=2\pi\,p^{\prime}_{s}(\theta)\sim 1/s$, and
(ii) due to telescopic cancellation of poles the differential scattering phase shifts $K_{s,s'}$ involve $2\,{\rm min}(s,s')$ terms.
Based on this, we can make the following estimate at large $s$
\begin{equation}
\rho^{\rm tot}_{s} \lesssim \frac{1}{s} + \sum_{s'=1}^{s}2s'n_{s'}\rho^{\rm tot}_{s'}
+ 2s\sum_{s'=s+1}^{\infty}n_{s'}\rho^{\rm tot}_{s'},
\end{equation}
where each term under the sum has been upper-bounded by a constant using that $1\star K_{s}=1$.
The second sum can be understood as the residual contribution coming from large $s^{\prime}$-strings with $s^{\prime}>s$,
which gets suppressed in the $\sim 1/s$ fashion. The first sum converges in the $s\to \infty$ limit provided
$\lim_{s\to 0}\rho^{\rm tot}_{s}=0$, i.e. the total density of states $\rho^{\rm tot}_{s}$ has to decay to zero at large $s$
in an algebraic fashion.
Using finally that the bulk recurrence relations pertaining to the algebraic dressing equations involve only rational functions 
of $s$, this implies $\rho^{\rm tot}_{s} \sim 1/s$ scaling at large $s$.

In more general integrable models (with nested spectra), one has to additionally account for the quasiparticles of different flavors,
whose mutual interaction are prescribed by the Cartan matrix $\mathcal{K}$. The above argument can be easily adapted
by including an extra sum over the flavor index.

\subsection{Analytic solutions}
\label{sec:NewSection2}

To solidify the conclusion of the preceding section, we now work out the exact closed-form solution to the algebraic equations.
We note that the basic case of the $\SU(2)$ spin-$S$ (Babujian--Takhtajan) quantum chains, the full solution has already been obtained
in ref.~\cite{NMKI19}. Our primary interest here are model that possess Lie symmetries of higher rank.

For definiteness, we specialize below to $\SU(n)$ quantum chains with fundamental onsite degrees of freedom.
While other cases can be treated in a similar fashion, for compactness of presentation we postpone a complete and comprehensive 
analysis for a separate technical study.

Let us consider, as our starting point, the algebraic limit of the group-theoretic dressing equations
\eqref{eqn:universal_dressing_equations}. In Fourier $k$-space, these turn into a system of coupled \emph{algebraic} equations
of the form
\begin{align}
s^{-1}\star F_{a,s} &- \ol{n}_{a,s-1}F_{a,s-1} - \ol{n}_{a,s+1}F_{a,s+1} \nonumber \\
&- n_{a-1,s}F_{a-1,s} - n_{a+1,s}F_{a+1,s} = \delta_{a,s},
\label{eqn:algebraic_dressing}
\end{align}
with $s^{-1}(k)=2\cosh{(k/2)}$. The first step is to solve the homogeneous part, which can be achieved with the following ansatz
\begin{equation}
F_{a,s}(k) = A_{a,s}(k)K_{a+s-1}(k) + B_{a,s}(k)K_{a+s+1}(k),
\end{equation}
where $K_{s}(k)=e^{-s|k|/2}$ are the elementary scattering kernels in Fourier space.
Plugging the ansatz into Eqs.~\eqref{eqn:algebraic_dressing}, Laplace transforming from the $k$-plane
to the $z$-plane, and demanding the null condition for all the residues located at $z_{i} \in \tfrac{1}{2}\mathbb{N}$,
we end up with following system of recurrence relations
\begin{align}
A_{a,s} &= \ol{n}_{a,s-1}A_{a,s-1} + n_{a-1,s}A_{a-1,s},\\
B_{a,s} &= \ol{n}_{a,s+1}B_{a,s+1} + n_{a+1,s}B_{a+1,s},\\
A_{a,s} - B_{a,s} &= \ol{n}_{a,s+1}A_{a,s-1} + n_{a+1,s}A_{a+1,s} \nonumber \\
&- \ol{n}_{a,s-1}B_{a,s-1} + n_{a-1,s}B_{a-1,s},
\label{eqn:sun_recurrences}
\end{align}
where the Fermi functions are given by $\mathfrak{su}(n)$ characters as per Eqs.~\eqref{eqn:algebraic_Fermi_functions}.
The solution has the form
\begin{align}
A_{a,s} &= A_{0}\frac{\chi_{a,s}\chi_{a-1,s-1}}{\chi_{a-1,s}\chi_{a,s-1}},\\
B_{a,s} &= B_{0}\frac{\chi_{a,s}\chi_{a+1,s+1}}{\chi_{a+1,s}\chi_{a,s+1}},
\end{align}
for some yet to determined coefficients $A_{0}$ and $B_{0}$. By plugging these back to Eqs.~\eqref{eqn:sun_recurrences}, we retrieve
the Hirota relation for classical characters \eqref{eqn:character_Hirota}. To fix the undetermined constants, we find
the particular solution of Eqs.~\eqref{eqn:algebraic_dressing}, with the source term attached at the first node at $(a,s)=(1,1)$,
yielding
\begin{equation}
A_{0} = -B_{0} = \frac{1}{\chi_{1,1}}.
\end{equation}
Transforming back to the $\theta$-plane, we finally find
\begin{align}
F_{a,s}(\theta) = \frac{\chi_{a,s}}{\chi_{1,1}}\Big[&\frac{\chi_{a-1,s-1}}{\chi_{a-1,s}\chi_{a,s-1}}K_{a+s-1}(\theta) \nonumber \\
-&\frac{\chi_{a+1,s+1}}{\chi_{a+1,s}\chi_{a,s+1}}K_{a+s+1}(\theta)\Big].
\end{align}
At the end we are interested in the $x_{i}\to 1$ limit, in which the $\mathfrak{su}(n)$ characters become dimensions $d_{a,s}$
or rectangular irreducible representations $\mathcal{V}_{s\,\omega_{a}}$ (see appendix \ref{app:characters} for details).
Specifically, the total state densities read
\begin{equation}
\rho^{\rm tot}_{a,s}(\theta) = \frac{s+n-a}{n}\left[\frac{K_{s+a-1}(\theta)}{s+a-1}-\frac{K_{s+a+1}(\theta)}{s+a+1}\right].
\end{equation}
There is an analogous expression (up to a multiplicative prefactor) for the dressed differential quasiparticles energies
$\varepsilon^{\prime}_{a,s}(\theta)$ which can be obtained by replacing $K_{s}(\theta)$ with $K^{\prime}_{s}(\theta)$.
In rapidity space, the latter is given by a discrete family of Cauchy distributions
\begin{equation}
K_{s}(\theta) = \frac{1}{2\pi \ii}\partial_{\theta}\log S_{s}(\theta) = \frac{1}{2\pi}\frac{s}{\theta^{2} + (s/2)^{2}}.
\end{equation}

Notice that any finite fixed rapidity $\theta$, the density of states $\rho^{\rm tot}_{a,s}$ decays as $\sim 1/s^{2}$
in the large-$s$ limit. However, when integrated against any dummy integrable function $f(\theta)$ as in GHD,
we have the following large-$s$ properties
\begin{align}
\label{eqn:asy1}
\int_{\mathbb{R}} \dd \theta\, f(\theta)\rho^{\rm tot}_{a,s}(\theta) &\sim \frac{1}{s},\\
\int_{\mathbb{R}} \dd \theta\, f(\theta) v^{\rm eff}_{a,s}(\theta) &\sim \frac{1}{s},
\label{eqn:asy2}
\end{align}
for every quasiparticle flavor $a$.

As announced in Sec.~\ref{sec:superdiffusion}, these properties allow us to infer the superdiffusive scaling of charge dynamics
with an algebraic dynamical exponent $z=3/2$.

\paragraph*{Non-fundamental onsite representations.}

The large-$s$ scaling properties \eqref{eqn:asy1} and \eqref{eqn:asy2} likewise hold in models with non-fundamental local
degrees of freedom, that is for any on-site, finite-dimensional unitary irreducible representations $\mathcal{V}_{\Lambda}$.
At the level of algebraic dressing equations, this amounts to moving the source terms to a generic position.
For a quick illustration, we derive the explicit solution to the  $\SU(3)$ chain in the adjoint representation $({\bf 8})$,
with Dynkin labels $\Lambda_{\rm p}=\omega_{1}+\omega_{2}\equiv [1,1]$. By virtue of $\mathbb{Z}_{2}$ invariance of the dressing 
equations (under interchanging the flavors) we have $F_{1,s}=F_{2,s}$, and therefore
\begin{equation}
s^{-1}\star F_{a,s} - I^{A_{\infty}}_{s,s'}\ol{n}_{s'}F_{a,s'} - n_{s}F_{a,s} = \delta_{s,1},
\end{equation}
where $I^{A_{\infty}}_{s,s'}=\delta_{s,s'-1}+\delta_{s,s'+1}$. The solution reads
\begin{equation}
F_{a,s} = \frac{1}{3}\left[\frac{s+2}{s}K_{s}+K_{s+1}-K_{s+2}-\frac{s+1}{s+3}K_{s+3}\right].
\end{equation}

\section{Semiclassical spectrum}
\label{sec:semiclassical}

Thus far we have established that the anomalous charge transport comes from giant quasiparticles (i.e. bound states carrying large
quanta $s$) immersed in a charge-neutral thermal background. To give a different angle and better elucidate their physical nature, we 
next discuss a semiclassical description of these giant quasiparticles. It is helpful to first review some results along these lines 
that were shown for $\SO(3)$ spin chains~\cite{NGIV20}. We then proceed by outlining how these arguments generalize to other symmetry 
groups.

We begin with a classical $\SO(3)$ ferromagnet in its ground state, i.e., all spins aligned in some direction. The elementary 
excitations of the ferromagnet are Goldstone modes in which the spin orientation changes slowly across the lattice. One can regard any 
sufficiently slowly varying spin texture as being composed of Goldstone modes, as it locally consists of slow modulations (rotations) 
of the vacuum orientation. In one dimension, the long-wavelength dynamics of these ferromagnetic, \emph{quadratically dispersing}, 
Goldstone modes is governed by the Landau--Lifshitz equation, which is nonlinear but integrable PDE possessing stable soliton solutions 
of any width $w$~\cite{lakshmanan1976}. The properties of these soliton solutions can be inferred by observing that they are 
wavepackets of Goldstone modes that are stabilized by nonlinearity. For example, a soliton of width $w$ is made up of Goldstone modes 
with characteristic momentum $1/w$ and thus characteristic energy density $1/w^2$; therefore its characteristic \emph{energy} scales as 
$1/w$ and its velocity also as $1/w$. On the other hand, its spin orientation is away by ${\cal O}(1)$ from the vacuum so it carries 
spin $\sim w$. These ``bare'' properties of classical solitons precisely match those of the large-$s$ Bethe ansatz strings, suggesting 
that there is some correspondence between them. Indeed, the exact correspondence between large-$s$ strings and large-$w$ solitons can 
be explicitly established, not only at the level of individual spin-wave configurations as usual, but \emph{even} in high-temperature
equilibrium ensembles~\cite{NGIV20}.

This classical quasiparticle construction generalizes to arbitrary quantum or classical ferromagnets. As discussed below, a 
distinguished feature of such ferromagnets are quadratically dispersing Goldstone modes. One can analogously construct slowly varying 
wavepackets of these Goldstone modes; by construction these have the same scaling as the large-$s$ strings, so it is tempting to 
identify the two types of excitations in the general case also. Although we have not yet attempted to analytically solve the classical 
problem in full generality, by noting that large-$s$ strings must correspond to objects in the semiclassical spectrum and that 
Goldstone-like excitations exhaust this spectrum, it is tempting to infer the existence of such solitons from the integrability of the 
quantum model.

However, one major puzzle arises when one tries to perform such an identification: the number of magnon flavors
(which is always set by the rank $r$ of the group $G$) is far fewer than that of Goldstone modes; the latter is given by one half
of the real dimension of a coset manifolds $G/H$, where $H$ is the residual symmetry group of the ferromagnetic state.
The ``missing'' Goldstone bosons \emph{must} in some fashion emerge out of the Bethe ansatz spectrum; however, they cannot
be identified directly with elementary (physical and auxiliary) magnons of integrable quantum chains.
Curiously, they rather correspond to `stacks' of magnons, which are long-lived wavepackets in a finite system, and only become sharp 
eigenstates in the infinite system limit.

The main result of this section is identifying the correspondence between Goldstone modes in the semiclassical spectrum and stacks of 
magnons (and condensates thereof, representing semi-classical Bethe $s$-strings) in the Bethe ansatz spectrum.
Using this correspondence, one can identify large-spin semiclassical excitations above the ferromagnetic vacuum precisely as stacked 
strings. These semiclassical excitations can then be identified as solitons that involve smooth maps from $\mathbb{R}$ to the coset 
space $G/H$. While it would be interesting to construct such solitons directly in the classical theory, this task is left to future 
work.

\subsection{Goldstone modes}

As a natural starting point to describe classical continuous ferromagnets, we being by characterizing the spectrum of linear
fluctuations of a ferromagnetic order parameter. This leads to the notion of Goldstone excitations which naturally arise in systems 
with spontaneously broken continuous internal symmetry. To this end, we begin by shortly revisiting the Goldstone theorem in a general 
context of non-relativistic field theories. Our objective here is mainly to characterize the (non-relativistic) Goldstone modes
that govern the low-energy spectrum for the class of ferromagnetic quantum spin chains introduced in Sec.~\ref{sec:Hamiltonians}.

First we recapitulate the main result on the counting of Goldstone modes systems without Lorentz invariance, independently worked out 
in Refs.~\cite{WM12,WM13} and \cite{Hidaka13} (optimizing the earlier Nielsen--Chadha inequality \cite{NielsenChadha76}).
The conventional setting concerns Hamiltonian systems invariant under a non-abelian Lie group $G$, possessing a
degenerate ferromagnetic ground state manifold. Spontaneous breaking of internal symmetry amounts to picking a particular vacuum 
polarization (order parameter), say $\Omega$, thereby breaking the symmetry of the isometry group $G$ down to stability subgroup
$H \subset G$ of $\Omega$, determined by condition
\begin{equation}
h\,\Omega\,h^{-1} = \Omega,\qquad h \in H.
\label{eqn:vacuum_stability}
\end{equation}
We are specifically interested in homogeneous (i.e. translational invariant) ferromagnetic chains (of length $L$),
where the global (pseudo)vacuum is simply a product state $\Omega_{L} = \bigotimes_{\ell=1}^{L}\Omega$.
It is therefore sufficient to carry out the analysis at the level of local Hilbert spaces and (with no loss of generality)
we can assume $\Omega\equiv \ket{0}\bra{0}$.

At the algebraic level, the symmetry breaking pattern $G\to H$ implies  the splitting
$\mathfrak{g} = \mathfrak{h} \oplus \mathfrak{m}$, with $\mathfrak{h}$ being the Lie algebra that generates $H$ (which contains
the Cartan subalgebra $\mathfrak{t}$) and and $\mathfrak{m}$ denoting the linear span of the `broken generators'
(these do not enclose an algebra). The number of broken generators is
\begin{equation}
n_{b} = {\rm dim}\,\mathfrak{m} = {\rm dim}(G) - {\rm dim}(H).
\end{equation}
Let $\rX^{\sigma}\in \mathfrak{m}$, $\sigma \in \{1,2,\ldots,{\rm dim}(G/H)\}$ be a hermitian basis of $\mathfrak{m}$.
By the Goldstone counting theorem, the number of independent Goldstone modes $n_{\rm GB}$ is
\begin{equation}
n_{\rm G} = n_{b} - \frac{1}{2}{\rm rank}\,\mathbf{V},
\end{equation}
with matrix
\begin{equation}
\mathbf{V}_{ab} = -\ii \bra{0}[\rX^{a},\rX^{b}]\ket{0},
\end{equation}
storing the vacuum expectation values of the commutators amongst the broken generators $X^{\sigma}$.

Goldstone modes come in two varieties, depending on the type of the dispersion law at long wave-lengths ($k\to 0$),
$\omega(k)\sim k^{\nu}$: type-I (or type-$A$) with $\nu$ \emph{odd}, and type-II (or type-$B$) with $\nu$ \emph{even}.
In general, we thus have
\begin{equation}
n_{\rm G} = n_{\rm I} + n_{\rm II},
\end{equation}
with
\begin{equation}
n_{\rm I} = n_{b} - 2\,n_{\rm II},\quad
n_{\rm II} = \frac{1}{2}{\rm rank}\,\mathbf{V}.
\end{equation}
In relativistic systems, Lorentz invariance forces $\mathbf{V}$ to be identically zero. In contrast, target spaces of
non-relativistic ferromagnets are symplectic manifolds. These have even (real) dimension and consequently $\mathbf{V}$ has full rank.
Therefore $n_{\rm I}=0$ and
\begin{equation}
n_{\rm G} \equiv n_{\rm II} = \frac{1}{2} n_{b}.
\end{equation}
In simple terms, every canonically-conjugate pair of broken generators contribute an independent \emph{magnon} branch,
a quadratically dispersing long-wavelength spin-wave excitation of the ferromagnetic order parameter.

\subsection{Symmetry breaking patterns}
\label{sec:patterns}

We now describe the spectra of Goldstone modes for the class of $G$-invariant ferromagnetic chains.
The structure of coset spaces depends on the choice of a finite-dimensional irreducible onsite representation
$\mathcal{V}_{\Lambda}$. Let us for simplicity suppose for the moment that the local degrees of freedom transform in the defining 
representation of $\mathfrak{g}$. The highest weight is the first fundamental weight $\Lambda = \omega_{1}=[1,0,\ldots,0]$
of $\mathfrak{g}$, and the stabilizer $H \subset G$ of the vacuum state $\Omega$ has the structure $\U(1)\times H^{\prime}$
(i.e. $H$ is not semi-simple). For the classical series, the coset structure is summarized in Table \ref{tabl:classical_target_spaces}.
\begin{center}
\centering
\begin{table}
\renewcommand{\arraystretch}{1.2}
\setlength{\tabcolsep}{0pt}
\begin{tabular}{|c|c|c|c|}
\hline
\rowcolor{pink}
$\mathfrak{g}$ & $\mathcal{V}_{\omega_{1}}$ & Target space $G/H$ & $n_{\rm G}$ \\
\hline \hline
$A_{n}$ & $({\bf n})$ & $\SU(n)/\U(1)\times \SU(n-1)$ & $n-1$ \\
$\,B_{n}\, $ & $\,({\bf 2n+1})\,$ & $\quad \SO(2n+1)/\U(1)\times \SO(2n-1)\quad$ & $\quad 2n-1 \quad$ \\
$C_{n}$ & $({\bf 2n})$ & $\USp(2n)/\U(1)\times \USp(2n-2)$ & $2n-1$ \\
$D_{n}$ & $({\bf 2n})$ & $\SO(2n)/\U(1)\times \SO(2n-2)$ & $2n-2$ \\
\hline
\end{tabular}
\caption{Target spaces associated with classical continuous ferromagnets, describing the low-energy sector of quantum ferromagnetic 
chains with local degrees of freedom realized in the defining representation $\mathcal{V}_{\omega_{1}}$ of a
Lie algebras $\mathfrak{g}$. $n_{G}$ is the number of Goldstone modes in the spectrum.}
\label{tabl:classical_target_spaces}
\end{table}
\end{center}

Next, we specialize to the family of fundamental representations with $\Lambda = \omega_{a}$.
In this case, the symmetry breaking pattern can be inferred directly from the enumerated Dynkin diagrams with use of a graphical 
recipe by simply breaking the bonds which connect to the node containing the non-zero label.
This is illustrated in Fig.~\ref{fig:breaking_Dynkin} on a two simple examples.
The list of possibilities for all the classical series is summarized in Table \ref{tab:symmetry_breaking_patterns}.
Employing the same trick allows us to determine the structure of coset spaces also for $G$-invariant classical ferromagnets 
corresponding to the exceptional groups $G_{2},F_{4},E_{6-8}$, specializing to the \emph{fundamental} representations
of $\mathfrak{g}$ (for completeness we include them in appendix \ref{app:exceptional}).

\begin{center}
\centering
\begin{table}[htb]
\renewcommand{\arraystretch}{1.2}
\setlength{\tabcolsep}{0pt}
\begin{tabular}{|c|c|c|c|}
\hline
\rowcolor{pink}
$\mathfrak{g}$ & irrep $\Lambda$ & Stabilizer $H$ & ${\rm dim}(G/H)$ \\
\hline \hline
$A_{n}$ & $\omega_{k}$ & $A_{k-1}\times \U(1)\times A_{n-k}$ & $2k(n+1-k)$ \\
\hline
$B_{n}$ & $\omega_{k\leq n-2}$ & $A_{k-1}\times \U(1)\times B_{n-k}$ & $k(4n+1-3k)$ \\
$B_{n}$ & $\omega_{n-1}$ & $A_{n-2}\times \U(1)\times A_{1}$ & $(n+4)(n-1)$ \\
$B_{n}$ & $\omega_{n}$ & $A_{n-1}\times \U(1)$ & $n(n+1)$ \\
\hline
$C_{n}$ & $\omega_{k\leq n-2}$ & $A_{k-1}\times \U(1)\times C_{n-k}$ & $k(4n+1-3k)$ \\
$C_{n}$ & $\omega_{n-1}$ & $A_{n-2}\times \U(1)\times A_{1}$ & $(n+4)(n-1)$ \\
$C_{n}$ & $\omega_{n}$ & $A_{n-1}\times \U(1)$ & $n(n+1)$ \\
\hline
$D_{n}$ & $\omega_{k\leq n-3}$ & $A_{k-1}\times \U(1)\times C_{n-k}$ & $\quad k(4n-1-3k) \quad$ \\
$D_{n}$ & $\omega_{n-2}$ & $\, A_{n-3}\times \U(1)\times A_{1} \times A_{1} \,$ & $(n+5)(n-2)$ \\
$\,D_{n}\,$ & $\,\omega_{n-1},\,\omega_{n}\,$ & $A_{n-1}\times \U(1)$ & $n(n-1)$ \\
\hline
\end{tabular}
\caption{Symmetry breaking patters for all the fundamental representations $\Lambda=\omega_{k}$ for the family of
classical Lie algebras $\mathfrak{g}$.}
\label{tab:symmetry_breaking_patterns}
\end{table}
\end{center}

The outlined recipe likewise applies for generic (non-fundamental) finite dimensional irreducible representations 
$\mathcal{V}_{\Lambda}$ of $\mathfrak{g}$ with Dynkin labels $\Lambda = [\omega_{1},\ldots,\omega_{r}]$; the rule is to
break all the bonds that connect to nodes with non-zero Dynkin labels. The resulting coset spaces are \emph{generalized flag manifolds}
$\mathscr{F}_{i_{1},i_{2},\ldots,i_{k}} \equiv G/H_{i_{1},i_{2},\ldots_{i_{k}}}$,
where the indices $i_{i}$ mark the nodes of the Dynkin diagram of $\mathfrak{g}$ with non-vanishing Dynkin labels
(that is $i_{i}=1$ if and only if $m_{i}\neq 0$).
Compact complex manifolds $\mathscr{F}_{i_{1},i_{2},\ldots,i_{k}}$
indeed exhaust all K\"{a}hler manifolds (i.e. possess compatible Riemannian metric and symplectic structures) which are homogeneous 
spaces with a transitive action of $G$.
Furthermore, points on such flag manifolds are in one-to-one correspondence with generalized coherent states in a 
specific representation of group $G$, namely
\begin{equation}
\ket{\psi} = \exp{\Big[\sum_{w^{-}\in \Delta_{+}}w^{-}_{\alpha}{\rm E}^{-\alpha}\Big]}\ket{\Lambda},
\label{eqn:coherent_state}
\end{equation}
for the highest weight state $\ket{\Lambda}$ of $\mathcal{V}_{\Lambda}$. If some $m_{a}\neq 0$ however, not all
coordinates $w^{-}_{\alpha}$ are independent and some can be eliminated.
For instance, in the unitary case (A-series), the family of generalized flag manifolds has the structure
\begin{equation}
SU(n)/S(U(n_{1})\times U(n_{2})\times \cdots \times U(n_{j})\times U(1)^{\times k}),
\label{eqn:unitary_flags}
\end{equation}
for integers $j,k\geq 1$ and $n_{1}\geq n_{2}\geq \ldots \geq n_{j}>1$, subjected to $k+\sum_{i=1}^{j}n_{i}=n$.
For example, for the case of $G=\SU(3)$, there are two possible isotropy groups: $H_{1,0}=H_{0,1}=U(2)$ (as e.g. for $({\bf 3})$, 
$(\bar{\bf 3})$ or $({\bf 6})$) or $H_{1,1}=U(1)^{\times 2}$ (as e.g. for $({\bf 8})$, $({\bf 15})$ or $({\bf 27})$).
The upshot of that is that physical degrees of freedom of the low-energy effective theories corresponding to quantum spin chains 
whose degrees of freedom transform in different irreducible representations can live on the same target coset manifold.
Below we explain this fact from another perspective by directly examining the structure of semi-classical spectrum.

\paragraph*{Hermitian symmetric spaces.}

An important subclass of flag manifolds are irreducible compact \emph{hermitian} symmetric spaces.
Amongst cosets \eqref{eqn:unitary_flags} one finds (in the special cases of $\Lambda = \omega_{a}$) complex Grassmannian manifolds
$SU(n)/S(U(k)\times U(n-k))$. The remaining infinite families of classical (i.e. non-exceptional) hermitian symmetric spaces comprise
(i) Lagrangian Grassmannians $\USp(2n)/\U(n)$ are associated to representations $\Lambda=\omega_{n}=[0,0,\ldots,1]$ from
the $C_{n}$-series, $\SO(2n)/U(n)$ of complex dimension $\frac{1}{2}n(n-1)$ obtained from the fundamental spinor representations
of $D_{n}$ (with Dynkin labels $\Lambda \in \{\omega_{n-1},\omega_{n}\}$), and (ii) orthogonal Grassmannians
$\SO(n)/\SO(n-2)\times \SO(2)$ or real dimension $2n$ obtained from the defining $\SO(n)$ representations $\Lambda = \omega_{1}$ of 
$B_{n}$ or $D_{n}$ (notice however that compact Riemannian symmetric spaces, such as e.g. $\SU(n)/\SO(n)$ and $\SU(2n)/\USp(2n)$,
do not appear). Concerning charge transport, we note that a recent numerical study \cite{MatrixModels} demonstrated that a class
of \emph{integrable} non-relativistic \emph{classical} sigma models (of the Landau--Lifshitz type) on complex and Lagrangian 
Grassmannians supports anomalous charge transport with dynamical exponent $z=3/2$ (and KPZ scaling profiles).

\begin{widetext}

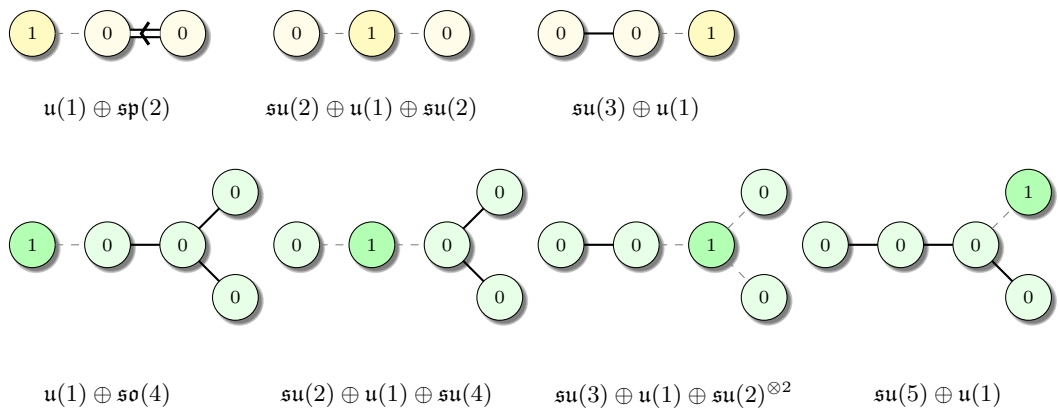
\begin{figure}[htb]
\centering
\begin{tikzpicture}[
	gcirc/.style = {circle, draw, fill=green!10, circular drop shadow, minimum size = 6pt,inner sep = 4pt, outer sep=0pt},
	ycirc/.style = {circle, draw, fill=yellow!10, circular drop shadow, minimum size = 6pt,inner sep = 4pt, outer sep=0pt}
]


\begin{scope}[xshift=-200pt]
\node (label1) at (1,-2) {$\mathfrak{u}(1)\oplus \mathfrak{so}(4)$};
\node[gcirc,fill=green!30] (0) at (0,0) {\scriptsize $1$};
\node[gcirc] (1) at (1,0) {\scriptsize $0$};
\node[gcirc] (2) at (2,0) {\scriptsize $0$};
\node[gcirc] (3) at (2.7,0.7) {\scriptsize $0$};
\node[gcirc] (4) at (2.7,-0.7) {\scriptsize $0$};
\draw[-,gray,dashed] (0) -- (1);
\draw[-,thick] (1) -- (2);
\draw[-,thick] (2) -- (3);
\draw[-,thick] (2) -- (4);
\end{scope}

\begin{scope}[xshift=-100pt]
\node (label2) at (1.2,-2) {$\mathfrak{su}(2)\oplus \mathfrak{u}(1)\oplus \mathfrak{su}(4)$};
\node[gcirc] (0) at (0,0) {\scriptsize $0$};
\node[gcirc,fill=green!30] (1) at (1,0) {\scriptsize $1$};
\node[gcirc] (2) at (2,0) {\scriptsize $0$};
\node[gcirc] (3) at (2.7,0.7) {\scriptsize $0$};
\node[gcirc] (4) at (2.7,-0.7) {\scriptsize $0$};
\draw[-,gray,dashed] (0) -- (1);
\draw[-,gray,dashed] (1) -- (2);
\draw[-,thick] (2) -- (3);
\draw[-,thick] (2) -- (4);
\end{scope}

\begin{scope}
\node (label3) at (1.5,-2) {$\mathfrak{su}(3)\oplus \mathfrak{u}(1)\oplus \mathfrak{su}(2)^{\otimes 2}$};
\node[gcirc] (0) at (0,0) {\scriptsize $0$};
\node[gcirc] (1) at (1,0) {\scriptsize $0$};
\node[gcirc,fill=green!30] (2) at (2,0) {\scriptsize $1$};
\node[gcirc] (3) at (2.7,0.7) {\scriptsize $0$};
\node[gcirc] (4) at (2.7,-0.7) {\scriptsize $0$};
\draw[-,thick] (0) -- (1);
\draw[-,gray,dashed] (1) -- (2);
\draw[-,gray,dashed] (2) -- (3);
\draw[-,gray,dashed] (2) -- (4);
\end{scope}

\begin{scope}[xshift=100pt]
\node (label3) at (1.5,-2) {$\mathfrak{su}(5)\oplus \mathfrak{u}(1)$};
\node[gcirc] (0) at (0,0) {\scriptsize $0$};
\node[gcirc] (1) at (1,0) {\scriptsize $0$};
\node[gcirc] (2) at (2,0) {\scriptsize $0$};
\node[gcirc,fill=green!30] (3) at (2.7,0.7) {\scriptsize $1$};
\node[gcirc] (4) at (2.7,-0.7) {\scriptsize $0$};
\draw[-,thick] (0) -- (1);
\draw[-,thick] (1) -- (2);
\draw[-,gray,dashed] (2) -- (3);
\draw[-,thick] (2) -- (4);
\end{scope}


\begin{scope}[xshift=-200pt,yshift=80]
\node (label1) at (1,-1) {$\mathfrak{u}(1)\oplus \mathfrak{sp}(2)$};
\node[ycirc,fill=yellow!30] (0) at (0,0) {\scriptsize $1$};
\node[ycirc] (1) at (1,0) {\scriptsize $0$};
\node[ycirc] (2) at (2,0) {\scriptsize $0$};
\draw[-,gray,dashed] (0) -- (1);
\draw[-,thick,double distance = 2pt,postaction = {decorate}] (1) -- (2);
\draw[-,very thick] (1.55,0.15) -- (1.45,0) -- (1.55,-0.15);
\end{scope}

\begin{scope}[xshift=-100pt,yshift=80]
\node (label2) at (1,-1) {$\mathfrak{su}(2)\oplus \mathfrak{u}(1) \oplus \mathfrak{su}(2)$};
\node[ycirc] (0) at (0,0) {\scriptsize $0$};
\node[ycirc,fill=yellow!30] (1) at (1,0) {\scriptsize $1$};
\node[ycirc] (2) at (2,0) {\scriptsize $0$};
\draw[-,gray,dashed] (0) -- (1);
\draw[-,gray,dashed] (1) -- (2);
\end{scope}

\begin{scope}[yshift=80]
\node (label3) at (1,-1) {$\mathfrak{su}(3)\oplus \mathfrak{u}(1)$};
\node[ycirc] (0) at (0,0) {\scriptsize $0$};
\node[ycirc] (1) at (1,0) {\scriptsize $0$};
\node[ycirc,fill=yellow!30] (2) at (2,0) {\scriptsize $1$};
\draw[-,thick] (0) -- (1);
\draw[-,gray,dashed] (1) -- (2);
\end{scope}

\end{tikzpicture}
\caption{Various symmetry breaking patters in fundamental ferromagnets with onsite irreduicible representations 
$\mathcal{V}_{\omega_{a}}$, depicted for Lie algebras $C_{4}=\mathfrak{sp}(4)$ (\emph{top})
and $D_{5}=\mathfrak{so}(10)$ (\emph{bottom}). The vacuum stability subgroup $H$ always includes a $\U(1)$ factor.
The $n$th Dynkin node with label $1$ corresponds to the $a$th fundamental irreducible representation with highest weight
$\Lambda=\omega_{a}$. Dashed lines designate the broken bonds.}
\label{fig:breaking_Dynkin}
\end{figure}

\end{widetext}

\subsection{Counting degrees of freedom}

Magnon modes can arise as plane-wave solutions to linear partial differential equations.
On the other hand, classical field theories are interacting systems governed by \emph{nonlinear} equations of motion which possess
a vastly richer spectrum of solutions. A hallmark feature of nonlinear completely integrable PDEs are soliton solutions.
Solitons refer to nonlinear ballistically propagating field configurations that behave as particles, whose stability is ensured
by integrability. In isotropic ferromagnets considered in this work, magnetic solitons assume a non-trivial internal structure.
Indeed, from the perspective of quantum spin chains, soliton emerge as certain macroscopic coherent superpositions of magnons.
Such states represent highly-excited semiclassical eigenstates in the the low-energy spectrum of the model and their time evolution
is generated and an effective \emph{classical} Hamiltonian. This fact can be established in various ways;
the most standard and direct is employ a standard path-integral technique to project the quantum many-body Hamiltonian onto the 
manifold of (generalized) coherent states $\ket{\psi}$, given formally by the `vacuum rotation' $\ket{\psi}=g\ket{\Lambda}$,
as prescribed by Eq.~\eqref{eqn:coherent_state}. This way one can deduce a classical evolution for the ferromagnetic order 
parameter that takes values on a quotient manifolds $G/H$, where $H$ is the vacuum stability subgroup defined above
in Eq.~\eqref{eqn:vacuum_stability} (see also e.g. \cite{Stefanski04,Stefanski07,MatrixModels}). Classical fields can be thus 
identified with coordinates of generalized (Perelomov) coherent states \cite{Perelomov86}.

Every admissible classical solution is a particular long-wavelength (i.e. low-momentum) solution of the Bethe Ansatz equations with
a macroscopically large number of quanta. This particular regime, often referred to as the `asymptotic Bethe Ansatz', has
been first studied by Sutherland \cite{Sutherland1975} and subsequently more thoroughly examined in \cite{DS00}.
Nevertheless, an exact identification with classical field solutions has however only been made precise afterwards
in Ref.~\cite{Kazakov2004} which provided a prescription to construct the associated classical spectral curve
(the characteristic equation associated with the classical monodromy matrix, see e.g.~\cite{Bargheer08,SchaferNameki11}) by 
transcribing the asymptotic Bethe equations as a Riemann--Hilbert problem. This method  permits to describe and classify the nonlinear 
modes for the class of so-called finite-gap solutions \cite{DKN01}.

In the remainder of this section we explain how the spectrum of linear and nonlinear mode of classical integrable ferromagnets
arises from the nested magnon spectra in integrable $G$-invariant ferromagnetic quantum chains. Complete characterization of 
classical algebraic curves that emerge in this picture is a technical matter exceeds our scope and we thus do not undertaking it here. 
We rather focus here on a deceptively innocent problem of identifying the relevant emergent classical degrees of freedom.
The task turns out to be quite subtle and in order to resolve we will need a basic description of the asymptotic Bethe Ansatz.

\subsubsection*{Asymptotic Bethe Ansatz}

We begin by recalling that every finite-volume (highest-weight) Bethe eigenstates in a quantum chain can be uniquely resolved
in terms of magnons excitations. Individual unbond magnons correspond to real-valued Bethe roots with rapidities
$\theta \sim \mathcal{O}(L^{0})$. As described earlier, complex-valued rapidities of constituent magnons which are a part of a bound 
state take the form of Bethe strings, see Eq.~\eqref{eqn:Bethe_string}. Semiclassical eigenstates in the spectrum belong to
highly-excited states with $\mathcal{O}(L)$ excitations and large rapidities $\theta \sim \mathcal{O}(L)$, which can be thought
of condensing into a finite number of coherent macroscopic modes. In this low-energy regime, magnon momenta are inversely proportional 
to the system size, $p(\theta)\sim \mathcal{O}(1/L)$. By accordingly introducing rescaled rapidities $\lambda^{(a)}_{j}$ via 
$\theta^{(a)}_{j}=L\,\lambda^{(a)}_{j}$, the Bethe equations (i.e. the logarithm of Eq.~\eqref{eqn:BetheEqs}) in the leading order 
$\mathcal{O}(L)$ assume the form
\begin{equation}
2\pi n^{(a)}_{j} + \frac{\hat{m}_{a}}{\lambda^{(a)}_{j}}
= \sum_{b=1}^{r}\sum_{k=1,k\neq j}^{N_{b}}\frac{t^{-1}_{a}\mathcal{K}_{ab}}{\lambda^{(a)}_{j}-\lambda^{(b)}_{k}}.
\label{eqn:ABA_equations}
\end{equation}
As usual, $n^{(a)}_{j}$ represent (integer) mode numbers due to multivaluedness of complex logarithm.

Classical nonlinear modes pertain to bound states of $N_{a}\sim \mathcal{O}(L)$ constituent low-momentum magons that all share
the same mode number. Crucially, rapidities of nearby magnons are separated by an amount $o(L)$; however, since
$\theta \sim \mathcal{O}(L)$ magnons need not be spaced equidistantly as the Bethe strings with $\theta \sim \mathcal{O}(1)$.
The situation one encounters is in general as follows: upon taking the large-$L$ limit, rapidities $\lambda^{(a)}_{j}$ distribute
densely along certain one-dimensional disjoint segments (i.e. contours) $\mathcal{C}^{(a)}=\bigcup_{i}\mathcal{C}^{(a)}_{i}$ in the 
spectral $\lambda$-plane, described by some smooth (in general non-uniform) densities $\varrho_{a}(\lambda)$ of Bethe roots 
$\lambda^{(a)}_{j}$ supported on $\mathcal{C}^{(a)}$. Such macroscopic low-energy configurations assume a purely classical description
as they correspond to interacting `nonlinear Fourier modes' of nonlinear classical field configurations.
The long-wavelength spectrum of excitations involves in addition non-macroscopic excitations, i.e. modes that carry an infinitesimal 
amount of conserved (classical) charges -- these are the aforementioned Goldstone modes (i.e. fluctuations above the ferromagnetic 
vacuum) and correspond to infinitesimally small contours (isolated poles) in the spectral complex plane.

\subsubsection*{Classical degrees of freedom}

Based on the above picture, it is tempting to readily conclude that the total number of internal (isotropic) degrees of freedom
of a generic classical field configurations coincides with the number of distinct flavors of the Bethe roots
(which equals $r={\rm rank}(\mathfrak{g})$). Before long one however realizes why such a na\"{i}ve identification is flawed.
Here we mention two apparent inconsistencies that arise:
\begin{enumerate}
\item Imagine first, for definiteness, the fundamental $\SU(n)$ ferromagnets. In this case the classical low-energy theories
are given by the higher-rank Landau--Lifshitz equation constrained to complex projective manifolds
$\mathbb{C}\mathbb{P}^{n-1}\cong \SU(n)/[\U(1)\times \SU(n-1)]$ (the derivation can be found in e.g. \cite{MatrixModels}),
with classical phase space of real dimensions ${\rm dim}(\mathbb{C}\mathbb{P}^{n-1})=2(n-1)$. The Goldstone theorem
predicts $n-1$ magnonic branches in this case, and indeed this number precisely agrees with rank of $\SU(n)$ which
also matches the total number of flavors in the $\SU(n)$ magnets. Although this appears to be a good sign, we nevertheless run into
a problem: in generic excited quantum eigenstates, auxiliary quasiparticles cannot be excited on their own without first exiting 
physical (i.e. momentum-carrying) excitations. This can also be readily deduced from the selection rule \eqref{eqn:inequality_rule}.\\
\item Consider next, as another example, the $\SO(5)$ chain of rank $r=2$ in the fundamental onsite representation
$\mathcal{V}_{\square}$.  The low-energy is associated with the coset space $\SO(5)/[\SO(2)\times \SO(3)]$ of real dimension $6$.
This time we expect to find $3$ Goldstone modes which this time obviously exceeds the total number of flavors. More generally,
one would expect the number of distinct types of macroscopic semiclassical strings to always equal $r={\rm rank}(\mathfrak{g})$, which 
is typically vastly smaller than the number of independent Goldstone modes (cf. Sec.~\ref{sec:patterns}).
\end{enumerate}

The only viable option to avoid an apparent paradox is to look for additional classical degrees of freedom that somehow managed to stay 
unnoticed. These clearly would have to be emergent as our enumeration of the genuinely quantum excitations spectrum is unquestionably 
correct. This brings us to the notion of \emph{stacks}. The existence of stack excitations has, to the best our knowledge,
been first encountered in Refs.~\cite{BKS06,BKSZ06,GV08} in the context of gauge-string duality, devoted specifically to the low-energy 
sector for the particular case of unitary supergroups related to the underlying symmetry algebra of the superconformal Yang--Mills 
theory \cite{MZ03,AdSCFT_review}. Outside of that, it appears there a general or systematic discussion devoted to these emergent 
inherently classical objects is still lacking in the literature. To this end, we proceed to explain how the notion of stacks naturally 
arises in the context of integrable quantum ferromagnets with continuously degenerate ground states.

In most general terms, stacks pertain to excitations with an internal multi-flavor structure. They are produced by merging together
a subset of Bethe roots of distinct flavors that share the same rapidity (no merely just the real part).
As already emphasized, in a quantum spin chain such a configuration is not permitted as rapidities must be pair-wise distinct.
In the low-momentum scaling limit described by the Asymptotic Bethe Ansatz this restriction is however lifted as Bethe roots
of different flavors may approach arbitrary close to each other before eventually recombining into an independent low-momentum 
fluctuation mode. Although a large number of stacks can form this way, only those stacks that involve momentum-carrying roots can
be regarded as physical fluctuations, whereas stacks made of solely from auxiliary excitations are likewise called auxiliary.
Similarly to the elementary excitations of a quantum chain, stacks can exert attractive interactions and therefore
can also condense into `large' (i.e. non-linear) modes which carry finite filling fractions and charge densities.
	
For the complete classification of classical modes additional quantum numbers are thus required to properly account for
the internal structure all the admissible stacks. This additional information, which can be though of various polarization directions
in which a classical configuration can vibrate, depends on both the symmetry algebra and on-site representation under consideration.
For illustration, let us have a look at our main example of the fundamental $\SU(n)$ chains.
The asymptotic Bethe Ansatz equations, see Eq.~\eqref{eqn:ABA_equations}, take place on an $n$-sheeted Riemann surface, whose
sheets are stitched together at a finite number of branch cuts (see e.g. \cite{Kazakov2004,Bargheer08} for additional details).
Standard branch cuts, forming along contours $\mathcal{C}^{(a)}_{i}$ along which the Bethe roots condense, are of the
square-root type. Creating a single excitation of flavor $a$ amount to connects the $a$th and $(a+1)$th sheet with an isolated pole 
(infinitesimal cut). Exciting a stack $(a,b)$ means joining two non-adjacent sheets indexed by $a$ and $b+1$.
Equations \eqref{eqn:ABA_equations} need to be appropriately amended to incorporate these extra stack excitations.
In Fig.~\ref{fig:stacks} we illustrate the process of stack formation on a four-sheeted Riemann surface associated
to the asymptotic Bethe equations of the $\SU(4)$ quantum chain made out of fundamental spins $({\bf 4})$.

\begin{widetext}

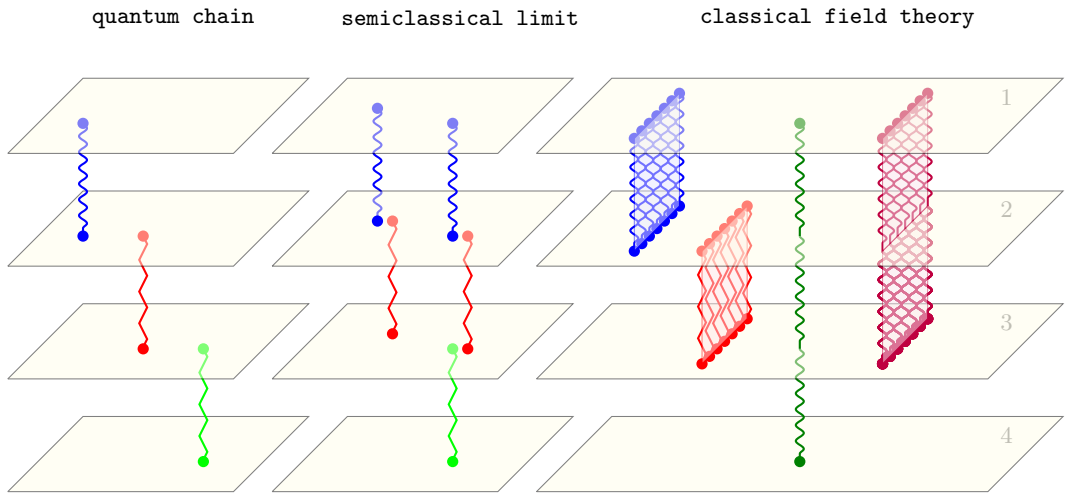
\begin{figure}[t]
\centering
\tikzset{
    physical/.style={decorate, decoration={snake, amplitude = 1.5pt, segment length = 6pt}},
    auxiliary/.style={decorate, decoration={zigzag, amplitude = 1.5pt, segment length = 12pt}},
    dot/.style={inner sep = 1.5pt,fill,circle}
}


\begin{tikzpicture}

\begin{scope}

\node[thick] (label1) at (2.2,1.8) {{\tt quantum chain}};

\draw[black,fill=yellow!10,opacity=0.5] (0,-4.5) -- (3,-4.5) -- (4,-3.5) -- (1,-3.5) -- (0,-4.5);
\path[draw=green, thick ,auxiliary] (2.6,-2.6) -- (2.6,-4.1);
\draw (2.6,-2.6) node[dot,green]{};
\draw (2.6,-4.1) node[dot,green]{};

\draw[black,fill=yellow!10,opacity=0.5] (0,-3) -- (3,-3) -- (4,-2) -- (1,-2) -- (0,-3);
\path[draw=red, thick, auxiliary] (1.8,-1.1) -- (1.8,-2.6);
\draw (1.8,-1.1) node[dot,red]{};
\draw (1.8,-2.6) node[dot,red]{};

\draw[black,fill=yellow!10,opacity=0.5] (0,-1.5) -- (3,-1.5) -- (4,-0.5) -- (1,-0.5) -- (0,-1.5);

\draw (1,0.4) node[dot,blue]{};
\path[draw=blue,thick, physical] (1,0.4) -- (1,-1.1);
\draw[black,fill=yellow!10,opacity=0.5] (0,0) -- (3,0) -- (4,1) -- (1,1) -- (0,0);
\draw (1,-1.1) node[dot,blue]{};

\end{scope}


\begin{scope}[xshift=100pt]

\node[thick] (label1) at (2.5,1.8) {{\tt semiclassical limit}};

\draw[black,fill=yellow!10,opacity=0.5] (0,-4.5) -- (3,-4.5) -- (4,-3.5) -- (1,-3.5) -- (0,-4.5);

\path[draw=green, thick, auxiliary] (2.4,-2.6) -- (2.4,-4.1);
\draw (2.4,-2.6) node[dot,green]{};
\draw (2.4,-4.1) node[dot,green]{};

\draw[black,fill=yellow!10,opacity=0.5] (0,-3) -- (3,-3) -- (4,-2) -- (1,-2) -- (0,-3);

\path[draw=red, thick, auxiliary] (2.6,-1.1) -- (2.6,-2.6);
\draw (2.6,-1.1) node[dot,red]{};
\draw (2.6,-2.6) node[dot,red]{};

\draw (1.6,-0.9) node[dot,red]{};
\draw (1.6,-2.4) node[dot,red]{};
\path[draw=blue,thick, physical] (1.4,0.6) -- (1.4,-0.9);
\path[draw=red, thick, auxiliary] (1.6,-0.9) -- (1.6,-2.4);
\draw[black,fill=yellow!10,opacity=0.5] (0,-1.5) -- (3,-1.5) -- (4,-0.5) -- (1,-0.5) -- (0,-1.5);

\draw (2.4,0.4) node[dot,blue]{};
\draw (2.4,-1.1) node[dot,blue]{};
\path[draw=blue,thick, physical] (2.4,0.4) -- (2.4,-1.1);

\draw (1.4,0.6) node[dot,blue]{};
\draw (1.4,-0.9) node[dot,blue]{};
\draw[black,fill=yellow!10,opacity=0.5] (0,0) -- (3,0) -- (4,1) -- (1,1) -- (0,0);

\end{scope}


\begin{scope}[xshift=200pt]

\node[thick] (label1) at (4,1.8) {{\tt classical field theory}};

\node[thick,gray] (1st) at (6.25,0.75) {$1$};
\node[thick,gray] (1st) at (6.25,-0.75) {$2$};
\node[thick,gray] (1st) at (6.25,-2.25) {$3$};
\node[thick,gray] (1st) at (6.25,-3.75) {$4$};

\draw[black,fill=yellow!10,opacity=0.5] (0,-4.5) -- (6,-4.5) -- (7,-3.5) -- (1,-3.5) -- (0,-4.5);

\path[draw=green!50!black, thick, physical] (3.5,-2.6) -- (3.5,-4.1);
\draw (3.5,-4.1) node[dot,green!50!black]{};

\draw[black,fill=yellow!10,opacity=0.5] (0,-3) -- (6,-3) -- (7,-2) -- (1,-2) -- (0,-3);
\foreach \x in {-2,...,4}
{
	\draw (2.4+0.1*\x,-1.1+0.1*\x) node[dot,red]{};
	\draw (2.4+0.1*\x,-2.6+0.1*\x) node[dot,red]{};
	\path[draw=red,thick,auxiliary] (2.4+0.1*\x,-1.1+0.1*\x) -- (2.4+0.1*\x,-2.6+0.1*\x);
}
\draw[red,fill=red!10,opacity=0.5] (2.2,-1.3) -- (2.2,-2.8) -- (2.8,-2.2) -- (2.8,-0.7) -- (2.2,-1.3);
\foreach \x in {-2,...,4}
\foreach \x in {-2,...,4}
{
	\draw (4.8+0.1*\x,-2.6+0.1*\x) node[dot,purple]{};
	\path[draw=purple,thick,physical] (4.8+0.1*\x,-1.1+0.1*\x) -- (4.8+0.1*\x,-2.6+0.1*\x);
}
\draw[purple,fill=purple!10,opacity=0.5] (4.6,-1.3) -- (4.6,-2.8) -- (5.2,-2.2) -- (5.2,-0.7) -- (4.6,-1.3);

\path[draw=green!50!black, thick, physical] (3.5,-1.1) -- (3.5,-2.6);

\draw[black,fill=yellow!10,opacity=0.5] (0,-1.5) -- (6,-1.5) -- (7,-0.5) -- (1,-0.5) -- (0,-1.5);
\foreach \x in {-2,...,4}
{
	\draw (1.5+0.1*\x,0.4+0.1*\x) node[dot,blue]{};
	\draw (1.5+0.1*\x,-1.1+0.1*\x) node[dot,blue]{};
	\path[draw=blue,thick, physical] (1.5+0.1*\x,0.4+0.1*\x) -- (1.5+0.1*\x,-1.1+0.1*\x);
}

\draw (3.5,0.4) node[dot,green!50!black]{};
\foreach \x in {-2,...,4}
{
	\draw (4.8+0.1*\x,0.4+0.1*\x) node[dot,purple]{};
	\path[draw=purple,thick, physical] (4.8+0.1*\x,0.4+0.1*\x) -- (4.8+0.1*\x,-1.1+0.1*\x);
}

\path[draw=green!50!black,thick, physical] (3.5,0.4) -- (3.5,-1.1);
\draw[blue,fill=blue!10,opacity=0.5] (1.3,0.2) -- (1.3,-1.3) -- (1.9,-0.7) -- (1.9,0.8) -- (1.3,0.2);
\draw[purple,fill=purple!10,opacity=0.5] (4.6,0.2) -- (4.6,-1.3) -- (5.2,-0.7) -- (5.2,0.8) -- (4.6,0.2);

\draw[black,fill=yellow!10,opacity=0.5] (0,0) -- (6,0) -- (7,1) -- (1,1) -- (0,0);

\end{scope}

\end{tikzpicture}
\caption{Graphical representation of elementary and auxiliary excitations on a Riemann surface belonging to the $\SU(4)$ quantum spin 
chain with onsite representation $({\bf 4})$: (\emph{left}) physical momentum-carrying magnons (blue) emanating from the upper sheet,
and two flavors of auxiliary magnons (red and green); (\emph{middle}) the semiclassical limit, showing approximate stacks excitations 
with nearly coinciding rapidities; (\emph{right}) the four-sheeted Riemann surface associated to the classical spectral curve 
describing an effective low-energy field theory, with two types of singularities involved: isolated \emph{stacks}, representing various
branches of Goldstone modes that constitute the spectrum of linear modes, and \emph{condensates} of stacks (square-root branch cuts) 
pertaining to classical interacting nonlinear modes carrying finite amount of energy. Physical stacks (dark green) and condensates blue 
and purple) involve momentum-carrying excitations. Condenstates of stacks of a uniform density are identified with classical solitons.}
\label{fig:stacks}
\end{figure}

\end{widetext}

In summary, \emph{the number of independent classical degrees of freedom equals the number of distinct physical stacks.}
In the next section, we carry out a systematic analysis to determine the admissible symmetry breaking patterns and
classify the corresponding physical stacks.

\paragraph*{Remark.}
We would like to re-emphasize that stack excitations are are not permitted in the spectrum of a quantum chain due to
kinematic constraints. In other words, upon including quantum fluctuations (even perturbatively at the level of asymptotic
equations \eqref{eqn:ABA_equations}) stacks are immediately forced disintegrate into individual magnons.

\subsubsection*{Giant quasiparticles as classical solitons}

We have thus far have not yet encountered or mentioned solition modes. As a matter fact, classical soliton modes are (strictly 
speaking) not a part of the nonlinear (finite-gap) spectrum. Nevertheless, they emerge as from elliptic waves in the limit of unit 
elliptic modulus, which invariably requires decompactification of the spatial dimension \cite{Kazakov2004,Vicedo07}.
This process, when observed in the spectral $\lambda$-plane, corresponds to a singular event in which two adjacent density 
contours merge together, resulting in a perfectly vertically oriented contour with constant unit density of Bethe roots.
Such object have attracted considerable attention and have most commonly been referred to as \emph{condensates},
see e.g. Refs.~\cite{Kazakov2004,BKSZ06,Vicedo07,Bargheer08} (note however that these are distinct from ordinary condensates
that form square-root branch cuts of finite-genus Riemann surfaces). From the algebraic curve viewpoint, such uniform
condensates arise from coalescing two square-root branch cuts into a logarithmic branch cut. The giant quasiparticles from the 
preceding discussion, i.e. low-momentum bound state excitations with macroscopic number of quanta, are semiclassically quantized 
soliton modes.

\subsubsection*{Counting stacks}

Now we return back to the `counting problem'. What is left is to extract the correct number of Goldstone modes from the
semiclassical spectrum of quantum ferromagnetic chains whose elementary (both physical and auxiliary) excitations carry $r$ distinct 
flavor numbers (one per each simple root $\alpha_{i}$ of $\mathfrak{g}$). One can generate different types of stacks simply by
forming linear combinations of $\alpha_{i}$ with non-negative integral coefficients. In other words, excitation operators 
associated with stacks correspond to \emph{positive} roots $\alpha \in \Delta_{+}(\mathfrak{g})$. Among those, the physical ones are 
those that involve at least one simple root, associated with physical (i.e. momentum-carrying) excitations.
In the fundamental models for example, that is only $\alpha_{1}$.

\begin{center}
\centering
\begin{table}[t]
\renewcommand{\arraystretch}{1.2}
\setlength{\tabcolsep}{0pt}
\begin{tabular}{|c|c|c|c|}
\hline
\rowcolor{pink}
$\mathfrak{g}$ & onsite irrep & Target space $G/H$ & $n_{\rm G}$ \\
\hline \hline
$A_{4}$ & $({\bf 5}),\,(\overline{\bf 5})$ & $\SU(5)/\U(1)\times \SU(4)$ & $4$ \\
 & $({\bf 10}),\,(\overline{\bf 10})$ & $\SU(5)/\U(1)\times \SU(2)\times \SU(3)$ & $6$ \\
\hline
$B_{4}$ & $({\bf 9})$ & $\SO(9)/\SO(2)\times \SO(7)$ & $7$ \\
 & $({\bf 36})$ & $\SO(9)/\U(1)\times \SU(2)\times \SO(5)$ & $11$ \\
 & $({\bf 84})$ & $\SO(9)/\U(1)\times \SU(2)\times \SU(3)$ & $12$ \\
 & $({\bf 16})$ & $\SO(9)/\U(1)\times \SU(4)$ & $10$ \\
\hline
$\,\, C_{4} \,\,$ & $({\bf 8})$ & $\USp(8)/\SO(2)\times \USp(6)$ & $7$ \\
 & $({\bf 27})$ & $\,\, \USp(8)/\U(1)\times \SU(2)\times \SO(5)\,\,$ & $\quad 11 \quad$ \\
 & $({\bf 48})$ & $\,\, \USp(8)/\U(1)\times \SU(2)\times \SU(3)\,\,$ & $12$ \\
 & $({\bf 42})$ & $\USp(8)/\U(1)\times \SU(4)$ & $10$ \\
\hline
$D_{4}$ & $\,({\bf 8}_{v}),\,({\bf 8}_{c}),\,({\bf 8}_{s})\,$ & $\SO(8)/\U(1)\times \SU(3)$ & $6$ \\
 & $({\bf 28})$ & $\SO(8)/\U(1)\times \SU(2)^{\times 3}$ & $9$ \\
\hline
\end{tabular}
\caption{Coset spaces for all the fundamental representations $\Lambda = \omega_{a}$
for classical Lie algebras of rank $4$.}
\end{table}
\end{center}

There indeed exist a nifty graphical representation for all the physical stacks in terms of partially ordered sets known
as \emph{Hasse diagrams}. Let us first outline the construction by specializing to fundamental irreducible onsite representations
$\Lambda = \omega_{a}$. One starts by picking one of the bottom nodes (associated to simple roots) and proceeds by `climbing'  in the 
direction of the top node (being the highest root) by successively adding one simple root at a time. With this rule, several 
inequivalent paths through the Hasse diagram can be generated. Since each vertex is an independent Goldstone mode, their total number 
$n_{\rm G}$ equals the total number of vertices visited along the way. More generally, that is in the case of non-fundamental irreducible onsite representations $\mathcal{V}_{\Lambda}$ with $\Lambda = \sum_{a=1}^{r}m_{a}\omega_{a}$ involving
multiple non-vanishing Dynkin labels $m_{a}$, there will be multiple  momentum-carrying magnon species in the spectrum.
Now one has to simply superimpose all the sublattices that emanate from base nodes $\alpha_{i}$ for all
$a\in \mathcal{I}_{r}$ for which $m_{a}\neq 0$. The `maximal' breaking occurs when none of the $m_{a}$'s vanishes, as then the isotropy 
group $H$ coincides with the maximal torus subgroup $T$ of $G$. This scenario results in the maximal number of Goldstone modes
$n_{G} = \tfrac{1}{2}{\rm dim}(G/T) = |\Delta_{+}|$, rendering all stacks physical excitations.

\begin{widetext}

\begin{figure}
\centering

\begin{tikzpicture}[
dot/.style = {circle, fill, darkgray, circular drop shadow, minimum size=#1,
              inner sep=0pt, outer sep=0pt},
dot/.default = 6pt
]

\node[thick] (caption1) at (1.5,3) {{\tt Hasse lattice of} $\Delta_{+}$};
\node[thick] (caption2) at (8,3) {{\tt fundamental irreps} $\Lambda = \omega_{a}$};
\node[thick] (caption2) at (14.5,3) {{\tt non-fundamental irrep} $\Lambda = [1,0,1]$};


\begin{scope}[rotate = -45]
\node[thick] (label1) at (-0.5,2.5) {$A_{3}=\mathfrak{su}(4)$};
\node[dot,blue,label=below:$1$] (n100) at (0,0) {};
\node[dot,red,label=below:$2$] (n010) at (1,1) {};
\node[dot,magenta,label=below:$3$] (n001) at (2,2) {};
\node[dot] (n110) at (0,1) {};
\node[dot] (n011) at (1,2) {};
\node[dot] (n111) at (0,2) {};
\draw[-,thick,red] (n100) -- node[above left] {$2$} (n110);
\draw[-,thick,red] (n001) -- node[above right] {$2$} (n011);
\draw[-,thick,blue] (n010) -- node[above right] {$1$} (n110);
\draw[-,thick,magenta] (n110) -- node[above left] {$3$} (n111);
\draw[-,thick,magenta] (n010) -- node[above left] {$3$} (n011);
\draw[-,thick,blue] (n011) -- node[above right] {$1$} (n111);
\end{scope}

\begin{scope}[xshift=150pt,rotate = -45]
\node[thick] (1st) at (-1,2) {$({\bf 4})$};
\node[dot,blue,label=below:$1$] (n100) at (0,0) {};
\node[dot] (n110) at (0,1) {};
\node[dot] (n111) at (0,2) {};
\draw[-,thick,red] (n100) -- node[above left] {$2$} (n110);
\draw[-,thick,magenta] (n110) -- node[above left] {$3$} (n111);
\end{scope}

\begin{scope}[xshift=200pt,rotate = -45]
\node[thick] (2nd) at (-0.5,2.5) {$({\bf 6})$};
\node[dot,red,label=below:$2$] (n010) at (1,1) {};
\node[dot] (n110) at (0,1) {};
\node[dot] (n011) at (1,2) {};
\node[dot] (n111) at (0,2) {};
\draw[-,thick,blue] (n010) -- node[above right] {$1$} (n110);
\draw[-,thick,magenta] (n010) -- node[above left] {$3$} (n011);
\draw[-,thick,magenta] (n110) -- node[above left] {$3$} (n111);
\draw[-,thick,blue] (n011) -- node[above right] {$1$} (n111);
\end{scope}

\begin{scope}[xshift=250pt,rotate = -45]
\node[thick] (2nd) at (0,3) {$(\overline{\bf 4})$};
\node[dot,magenta,label=below:$3$] (n001) at (2,2) {};
\node[dot] (n011) at (1,2) {};
\node[dot] (n111) at (0,2) {};
\draw[-,thick,red] (n001) -- node[above right] {$2$} (n011);
\draw[-,thick,blue] (n011) -- node[above right] {$1$} (n111);
\end{scope}

\begin{scope}[xshift=375pt,rotate = -45]
\node[thick] (2nd) at (-0.5,2.5) {$({\bf 15})$};
\node[dot,magenta,label=below:$3$] (n001) at (2,2) {};
\node[dot] (n011) at (1,2) {};
\node[dot] (n111) at (0,2) {};
\node[dot,blue,label=below:$1$] (n100) at (0,0) {};
\node[dot] (n110) at (0,1) {};
\draw[-,thick,red] (n100) -- node[above left] {$2$} (n110);
\draw[-,thick,magenta] (n110) -- node[above left] {$3$} (n111);
\draw[-,thick,red] (n001) -- node[above right] {$2$} (n011);
\draw[-,thick,blue] (n011) -- node[above right] {$1$} (n111);
\end{scope}


\begin{scope}[yshift=-150pt,rotate = -45]
\node[thick] (label1) at (-1.5,3.5) {$B_{3}=\mathfrak{so}(7)$};
\node[dot,blue,label=below:$1$] (n100) at (0,0) {};
\node[dot,red,label=below:$2$] (n010) at (1,1) {};
\node[dot,magenta,label=below:$3$] (n001) at (2,2) {};
\node[dot] (n110) at (0,1) {};
\node[dot] (n011) at (1,2) {};
\node[dot] (n111) at (0,2) {};
\node[dot] (n012) at (1,3) {};
\node[dot] (n112) at (0,3) {};
\node[dot] (n122) at (0,4) {};
\draw[-,thick,red] (n100) -- node[above left] {$2$} (n110);
\draw[-,thick,blue] (n010) -- node[above right] {$1$} (n110);
\draw[-,thick,magenta] (n010) -- node[above left] {$3$} (n011);
\draw[-,thick,blue] (n011) -- node[above right] {$1$} (n111);
\draw[-,thick,magenta] (n011) -- node[above left] {$3$} (n012);
\draw[-,thick,blue] (n012) -- node[above right] {$1$} (n112);
\draw[-,thick,red] (n001) -- node[above right] {$2$} (n011);
\draw[-,thick,magenta] (n110) -- node[above left] {$3$} (n111);
\draw[-,thick,magenta] (n111) -- node[above left] {$3$} (n112);
\draw[-,thick,red] (n112) -- node[above left] {$2$} (n122);
\end{scope}

\begin{scope}[xshift=125pt,yshift=-150pt,rotate = -45]
\node[thick] (1st) at (-1.5,3.5) {$({\bf 7})$};
\node[dot,blue,label=below:$1$] (n100) at (0,0) {};
\node[dot] (n110) at (0,1) {};
\node[dot] (n111) at (0,2) {};
\node[dot] (n112) at (0,3) {};
\node[dot] (n122) at (0,4) {};
\draw[-,thick,red] (n100) -- node[above left] {$2$} (n110);
\draw[-,thick,magenta] (n110) -- node[above left] {$3$} (n111);
\draw[-,thick,magenta] (n111) -- node[above left] {$3$} (n112);
\draw[-,thick,red] (n112) -- node[above left] {$2$} (n122);
\end{scope}

\begin{scope}[xshift=175pt,yshift=-150pt,rotate = -45]
\node[thick] (2nd) at (-1,4) {$({\bf 21})$};
\node[dot,red,label=below:$2$] (n010) at (1,1) {};
\node[dot] (n110) at (0,1) {};
\node[dot] (n011) at (1,2) {};
\node[dot] (n111) at (0,2) {};
\node[dot] (n012) at (1,3) {};
\node[dot] (n112) at (0,3) {};
\node[dot] (n122) at (0,4) {};
\draw[-,thick,blue] (n010) -- node[above right] {$1$} (n110);
\draw[-,thick,magenta] (n010) -- node[above left] {$3$} (n011);
\draw[-,thick,blue] (n011) -- node[above right] {$1$} (n111);
\draw[-,thick,magenta] (n011) -- node[above left] {$3$} (n012);
\draw[-,thick,blue] (n012) -- node[above right] {$1$} (n112);
\draw[-,thick,magenta] (n110) -- node[above left] {$3$} (n111);
\draw[-,thick,magenta] (n111) -- node[above left] {$3$} (n112);
\draw[-,thick,red] (n112) -- node[above left] {$2$} (n122);
\end{scope}

\begin{scope}[xshift=250pt,yshift=-150pt,rotate = -45]
\node[thick] (2nd) at (-1,4) {$({\bf 8})$};
\node[dot,magenta,label=below:$3$] (n001) at (2,2) {};
\node[dot] (n011) at (1,2) {};
\node[dot] (n111) at (0,2) {};
\node[dot] (n012) at (1,3) {};
\node[dot] (n112) at (0,3) {};
\node[dot] (n122) at (0,4) {};
\draw[-,thick,blue] (n011) -- node[above right] {$1$} (n111);
\draw[-,thick,magenta] (n011) -- node[above left] {$3$} (n012);
\draw[-,thick,blue] (n012) -- node[above right] {$1$} (n112);
\draw[-,thick,red] (n001) -- node[above right] {$2$} (n011);
\draw[-,thick,magenta] (n111) -- node[above left] {$3$} (n112);
\draw[-,thick,red] (n112) -- node[above left] {$2$} (n122);
\end{scope}

\begin{scope}[xshift=375pt,yshift=-150pt,rotate = -45]
\node[thick] (1st) at (-1.5,3.5) {$({\bf 48})$};
\node[dot,blue,label=below:$1$] (n100) at (0,0) {};
\node[dot,magenta,label=below:$3$] (n001) at (2,2) {};
\node[dot] (n110) at (0,1) {};
\node[dot] (n011) at (1,2) {};
\node[dot] (n111) at (0,2) {};
\node[dot] (n012) at (1,3) {};
\node[dot] (n112) at (0,3) {};
\node[dot] (n122) at (0,4) {};
\draw[-,thick,red] (n100) -- node[above left] {$2$} (n110);
\draw[-,thick,blue] (n011) -- node[above right] {$1$} (n111);
\draw[-,thick,magenta] (n011) -- node[above left] {$3$} (n012);
\draw[-,thick,blue] (n012) -- node[above right] {$1$} (n112);
\draw[-,thick,red] (n001) -- node[above right] {$2$} (n011);
\draw[-,thick,magenta] (n110) -- node[above left] {$3$} (n111);
\draw[-,thick,magenta] (n111) -- node[above left] {$3$} (n112);
\draw[-,thick,red] (n112) -- node[above left] {$2$} (n122);
\end{scope}

\end{tikzpicture}

\caption{(\emph{left}) Hasse diagrams ascribed to the positive root lattice $\Delta_{+}$ of $\mathfrak{g}$, depicted
for Lie algebras $\mathfrak{su}(4)$ (\emph{top}) and $\mathfrak{so}(7)$ (\emph{bottom}) of rank $3$.
The nodes represent positive roots $\alpha \in \Delta_{+}$, amongst which the simple ones (in the bottom level) are marked with 
different colors. Moving upwards along an colored edge amounts to add a simple roots of the corresponding color.
Positive roots are separated into $\ell$ levels, determined by the sum of all positive integral coefficients in the expansion
of $\alpha$ over the basis of simple roots $\alpha_{i}$. The total number of positive roots is the maximal
number of Goldstone modes $n_{G}=|\Delta_{+}|=\tfrac{1}{2}{\rm dim}(G/T)$, pertaining to the maximally broken symmetry
$G\to T$. (\emph{middle}) Hasse sublattices pertaining to all the distinct fundamental (i.e. single column Young diagram) irreducible
representations $\Lambda = \omega_{a}$, emanating from the base node $\alpha_{a}$ to the upper node (maximal root).
The total number of nodes in a given sublattice equals the number of distinct branches of magnon excitations (Goldstone bosons).
(\emph{right}) Hasse sublattices for non-fundamental onsite representations (shown for $\Lambda=[1,0,1]$) are obtained by
superimposition of the fundamental sublattices, each belonging to a single non-vanishing Dynkin label.}
\end{figure}
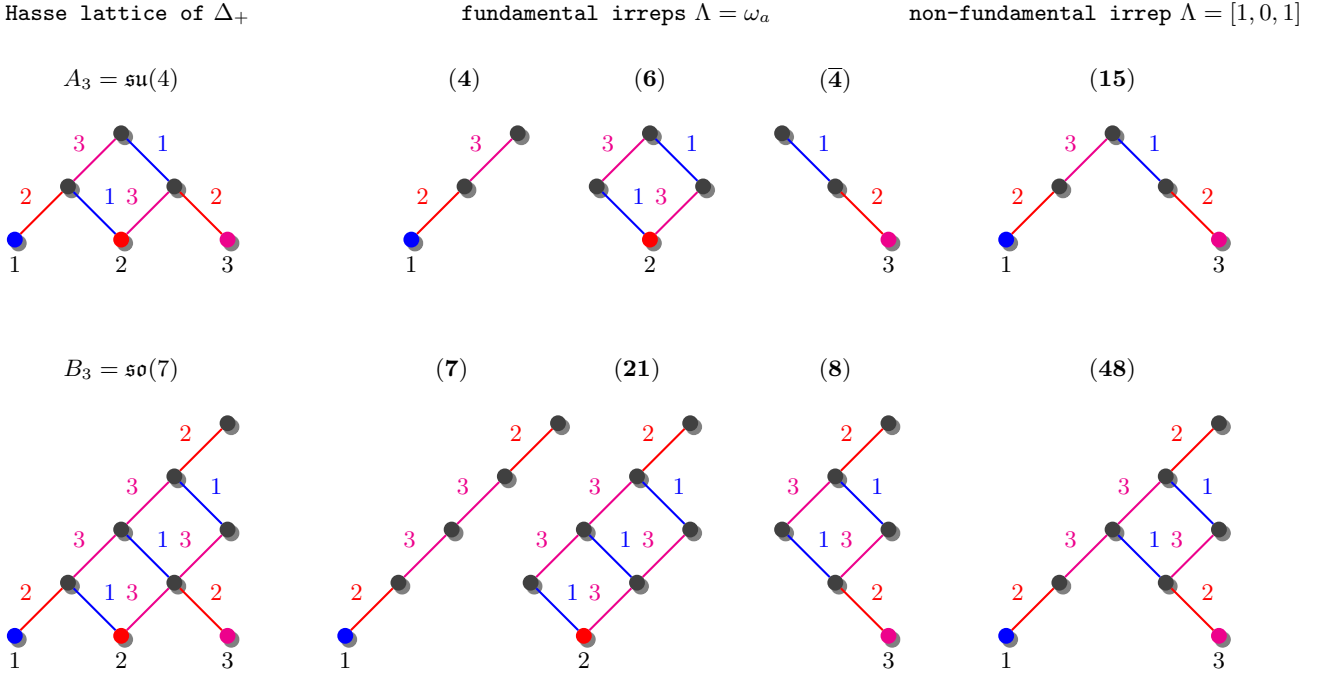

\end{widetext}

\begin{table}
\setlength{\tabcolsep}{0pt}
\renewcommand{\arraystretch}{1.2}
\begin{tabular}{|c|c|c|c}
\hline
\rowcolor{yellow}
$\alpha$-basis & $\omega$-basis & level \\
\hline \hline
$\quad(1,1,1)\quad$ & $\quad [1,0,1] \quad$ & $\quad 3 \quad$ \\
\hline
$(0,1,1)$ & $[-1,1,1]$ & $2$ \\
$(1,1,0)$ & $[1,1,-1]$ & $2$ \\
\hline
$(0,0,1)$ & $[0,-1,2]$ & $1$ \\
$(0,1,0)$ & $[-1,2,-1]$ & $1$ \\
$(1,0,0)$ & $[2,-1,0]$ & $1$ \\
\hline
\end{tabular}
\hspace{0.3em}
\begin{tabular}{|c|c|c|c}
\hline
\rowcolor{yellow}
$\alpha$-basis & $\omega$-basis & level \\
\hline \hline
$\quad (1,2,2)\quad $ & $\quad [1,0,1] \quad$ & $\quad 5 \quad$ \\
\hline
$(1,1,2)$ & $[1,-1,2]$ & $4$ \\
\hline
$(0,1,2)$ & $[-1,0,2]$ & $3$ \\
$(1,1,1)$ & $[1,0,0]$ & $3$ \\
\hline
$(0,1,1)$ & $[-1,1,0]$ & $2$ \\
$(1,1,0)$ & $[1,1,-2]$ & $2$ \\
\hline
$(0,0,1)$ & $[0,-1,2]$ & $1$ \\
$(0,1,0)$ & $[-1,2,-2]$ & $1$ \\
$(1,0,0)$ & $[2,-1,0]$ & $1$ \\
\hline
\end{tabular}
\caption{Positive roots systems for Lie algebras $A_{3}=\mathfrak{su}(4)$ and $B_{3}=\mathfrak{so}(7)$ of rank $3$: in $\alpha$-basis, 
any positive root $\alpha \in \Delta_{+}$ is decomposes as a integral linear combination of simple roots $\alpha_{i}$ (level equals the 
sum of coefficients), whereas $\omega$-basis provides an expansion in terms of the fundamental weights (with coefficients being the 
Dynkin labels).}
\end{table}

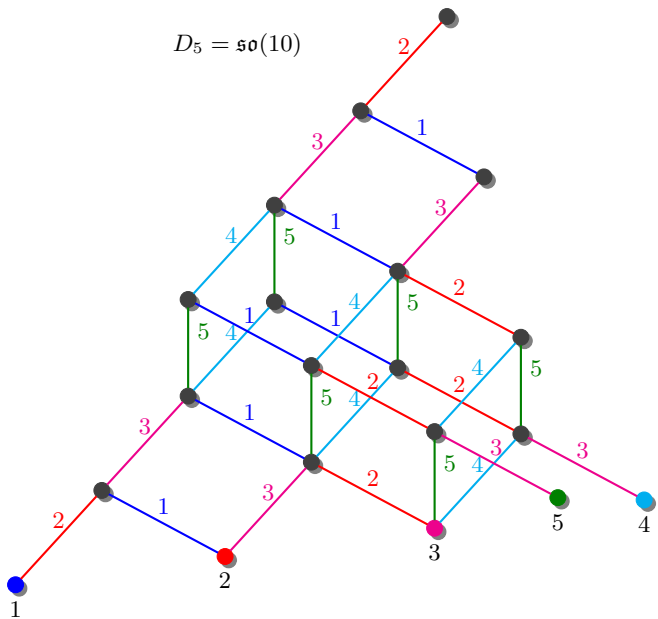
\begin{figure}[htb]
\centering

\tdplotsetmaincoords{40}{-10} 
\tdplotsetrotatedcoords{-45}{0}{0} 

\begin{tikzpicture}[scale=2,tdplot_rotated_coords,
dot/.style = {circle, draw, fill, darkgray, circular drop shadow, minimum size=#1,
              inner sep=0pt, outer sep=0pt},
dot/.default = 6pt
]


\begin{scope}
\node[thick] (label1) at (-1,4,1) {$D_{5}=\mathfrak{so}(10)$};
\node[dot,blue,label=below:$1$] (n10000) at (0,0,0) {};
\node[dot,red,label=below:$2$] (n01000) at (1,1,0) {};
\node[dot,magenta,label=below:$3$] (n00100) at (2,2,0) {};
\node[dot,cyan,label=below:$4$] (n00010) at (3,3,0) {};
\node[dot,green!50!black,label=below:$5$] (n00001) at (3,2,1) {};
\node[dot] (n11000) at (0,1,0) {};
\node[dot] (n01100) at (1,2,0) {};
\node[dot] (n00110) at (2,3,0) {};
\node[dot] (n00101) at (2,2,1) {};
\node[dot] (n11100) at (0,2,0) {};
\node[dot] (n00111) at (2,3,1) {};
\node[dot] (n01101) at (1,2,1) {};
\node[dot] (n01110) at (1,3,0) {};
\node[dot] (n11110) at (0,3,0) {};
\node[dot] (n01111) at (1,3,1) {};
\node[dot] (n11101) at (0,2,1) {};
\node[dot] (n11111) at (0,3,1) {};
\node[dot] (n01211) at (1,4,1) {};
\node[dot] (n11211) at (0,4,1) {};
\node[dot] (n12211) at (0,5,1) {};
\draw[-,thick,red] (n10000) -- node[above] {$2$} (n11000);
\draw[-,thick,blue] (n01000) -- node[above] {$1$} (n11000);
\draw[-,thick,magenta] (n01000) -- node[above] {$3$} (n01100);
\draw[-,thick,red] (n00100) -- node[above] {$2$} (n01100);
\draw[-,thick,cyan] (n00100) -- node[above] {$4$} (n00110);
\draw[-,thick,magenta] (n00010) -- node[above] {$3$} (n00110);
\draw[-,thick,green!50!black] (n00100) -- node[above right] {$5$} (n00101);
\draw[-,thick,magenta] (n00001) -- node[above] {$3$} (n00101);
\draw[-,thick,magenta] (n11000) -- node[above] {$3$} (n11100);
\draw[-,thick,green!50!black] (n00110) -- node[above right] {$5$} (n00111);
\draw[-,thick,green!50!black] (n01100) -- node[above right] {$5$} (n01101);
\draw[-,thick,blue] (n01100) -- node[above] {$1$} (n11100);
\draw[-,thick,cyan] (n01100) -- node[above] {$4$} (n01110);
\draw[-,thick,red] (n00110) -- node[above] {$2$} (n01110);
\draw[-,thick,red] (n00101) -- node[above] {$2$} (n01101);
\draw[-,thick,cyan] (n00101) -- node[above] {$4$} (n00111);
\draw[-,thick,cyan] (n11100) -- node[above] {$4$} (n11110);
\draw[-,thick,green!50!black] (n11100) -- node[above right] {$5$} (n11101);
\draw[-,thick,blue] (n01110) -- node[above] {$1$} (n11110);
\draw[-,thick,green!50!black] (n01110) -- node[above right] {$5$} (n01111);
\draw[-,thick,red] (n00111) -- node[above] {$2$} (n01111);
\draw[-,thick,blue] (n01101) -- node[above] {$1$} (n11101);
\draw[-,thick,cyan] (n01101) -- node[above] {$4$} (n01111);
\draw[-,thick,cyan] (n11101) -- node[above] {$4$} (n11111);
\draw[-,thick,green!50!black] (n11110) -- node[above right] {$5$} (n11111);
\draw[-,thick,blue] (n01111) -- node[above] {$1$} (n11111);
\draw[-,thick,magenta] (n01111) -- node[above] {$3$} (n01211);
\draw[-,thick,magenta] (n11111) -- node[above] {$3$} (n11211);
\draw[-,thick,blue] (n01211) -- node[above] {$1$} (n11211);
\draw[-,thick,red] (n11211) -- node[above] {$2$} (n12211);
\end{scope}

\end{tikzpicture}
\caption{Hasse diagram for the positive root lattice $\Delta_{+}(\mathfrak{so}(10))$. There are $20$ roots in total (black dots),
arranged into $7$ levels. The lattice extends into three dimensions as simple root $\alpha_{3}$ has connectivity $3$.
To find the $\alpha$-pattern corresponding to a simple roots $\alpha_{i}$, one begins at $\alpha_{i}$ and collects also the nodes
by always proceeding forward in the direction of the highest root. The $\mathbb{Z}_{2}$ automorphism of the Dynkin graph
reveals itself as the reflection symmetry with respect to the main diagonal of the cube passing though nodes $4$ and $5$.}
\label{fig:D5}
\end{figure}

Finally, in cases when $\mathfrak{g}$ contains a node of connectivity $3$ (that is types $D_{r}$ and $E$-family),
the Hasse lattices extend into the third dimension (this is illustrate in Fig.~\ref{fig:D5} on the example of $\mathfrak{so}(10)$).

\pagebreak
\section{Integrable quantum field theories}
\label{sec:QFTs}

Integrable ferromagnetic quantum chains discussed thus far \emph{do not} exhaust the list of integrable quantum models which possess 
global symmetries of non-abelian Lie group. There are several other non-relativistic systems, such as most prominently
integrable fermionic models that exhibit Lie supersymmetries ($\mathbb{Z}_{2}$-graded Lie algebras). It should therefore not come as a
surprise that the phenomenon of charge superdiffusion with exponent $z=3/2$ likewise occurs in there too (see \cite{Ilievski18,Fava20} 
for concrete models). What is more remarkable however is that that anomalous charge dynamics also emerges in relativistic invariant 
integrable quantum field theories in two space-time dimensions, provided that elementary particles carry 
internal isotropic degrees of freedom (which classically takes values in $G$ or coset manifolds $G/H$).
Two prominent examples of integrable quantum field theories (IQFT) are the class of $\rO(n)$ nonlinear sigma models (NLSM)
or $G\times G$ principal chiral fields (PCF), for more details see e.g. Refs.~\cite{PW83,WiegmannPCF,Wiegmann84,FR86,ORW87,KL16}.

Even though Goldstone modes of Lorentz invariant systems always comprise of \emph{linearly} dispersing (i.e. type-I) modes, one should
keep in mind that we are interested here exclusively in the time-evolution of the Noether currents and not correlation functions
amongst individual components of quantum fields. A crucial observation in this respect is that the temporal components of the Noether 
two-currents may be understood as (quantized) spin waves. Moreover, in finite temperature ensembles excitations of these internal 
interacting degrees of freedom undergo non-trivial dressing. The situation in fact mirrors that of the Heisenberg spin chains or its higher-rank analogues we investigated earlier. We shall not attempt to give a comprehensive exposition but rather demonstrate the basic 
principles on the $\rO(3)$ NLSM, representing a paradigmatic integrable interacting QFT with non-diagonal scattering. We shall
briefly outline the differences in a few other IQFTs that display $\SU(2)$ symmetry, while postponing a detailed quantitative study
of other integrable nonlinear sigma models on Riemannian symmetric spaces $G/H$ \cite{Fendley01} and their classical limits for
another study.

\subsubsection*{$\rO(3)$ nonlinear sigma model}

To begin, we spell out some the main properties of the $\rO(3)$ NLSM quantum field theory, employing the Hamitonian formalism.
The quantum field $\vec{n}=(n^{x},n^{y},n^{z})$, subjected to the nonlinear constraint $\vec{n}\cdot \vec{n}=1$,
transforms in the vector representation of $\rO(3)$. Since the vacuum stability subgroup w.r.t. polarization axis (say $n^{z}=1$)
is $\rO(2)$, the target manifold for $\rO(3)$ NLSM is $\rO(3)/\rO(2)\cong S^{2}$. Introducing the momentum $\vec{\pi}$ conjugate
to $\vec{n}$ and the angular momentum field $\vec{m}=\vec{n}\times \vec{\pi}$ perpendicular to $\vec{n}$ ($\vec{m}\cdot \vec{n}=0$), 
the Hamiltonian density (here without the topological $\varTheta$-term) has the form
\begin{equation}
H_{\rO(3)} = \frac{v}{2}\int \dd x \left[g\, \vec{m}^{2} + g^{-1}\vec{n}^{2}_{x})\right].
\label{eqn:O3}
\end{equation}
The conserved Noether two-current associated to global $\rO(3)$ rotations is given by
\begin{equation}
\partial_{\mu}\vec{j}_{\mu}=0,\qquad
\vec{j}_{\mu}=g^{-1}\vec{n}\times \partial_{\mu}\vec{n},
\end{equation}
for $\mu\in \{x,t\}$. The integrated magnetization density provides a local
conservation law, $(\dd/\dd t)\int \dd x\, \vec{m}(x)=0$.

The $\rO(3)$ NLSM is well-known to arise as the effective low-energy theory of (Haldane) spin-$S$ antiferromagnets, where
one identifies $v=2J\,S$ and coupling strength is given by $g=2/S$, assuming large $S$; the $S^{2}$-fields $\vec{n}$ governs 
fluctuations of the staggered order parameter above the antiferromagnetic (N\'{e}el) ground state, whereas $\vec{m}$ pertains
to ferromagnetic (spin wave) fluctuations. Elementary excitations of Eq.~\eqref{eqn:O3} are an $\SU(2)$ triplet of massive spinfull bosons with relativistic (bare) dispersion $e(p)=\sqrt{p^{2}+m^{2}}$. As customary, relativistic dispersion of massive particles can be 
parametrized by a rapidity variable $\theta$
\begin{equation}
e(\theta)=m\,\cosh{(\theta)},\qquad p(\theta)=m\,\sinh{(\theta)}.
\label{eqn:relativistic_dispersion}
\end{equation}
Mass $m$ gets generated dynamically through dimensional transmutation and vanishes in the large-$S$ limit.
At weak coupling ($g\to 0$),  the theory becomes effectively free with ${\rm dim}(G/H)=2$ massless (Goldstone) bosons.

The integrability of the model manifested itself though (i) infinitely many ``hidden'' conserved currents and (ii)
a completely factorizable many-body ${\bf S}$-matrix. Notice however that exchange of isotropic degrees of freedom upon elastic 
interparticle interactions renders the ${\bf S}$-matrix \emph{non-diagonal}, namely particles do not only pick up a $\U(1)$ phase but 
change their state (internal quantum numbers). This feature bears direct analogy with the nested spin chains discussed previously
in Sec.~\ref{sec:nested}.

Algebraic diagonalization using Bethe Ansatz has been first carried out in Ref.~\cite{ZZ92} (see also \cite{Fendley01,FendleyISM} for a 
more comprehensive discussion). The usual trick is to resolve the exchange of isotropic degrees of freedom at expense of introducing 
additional, called auxiliary, spin wave excitations (magnons); the extended basic of excitations which renders the scattering 
\emph{diagonal}. These auxiliary magnons are, despite being massless, regarded as independent quasiparticles.
Imposing finite-volume boundary condition, one arrive at the following the NBA equations \cite{ZZ92}
\begin{align}
e^{\ii\,p(\theta_{j})L}\prod_{k=1}^{N_{\theta}}S(\theta_{j},\theta_{k})\prod_{l=1}^{N_{\lambda}}S^{-1}(\theta_{j},\lambda_{l})
&= -1,\\
\prod_{k=1}^{N_{\theta}}S^{-1}(\lambda_{j},\theta_{k})\prod_{l=1}^{N_{\lambda}}S(\lambda_{j},\lambda_{l}) &= -1,
\label{eqn:O3_NBA}
\end{align}
belonging to the sector with $N_{\theta}$ momentum-carrying physical excitations, $N_{\lambda}$ magnons (with associated rapidity 
$\lambda$). There is a single rational scattering amplitude reading $S(\theta)=(\theta-\ii \pi/2)/(\theta + \ii \pi/2)$.
Equations \eqref{eqn:O3_NBA} already make it transparent that interactions among magnons are identical to those in the isotropic 
Heisenberg $\SU(2)$ chain, the essential difference being that here spin excitations are pinned onto an underlying field
(a kind of inhomogeneous background). Consequently, inter-particle interaction lead to formation of bound states (Bethe strings).

Taking the standard thermodynamic limit (that is $L\to \infty$ with ratios $N_{\theta}/L$ and $N_{\lambda}/L$ finite) and introducing
physical density $\rho_{0}(\theta)$ and densities of auxiliary magnonic Bethe $s$-strings $\rho_{s\geq 1}(\theta)$,
one finds following (decoupled) Bethe--Yang equations
\begin{align}
\rho^{\rm tot}_{0} &= \frac{p^{\prime}}{2\pi} + s\star \ol{\rho}_{2},\\
\rho^{\rm tot}_{s} &= \delta_{s,2}\,s\star \rho_{0} + s\star I^{A_{\infty}}_{s,s'}\ol{\rho}_{s'}.
\label{eqn:O3_BY}
\end{align}
Performing an extra particle-transformation on the `physical node', $\calY_{0}\mapsto 1/\calY_{0}$ (i.e. identifying
$\calY_{0}\equiv \rho_{0}/\ol{\rho}_{0}$), the TBA equations can be once again brought into the canonical $\calY$-system format
\begin{equation}
\log \calY_{s} = -\delta_{s,0}\,\beta\,e + s\star I^{D_{\infty}}_{s,s'}\log(1+\calY_{s'}).
\label{eqn:O3_TBA}
\end{equation}
with the incidence matrix of $D_{\infty}$ Dynkin diagram (see Fig.~\ref{fig:QFT_TBA_diagrams}).
Notice that the $\U(1)$ chemical potential $h$ enters only implicitly via asymptotics of the $\calY$-functions,
$\lim_{s\to \infty}\log \tfrac{1}{s} \log \calY_{s} = h$.
The $\rO(3)$ model is fact belongs to an infinite family of integrable $\SU(n)/\SO(n)$ sigma models with $n-1$ massive nodes whose 
thermodynamic particle content gets arranged according to $A_{n-1}\times D_{\infty}$ Dynkin diagrams \cite{Fendley01}.

The limit of infinite temperature is no longer physically meaningful in IQFTs. Although there is no issue with computing
the thermodynamic $\calY$-functions in the $\beta \to 0$ limit at half filling $h=0$ (which are given by $\scrY_{0} = \scrY_{1} = 2$ 
and $\scrY_{s\geq 2} = s^{2} - 1$), the issue arises when solving for the densities Eqs.~\eqref{eqn:O3_BY} which yield
divergent rapidity integrals in the UV regime. We purposely avoid imposing a momentum cut-off (e.g. as in Ref.~\cite{Fujimoto99})
as in general, it would break integrability and thereby spoil the late-time decay of charge correlations.
To ensure convergence of rapidity integrals temperature $\beta > 0$ has to be taken into account in a non-perturbative way.
Another possibility to overcome the issue is to consider instead integrable lattice regularizations, i.e. realizing
IQFTs are certain continuum scaling limit of inhomogeneous (staggered) quantum spin chains
(as exemplified for the case of $\SU(2)$ PCF in \cite{FR86}).

In the scope of physics applications, it is more meaningful to perform a low-temperature expansion of 
Eqs.~\eqref{eqn:O3_BY} and Eqs.~\eqref{eqn:O3_TBA}. We refer the reader to \cite{NMKI19} and only provide a few remarks here.
In order to properly account for the effects of thermal fluctuations in a \emph{finite-density} state (that means even at arbitrarily
small temperatures $T>0$), it is crucial to retain all the contributions of the spin-wave excitations.
Magnonic excitations can be safely discarded only in the regime $h/T \gg 1$ which can be well approximated by the semiclassical 
description developed in refs. \cite{SD97,Sachdev99,Sachdev_book}. In a close analogy to the Heisenberg spin chain, magnons propagating
in an unmagnetized (or half-filled) finite-density background carry vanishingly small amount of dressed magnetization,
and one accordingly finds a vanishing spin Drude weight with a diverging spin diffusion constant $D_{s} \sim 1/h$.

\paragraph*{Topological term.}
It is well known that the $\rO(3)$ NLSM admits an integrable deformation with the inclusion of the topological $\varTheta$-term
with $\varTheta = \pi$ \cite{ZZ92,Fendley_topo} (describing the low energy limit of $SU(2)$-invariant spin chains with odd spin $S$;
in general $\varTheta = 2 \pi S$~\cite{Haldane83}). This appreciably modifies the spectrum as instead of a massive triplet one now 
finds an $\SU(2)$ doublet of massless elementary excitations with bare dispersion
\begin{equation}
e_{\pm}(\theta) = \pm p(\theta) = \frac{M}{2}e^{\pm \theta},
\label{eqn:massless_dispersion}
\end{equation}
where here $\pm$ designate the right ($p>0$) and left ($p<0$) moving components, and $M$ is an arbitrary mass scale (which is 
inessential as far as only the left-left and right-right scattering matters). At the level of TBA description, the physical species 
comprise  of the left and right movers with densities $\rho_{\pm}(\theta)$ \cite{ZZ92}. The Internal magnon structure however
remains intact \cite{ZZ92}. For instance, the TBA equations in the quasi-local form are now of the form
\begin{align}
\log \calY_{\pm} &= \beta\frac{M}{2}e^{\pm \theta} + s\star \log(1+\calY_{1}),\\
\log \calY_{s} &= \delta_{s,1}\,s \star (1+\calY_{+})(1+\calY_{-}) \nonumber \\
&+ s\star I^{A_{\infty}}_{s,s'}\log(1+\calY_{s'}).
\end{align}
For further details we again refer to ref.~\cite{NMKI19} and references therein.

\paragraph*{Classical picture.}

The origin of magnon excitations in relativistic systems can be exhibited already at the classical level. To this end
we briefly consider the $\rO(3)$ NLSM as an integrable \emph{classical} field theory \cite{Mikhailov82}.
The Euler--Lagrange equation for the classical field $\vec{n}\in S^{2}$ reads (in dimensionless units) \cite{Faddeev1987}
\begin{equation}
\vec{n}_{tt} - \vec{n}_{xx} + (\vec{n}^{2}_{t} - \vec{n}^{2}_{x})\vec{n} = 0.
\end{equation}
Owing to Lorentz invariance, linear fluctuations above a degenerate ground state comprise of two transversal
linearly-dispersing (type-I) Goldstone modes. We are however interested the time-evolution of the temporal component of the
conserved Noether two-current which presently corresponds to the ferromagnetic order parameter $\vec{m}=\vec{n}\times \vec{\pi}$.
Its equation of motion reads simply $\vec{m}_{t}=\vec{n}\times \vec{\pi}_{t}$. In terms of the Hamiltonian equations,
\begin{equation}
\vec{n}_{t} = \vec{\pi},\qquad
\vec{\pi}_{t} = \vec{n}_{xx} + (\vec{n}^{2}_{x} - \vec{\pi}^{2})\vec{n},
\end{equation}
one readily finds that magnetization field $\vec{m}$ satisfies
\begin{equation}
\vec{m}_{t} = \vec{n}\times \vec{n}_{xx},\qquad \vec{m}^{2} = \vec{\pi}^{2}.
\end{equation}
thereby revealing the mechanism for how spatial fluctuations of $\vec{n}(x,t)$ generate the dynamics of the angular momentum 
field $\vec{m}(x,t)$. In the quantum $\rO(3)$ NLSM, fluctuations of these magnetization waves carrying integer quanta scatter 
completely elastically of one another, as described by Eqs.~\eqref{eqn:O3_NBA}.

\subsubsection*{Other integrable QFTs}

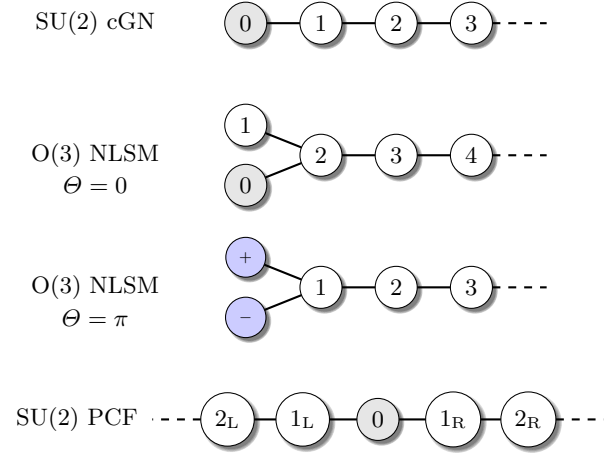
\begin{figure}[htb]
\centering
\begin{tikzpicture}[
	white/.style = {circle, draw, fill=white, circular drop shadow, minimum size = 6pt,inner sep = 3pt, outer sep = 0pt},
	black/.style = {circle, draw, fill=black!10, circular drop shadow, minimum size = 6pt,inner sep = 3pt, outer sep = 0pt},
	blue/.style = {circle, draw, fill=blue!20, circular drop shadow, minimum size = 6pt,inner sep = 3pt, outer sep = 0pt}
]

\begin{scope}
\node (label1) at (-2,0) {$\SU(2)$ cGN};
\node[black] (0) at (0,0) {$0$};
\node[white] (1) at (1,0) {$1$};
\node[white] (2) at (2,0) {$2$};
\node[white] (3) at (3,0) {$3$};
\draw[-,thick] (0) -- (1);
\draw[-,thick] (1) -- (2);
\draw[-,thick] (2) -- (3);
\draw[-,thick,dashed] (3) -- (4,0);
\end{scope}

\begin{scope}[yshift=-50pt]
\node (label1) at (-2,0) {${\rm O}(3)$ NLSM};
\node (topo) at (-2,-0.4) {$\varTheta = 0$};
\node[white] (0u) at (0,0.4) {$1$};
\node[black] (0d) at (0,-0.4) {$0$};
\node[white] (1) at (1,0) {$2$};
\node[white] (2) at (2,0) {$3$};
\node[white] (3) at (3,0) {$4$};
\draw[-,thick] (0u) -- (1);
\draw[-,thick] (0d) -- (1);
\draw[-,thick] (1) -- (2);
\draw[-,thick] (2) -- (3);
\draw[-,thick,dashed] (3) -- (4,0);
\end{scope}

\begin{scope}[yshift=-100pt]
\node (label1) at (-2,0) {${\rm O}(3)$ NLSM};
\node (topo) at (-2,-0.4) {$\varTheta = \pi$};
\node[blue] (0u) at (0,0.4) {\tiny $+$};
\node[blue] (0d) at (0,-0.4) {\tiny $-$};
\node[white] (1) at (1,0) {$1$};
\node[white] (2) at (2,0) {$2$};
\node[white] (3) at (3,0) {$3$};
\draw[-,thick] (0u) -- (1);
\draw[-,thick] (0d) -- (1);
\draw[-,thick] (1) -- (2);
\draw[-,thick] (2) -- (3);
\draw[-,thick,dashed] (3) -- (4,0);
\end{scope}

\begin{scope}[xshift=50pt,yshift=-150pt]
\node[white] (1m) at (-1,0) {$1_{\rm L}$};
\node[white] (2m) at (-2,0) {$2_{\rm L}$};
\node (label1) at (-4,0) {$\SU(2)$ PCF};
\node[black] (0) at (0,0) {$0$};
\node[white] (1) at (1,0) {$1_{\rm R}$};
\node[white] (2) at (2,0) {$2_{\rm R}$};
\draw[-,thick] (0) -- (1);
\draw[-,thick] (1) -- (2);
\draw[-,thick] (0) -- (1m);
\draw[-,thick] (1m) -- (2m);
\draw[-,thick,dashed] (2) -- (3,0);
\draw[-,thick,dashed] (2m) -- (-3,0);
\end{scope}

\end{tikzpicture}
\caption{Complete thermodynamic quasiparticle spectra for a number of paradigmatic examples of massive and massless IQFTs with 
rank-$1$ isometry groups, represented by their TBA incidence diagrams: $\SU(2)$ chiral Gross--Neveu model, $\rO(3)$ nonlinear sigma 
model, with and without the topological $\varTheta$-term and $\SU(2)$ principal chiral field, appearing from top to bottom in the 
respective order. Massive physical excitations, with bare dispersion \eqref{eqn:relativistic_dispersion} are designated by gray nodes. 
Massless physical excitations, with bare dispersion \eqref{eqn:massless_dispersion}, are marked in blue.
The structure of TBA equations is a direct product of two types of Dynkin diagrams; a finite one associated with the isometry 
group of the elementary physical excitations (marked with color nodes), and an infinite one of type $A$ associated with a tower
of auxiliary magnons (white nodes) which bijective correspond to finite-dimensional irreducible $\mathfrak{su}(2)$ representations.}
\label{fig:QFT_TBA_diagrams}
\end{figure}

The above basic example of the $\rO(3)$ NLSM already displays all the characteristic features of IQFTs possessing isotropic degrees
of freedom. This means that analogous types of TBA equations are found in other integrable QFTs with non-abelian
isometry groups \cite{GKSV07,KL16}. The best studied examples are $\rO(n)$ NLSMs on $\rO(n)/\rO(n-1)\cong S^{n-1}$ target spaces, 
described by Lagrangians
\begin{equation}
\mathcal{L}_{\rO(n)}=\frac{1}{2g}\int \dd x\,(\partial_{\mu}\vec{n})^{2},\qquad \vec{n}^{2}=1.
\end{equation}
In the simply-laced cases $\rO(2r)$ with $r\geq 4$ \cite{Balog01}, the thermodynamic spectrum of excitations comprises $r$ flavors
of auxiliary quasi-particles (one per node in the $D_{r}$ Dynkin diagram), each forming an infinite tower of magnonic bound states 
($s$-strings). The $\rO(4)$ NSLM (in the vector representation) is special as it can be viewed as $\SU(2)_{\rm L}\times \SU(2)_{\rm R}$ 
PCF with particles transforming in the bifundamental representation of $\mathfrak{su}(2)$ \cite{GKV09}. Its spectrum involves
massive spinfull particles with two types of $\SU(2)$ spins -- these are elementary excitations above the Fermi sea
(anti-ferromagnetic ground state) in an $\SU(2)$ spin-$S$ chain. Its thermodynamic particle content is depicted in
Fig.~\ref{fig:QFT_TBA_diagrams}. From this perspective, the $\SU(2)$ PCF can be, loosely speaking, perceived as the QFT bosonic 
counterpart of the Fermi--Hubbard model \cite{Takahashi72,Hubbard_book}.
In the strong-coupling limit, the $\SU(2)$ PCF model splits up into two copies of the isotropic Heisenberg chain.
The classical limit and the associated Riemann--Hilbert equations can once again be derived from the asymptotic
Bethe equations \cite{GKSV07}, governing the regime with $N\to \infty$ particles with large rapidities
$\theta \sim \mathcal{O}(N)$ (with quantity $m\,L = \exp{(-2\pi\,N)}$ playing the role of a small parameter).

Another prominent class of IQFTs are $\SU(n)$ chiral Gross--Neveu models (cGN) \cite{ZM80}.
Let us have a quick look at the simplest $\SU(2)$ case, describing two interacting Dirac fermions
expressed in terms of two-component spinors $\psi_{a}$ ($a=1,2$) with Lagrangian density
\begin{align}
\mathcal{L}_{\rm cGN} = \ii\ol{\psi}_{a}\slashed{\partial}\psi^{a}
&+ \frac{1}{2}g^{2}_{s}\big((\ol{\psi}_{a}\psi^{a})^{2}-(\ol{\psi}_{a}\gamma_{5}\psi^{a})^{2}\big) \nonumber \\
&-\frac{1}{2}g^{2}_{v}\big(\ol{\psi}_{a}\gamma_{\mu}\psi^{a}\big)^{2},
\label{eqn:GN_model}
\end{align}
with $\slashed{\partial}\equiv \gamma_{\mu}\partial^{\mu}$ and $\gamma$-matrices $\gamma_{0}=\sigma_{1}$, $\gamma_{1}=\ii \sigma_{2}$, 
$\gamma_{5}=\gamma_{0}\gamma_{1}$ obeying thhe Clifford algebra ${\gamma_{\mu},\gamma_{\nu}}=2\eta^{\mu \nu}$ with
metric $\eta = {\rm diag}(1,-1)$. Lagrangian \eqref{eqn:GN_model} is symmetric under $\U(2)\times \U(1)_{c}$; spinors transform
in the fundamental representation of $\U(2)$, whereas $\U(1)_{c}$ is associated with the chiral symmetry
$\psi \to e^{\ii \theta \gamma_{5}} \psi$.
The spectrum of the model involves a single $\SU(2)$ multiplet of massive fermions, with relativistic dispersion
$e(\theta)=m\,\cosh{(\pi \theta/2)}$ and $p(\theta)=m\,\sinh{(\pi \theta/2)}$ (aside of the massless excitation charged under 
$\U(1)_{c}$ that completely decouples). The amplitude of a two-fermion scattering is given by \cite{Tongeren_Intro,Fedor16}
\begin{equation}
S_{ff}(\theta)=-\frac{\Gamma(1-\tfrac{\theta}{4\ii})\Gamma(\tfrac{1}{2}+\tfrac{\theta}{4\ii})}{
\Gamma(1+\tfrac{\theta}{4\ii})\Gamma(\tfrac{1}{2}-\tfrac{\theta}{4\ii})}.
\end{equation}
Scattering of fermions carrying opposite spin is once again identical to the spin exchange of the $\SU(2)$ Heisenberg chain.
Performing the particle-hole transformation on the $0$th node assigned to physical (i.e. fermionic) density, i.e.
$\calY_{0}\mapsto 1/\calY_{0}$, the resulting Dynkin TBA equations have 
the structure of the $A_{1}\times A_{\infty}$ diagram, namely are of the form \cite{Tongeren_Intro}
\begin{align}
\log \calY_{s} = - \delta_{s,0}\beta\,e +  s\star \log(1+\calY_{s-1})(1+\calY_{s+1}),
\label{eqn:cGN_TBA}
\end{align}
Nodes with $s\geq 1$ have been assigned to magnonic $s$-strings.
Apart from an extra massive particle assigned to the initial node $s=0$, the obtained equations are structurally strongly reminiscent
of those of the isotropic Heisenberg spin-$1/2$ chain (see Fig.~\ref{fig:QFT_TBA_diagrams}).
A family of integrable Gross--Neveu models with $\SU(n)$ symmetry with $n-1$ coupled copies of Eqs.~\eqref{eqn:cGN_TBA}
($A_{n-1}\times A_{\infty}$ TBA incidence matrices) and Gross--Neveu models with $\rO(2n)$ symmetry have been described
in Ref.~\cite{Fendley01}.

\paragraph*{Numerical analysis.}
In order to extract the large-$s$ asymptotic properties of thermodynamic state functions, we have numerically solved the TBA equations 
for the above $\SU(2)$-invariant IQFTs with a common magnon structure. The deduced scaling properties match those of their
spin-chain counterparts, as specified by Eqs.~\eqref{eqn:asy1} and \eqref{eqn:asy2}, indicating once again superdiffusive transport 
with dynamical exponent $z=3/2$.

\section{Conclusion}
\label{sec:conclusion}

We devoted this work to a systematic study of anomalous charge diffusion in equilibrium at finite temperature,
found recently in certain quantum integrable systems with isotropic interactions. We aimed to settle, from a group-theoretical 
perspective, whether all integrable models exhibiting non-abelian continuous Lie group symmetries reveal the same type of
transport anomaly. By establishing the link between the thermodynamic quasiparticle content and representations of the corresponding 
quantum group, we obtained universal algebraic dressing equations --- a scaling analysis of these dressing equations, within the 
framework of generalized hydrodynamics, led to our conclusion that  anomalous charge transport with a superdiffusive dynamical exponent 
$z=3/2$ is generic to all integrable lattice ferromagnets described by Hamiltonians invariant under global non-abelian continuous 
symmetry, irrespective of the symmetry group or unitary irreducible representations associated with (local) physical degrees of 
freedom. Each such subclass indeed constitutes an infinitely large family of commuting integrable Hamiltonians, all of which exhibit
$z = 3/2$ superdiffusion. We have argued that any other anomalous exponent is incompatible with the computed quasiparticle structure. 
In addition, the same type anomalous charge transport persists even in Lorentz invariant integrable quantum 
field theories with group-valued Noether currents.
Due to its remarkable level of robustness, we dubbed this phenomenon as \emph{superuniversal}.

Let us briefly restate the key steps that lead to this general conclusion. 
Recall, first, that anomalous transport occurs only when equilibrium density matrices possess full $G$-invariance: if one considers
generic polarized states (realized by applying chemical potentials to the Cartan charges) the anomalous conductivity gets
regularized and one recovers ballistic transport with subleading diffusive corrections, as one generically expects within generalized 
hydrodynamics. If one starts slightly away from the $G$-symmetric state and approaches it, one finds that the Drude weight for 
ballistic transport vanishes, the diffusion constant diverges, and the
quasiparticles that dominate the magnetic susceptibility and transport are macroscopically large coherent bound states of magnons, 
which we have called \emph{giant quasiparticles}. Non-giant quasiparticles are effectively depolarized and do not contribute either to 
transport or to susceptibility. Transport of the Noether charges is in effect described by a dense gas of interacting giant 
quasiparticles. Each of these giant quasiparticles moves with a characteristic velocity inverse to its width: we established 
this result by explicit analysis of the GHD equations, and interpreted it as indicating that these giant quasiparticles are solitons 
made up of nonlinearly interacting, quadratically dispersing Goldstone modes. When these modes are present at finite density, they 
dress each other's charge in a nonperturbative way, and this nontrivial dressing leads to the fractional dynamical exponent $z = 3/2$.

To further elucidate the nature of the giant quasiparticles, we carefully examined the structure of semiclassical eigenstates
from the viewpoint of an effective low-energy theory with respect to a (continuously-degenerate) ferromagnetic vacuum.
At the level of an integrable classical field theory, giant quasiparticles can be identified with soliton modes, representing
the nonlinear counterparts of quadratically dispersing (type-II) Goldstone bosons resolving linear fluctuations above the
ferromagnetic vacuum. The number of internal (polarization) degrees of freedom of classical soliton modes is not 
equal to number of distinct flavour of the elementary excitations in the quasiparticle content of an integrable quantum chain
as one would na\"{i}vely expect, but indeed exceeds the rank of the group $G$. This owes to formation of emergent classical 
multiflavored degrees of freedom called stacks produced by gluing together magnons of different flavors (which dissolve upon 
introducing quantum corrections). Stacks that contain momentum-carrying magnons should be regarded as independent physical excitations 
and we outlined how they can be naturally arranged on vertices of the Hasse diagrams. The total number of physical stacks is found to 
be in perfect agreement with the prediction of the non-relativistic variant of the Goldstone theorem. As a byproduct of our 
work, we provided a full classification of ferromagnets with the global symmetry of a simple Lie algebra and fundamental onsite
degrees of freedom.	

The anomalous nature of charge transport can be explained, at the most formal level, by invoking fusion identities amongst 
characters of Yangian symmetry algebras, which at thermodynamic quasiparticle spectra translates to a particular rigid algebraic 
structure. We thus wish to stress that the list of models discussed in this study is unlikely to be exhaustive and
other classes of integrable models not included in our study are likewise expected to host a superdiffusive charge transport with 
characteristic exponent $z=3/2$: we anticipate that the phenomenon extends beyond Yangians associated with simple Lie algebras to other 
classes of models based on \emph{rational} solutions to the Yang--Baxter equations which includes (but is necessarily limited to)
the Fermi--Hubbard model \cite{Fava20} and to other supersymmetric $\mathfrak{su}(n|m)$ (or orthosymplectic $\mathfrak{osp}(n|2m)$) 
integrable spin chains that realize Yangian symmetries associated with Lie superalgebras (see e.g. \cite{KSZ08,Frassek11,I17}) whose 
bosonic Noether charges are known to display divergent diffusion constants \cite{Ilievski18}, or integrable models based on
infinite-dimensional affine symmetry algebras \cite{Jimbo86} (such as the Izergin--Korepin model \cite{IK81}).

To benchmark our predictions, we have performed numerical simulations on a handful of representative instances of integrable
quantum chain with local ferromagnetic exchange interaction, confining ourselves to classical simple Lie algebras and only to
local degrees of freedom  thattransform in the defining representation. We confirmed the predicted anomalous algebraic dynamical 
exponent $z=3/2$ with great numerical precision. On the flip side, we did not manage to reliably discern the anticipated KPZ scaling 
profiles. We note that spin chains with symmetry of exceptional Lie algebras have been deliberately excluded from the present numerical study largely for technical reasons: on the one hand, algebraic construction of integrable quantum chains based on the `quantized 
symmetry' of an exceptional Lie algebra is quite laborious (see e.g. \cite{KunibaG2}) and thus we find it better suited for a separate
technical consideration. Another practical obstacle is that dimensions of the defining irreducible representations are
(possibly with exception of $({\bf 7})$ and $({\bf 14})$ of $G_{2}$) conceivably too large to be efficiently simulated
with current DMRG techniques.

Finally, a central piece of the full puzzle is still missing: despite integrability, a first-principles proof of the KPZ 
scaling profiles of dynamical structure factors remains elusive. At this time, there only exists a phenomenological
picture which, under certain plausible assumptions, justifies the emergence of a noisy Burgers equation for the basic $\SU(2)$
case \cite{Vir20}. One obvious drawback of such an approach is that it cannot yield quantitative predictions such as the value of the 
KPZ coupling. Understanding quantitatively, from ab-initio principles, how KPZ universality emerges (and not only $z=3/2$) remains a 
major challenge for future works, even in the case of the $\SU(2)$ symmetric Heisenberg XXX spin chain.
Another open problem is to appropriately extend the proposal of \cite{NGIV20} to systems invariant under symmetries of higher-rank in 
order to extract the KPZ nonlinearity constant, which likely requires one to derive a semiclassical scaling limit for the NBA equations 
by fully accounting for presence of stack condensates.


{\textbf{Acknowledgments}}.
We thank Michele Fava, \v{Z}iga Krajnik, Fedor Levkovich-Maslyuk, Sid Parameswaran, Toma\v{z} Prosen, and Marko \v{Z}nidari\v{c} for 
comments, stimulating discussions and collaboration on related topics. This work was supported by the National Science Foundation under 
NSF Grant No. DMR-1653271 (S.G.), the US Department of Energy, Office of Science, Basic Energy Sciences,
under Early Career Award No. DE-SC0019168 (R.V.), the Alfred P. Sloan Foundation through a Sloan Research Fellowship (R.V.),
the Research Foundation Flanders (FWO, J.D.N.), and the Slovenian Research Agency (ARRS) program P1-0402 (E.I.).

\bibliography{SSD}

%

\appendix

\section{Simple Lie algebras}
\label{app:Lie}

For the reader's convenience, we briefly recap the basic concepts of Lie algebra.
For a more thorough presentation and other details the reader is referred to any of the standard textbook on the subject,
see e.g. \cite{Weyl_book,FultonHarris_book}.

A simple Lie algebra $\mathfrak{g}$ of rank $r={\rm rank}(\mathfrak{g})$ can be geometrically described
in terms of its root lattice $\Delta$ which is formed by $r$-dimensional real vectors $\alpha\in \Delta$.
The latter further splits into two disjoint sets $\Delta=\Delta_{+}\cup \Delta_{-}$, where $\Delta_{\pm}$ designate sublattices
of positive and negative roots.
Given a root lattice $\Delta$, there always exist a subset of $r$ roots $\alpha_{i}\in \Delta_{+}$, called \emph{simple roots};
every root can be expressed as an integral linear combination of the simple ones. Simple roots form a non-orthogonal basis in Euclidean 
space $\mathbb{R}^{r}$ (there is no unique choice of simple roots; different bases are related by the Weyl group transformations).
The root space is endowed with an Euclidean inner product $(\cdot,\cdot)$.

Below we employ the Cartan--Weyl basis of $\mathfrak{g}$, defined with respect to the Cartan decomposition
\begin{equation}
\mathfrak{g} = \mathfrak{t}\oplus \bigoplus_{\gamma \in \Delta_{\pm}}\mathfrak{g}_{\gamma},
\end{equation}
where we have denoted by
\begin{equation}
\mathfrak{t} = {\rm span}_{\mathbb{C}}\{H^{i};1\leq i\leq r\},
\end{equation}
the maximal abelian subalgebra spanned by Cartan generators, $[\rH^{i},\rH^{j}]=0$,
while $\mathfrak{g}_{\pm \alpha} \equiv {\rm span}_{\mathbb{C}}(X^{\pm \alpha})$ are one-dimensional spaces
spanned by Weyl generators $\rE^{\pm \alpha}$ for $\pm \alpha \in \Delta_{\pm}$.
In the Cartan--Weyl basis, commutation relations involving Weyl generators read
\begin{align}
[\rH^{i},\rE^{\pm \alpha}] &= \pm \alpha^{i}\rE^{\pm \alpha},\\
[\rE^{\alpha},\rE^{-\alpha}] &= \kappa_{\alpha,-\alpha}\sum_{i,j=1}^{r}\alpha^{i}(\kappa^{-1})_{i j}\rH^{j},\\
[\rE^{\alpha},\rE^{\beta \neq -\alpha}] &= {\rm N}_{\alpha,\beta}\rE^{\alpha+\beta},
\end{align}
where ${\rm N}_{\alpha \beta}=0$ when $\alpha + \beta \notin \Delta$.
Structure constants depend on normalization convention, and can be fixed by setting
\begin{equation}
\kappa_{ij}\equiv {\rm Tr}(\rH^{i}\rH^{j}),\qquad \kappa_{\alpha,-\alpha}\equiv {\rm Tr}(\rE^{\alpha}\rE^{-\alpha}),
\end{equation}
corresponding to the non-zero elements of the renormalized Killing metric restricted to the Cartan and Weyl sectors, respectively
(with the generators evaluated in the defining representation of $\mathfrak{g}$).

Let $\mathcal{V}$ be a representation of $\mathfrak{g}$ and $\ket{\lambda_{j}}$ denote a basis which diagonalizes the
Cartan elements, $\rH^{i}\ket{\lambda_{j}}=\lambda^{i}\ket{\lambda_{j}}$.
Eigenvalues $\lambda^{i}$, called \emph{weights}, are vectors from the dual vector space $\mathfrak{t}^{*}$.
In the case of unitary representations, generators $\rH^{i}$ are hermitian and weights are real, $\lambda^{i}\in \mathbb{R}$.
Weights of finite-dimensional representations obey the integrality constraint $2(\lambda,\alpha)/(\alpha,\alpha)\in \mathbb{Z}$ for
$\alpha \in \Delta$. Weights associated with the adjoint representation of $\mathfrak{g}$ are identified with \emph{roots},
that is $E^{\alpha}$ are eigenvectors of the adjoint action ${\rm ad}_{H}$,
$[\rH^{i},\rE^{\alpha}]=\alpha(\rH^{i})\,\rE^{\alpha}$ with $\alpha^{i}\equiv \alpha(H^{i})$ denoting the $i$th component of
a positive root vector $\alpha \in \Delta_{+}$. Roots are thus linear functional of $\mathfrak{t}$, and since every root is a linear 
combination of the simple ones $\alpha_{i}$, the latter provide the basis of $\mathfrak{t}^{*}$ dually paired to $\mathfrak{t}$.

\paragraph*{Classification.}
Geometric structure of the root lattice is encoded in the $r$-dimensional \emph{Cartan matrix} $\mathcal{K}$, defined as
\begin{equation}
\mathcal{K}_{ab} = t_{a}(\alpha_{a},\alpha_{b}),
\end{equation}
with $t_{a} \equiv 2/(\alpha_{a},\alpha_{a}) \in \{1,2,3\}$. The matrix elements of $\mathcal{K}$ are always integers.
For simply-laced $\mathfrak{g}$ we have $t_{a}=1$, whereas for non-simply laced algebra $t_{a}=1$ ($t_{a}=t\in \{2,3\}$) corresponds
to long (short) simple roots. The Cartan matrices can be visualized by \emph{Dynkin diagrams}
(as depicted in e.g. Fig.~\ref{fig:Dynkin_empty}): we draw an node (circle) for each diagonal elements $\mathcal{K}_{aa}=2$,
$a=1,\ldots r$; each two nodes are connected by $0\leq {\rm max}(|\mathcal{K}_{ab}|,|\mathcal{K}_{ba}|)\leq 3$ lines (i.e. not 
connected when $\mathcal{K}_{ab}=\mathcal{K}_{ba}=0$); when $|\mathcal{K}_{ab}|>|\mathcal{K}_{ba}|$ we put an arrow pointing towards 
node $b$.

Compete classification of simple Lie algebras is due to Cartan.
There are \emph{four} infinite families of so-called classical Lie algebras, isomorphic to the following
matrix algebras
\begin{align}
\mathfrak{su}(r) &=\! \{X\in {\rm End}(\mathbb{C}^{r})|X + X^{\dagger}=0,{\rm Tr}(X)=0\},\\
\mathfrak{so}(r) &=\! \{X\in {\rm End}(\mathbb{C}^{r})|X + X^{\rm T} = 0\},\\
\mathfrak{sp}(2r) &=\! \{X\in {\rm End}(\mathbb{C}^{2r})|X + J^{-1}\,X^{\rm T}\,J = 0\},
\end{align}
with symplectic unit matrix $J = \mathds{1}_{r}\otimes (-\ii\,\sigma^{2})$; the standard notations is $A_{r}\equiv \mathfrak{su}(r+1)$,
$C_{r}\equiv \mathfrak{sp}(2r)$, whereas orthogonal algebras are of two types: $B_{r}\equiv \mathfrak{so}(2r+1)$
and $D_{r}\equiv \mathfrak{so}(2r)$ ($r\in \mathbb{N}$).
The dimensions are
\begin{align}
{\rm dim}\,\mathfrak{su}(r) &= (r-1)(r+1),\\
{\rm dim}\,\mathfrak{so}(r) &= \tfrac{1}{2}(r-1)r,\\
{\rm dim}\,\mathfrak{sp}(r) &= r(2r+1).
\end{align}
In addition, there exist five exceptional finite dimensional Lie algebras.

\paragraph*{Irreducible representations.}
Let $\{\omega_{a}\}_{a=1}^{r}$ denote the set of \emph{fundamental weights}; they are dual to the co-roots associated to simple roots
\begin{equation}
(\alpha_{a},\omega_{b}) = \frac{\delta_{ab}}{t_{a}},\qquad \omega_{a} = \left(\mathcal{K}^{-1}\right)_{ba}\alpha_{b}.
\end{equation}
Any weight $\lambda$ can be expressed as a linear combination of the fundamental weights,
\begin{equation}
\lambda = m_{1}\omega_{1} + m_{2}\omega_{2} + \ldots + m_{r}\omega_{r},\quad m_{i} = 2t_{i}(\lambda,\alpha_{i}).
\end{equation}
Any finite-dimensional $\mathfrak{g}$-module $\mathcal{V}_{\Lambda}$ is characterized by the highest weight
\begin{equation}
\Lambda = \sum_{a=1}^{r}m_{a}\omega_{a},\qquad m_{a}\geq 0.
\end{equation}
The corresponding highest-weight vector is annihilated by all the raising (Chevalley) generators $\rE^{\alpha}$
for $\alpha \in \Delta_{+}$. Acting with the lowering generators $\rE^{-\alpha}$ on a state with weight $\lambda$,
its weight gets reduced by $\alpha$.

\section{Dressed charge fluctuations}
\label{app:magic}

Consider a general equilibrium macrostate characterized by quasiparticle densities $\rho_{A}(\theta)$.
Dressed charge susceptibilities within the Cartan sector can be linked to density fluctuations via
\begin{align}
\langle q^{(i){\rm dr}}_{A}q^{(j){\rm dr}}_{A} \rangle &= \sum_{B,B^{\prime}}\int \dd \theta \dd \theta^{\prime}
\frac{\delta q^{(i){\rm dr}}_{A}}{\delta \rho_{B}(\theta)}\frac{\delta q^{(j){\rm dr}}_{A}}{\delta \rho_{B^{\prime}}(\theta)}
\nonumber \\
&\times \langle \delta \rho_{B}(\theta) \delta \rho_{B^{\prime}}(\theta^{\prime}) \rangle.
\end{align}
Variations of $q^{(i){\rm dr}}_{A}$ with respect to quasiparticle densities $\rho_{A}(\theta)$ reads
\begin{equation}
\frac{\delta q^{(i){\rm dr}}_{A}}{\delta \rho_{B}(\theta)} = \sum_{B} q^{(i){\rm dr}}_{B}K^{\rm dr}_{AB}(\theta),
\end{equation}
where $K^{\rm dr}_{AB}(\theta)$ are the dressed scattering differential phases.
Using that~\cite{PhysRevB.54.10845}
\begin{equation}
\langle \delta \rho_{A}(\theta) \delta \rho_{A^{\prime}}(\theta^{\prime}) \rangle
= \frac{1}{\ell}C_{A A^{\prime}}(\theta,\theta^{\prime}),
\end{equation}
and employing for convenience the following compact matrix notations
\begin{align}
{\bf C} &= \boldsymbol{\Omega}[{\bf n}]^{-1}\boldsymbol{\rho}({\bf 1}-{\bf n})\boldsymbol{\Omega}[{\bf n}],\\
{\bf K}^{\rm dr} &= \boldsymbol{\Omega}[{\bf n}]{\bf K}.
\end{align}
where $\boldsymbol{\Omega}[{\bf n}]\equiv ({\bf 1}+{\bf K}\,{\bf n})^{-1}$, we arrive at
\begin{equation}
\langle q^{(i){\rm dr}}_{A}q^{(j){\rm dr}}_{A} \rangle
= \frac{1}{\ell}\Big({\bf K}^{\rm dr}{\bf q}^{\rm dr}\,{\bf C}\,{\bf q}^{\rm dr}{\bf K}^{\rm dr}\Big)_{A A},
\end{equation}
where ${\bf q}^{\rm dr}$ stands for the diagonal matrix of dressed charges $q^{\rm dr}_{A}$.

For the Cartan charges $Q^{(i)}$ considered in this paper, in an unpolarized background (i.e. $h_{i}=0$ for all $i=1,2,\ldots,r$),
we have $q^{(i)\rm dr}_{a,s}=0$, with exception in the limit $s \to \infty$ where the above expression simplifies
to give Eq.~\eqref{eqn:MagicFormula}.

\section{Kirillov--Reshetikhin modules}
\label{app:modules}

We consider an integrable quantum chain of length $L$ with a nested spectrum. In order to find the number of distinct Bethe 
eigenstates in the finite-volume spectrum we consider an $L$-fold product representation of $\mathcal{W}$-modules 
$\mathcal{W}_{\Lambda_{\rm p}}$ and its decomposition into $\mathfrak{g}$-modules $\mathcal{V}_{\Lambda}$
(including non-trivial multiplicities)
\begin{equation}
\mathcal{W}^{\otimes L}_{\Lambda_{\rm p}}\Big|_{\mathfrak{g}} = \bigoplus_{\Lambda}
Z_{\Lambda_{\rm p}}(\{N_{i}\})\mathcal{V}_{\Lambda}.
\label{eqn:W_decomposition}
\end{equation}
In the decomposition one finds multiplets with the highest weights
\begin{equation}
\Lambda = L\,\Lambda_{\rm p} - \sum_{i=1}^{r}N_{i}\alpha_{i},
\end{equation}
which can occur with higher multiplicities. Here integers $\{N_{i}\}_{i=1}^{r}$ denote the total number of Bethe roots of each flavor 
$i\in \mathcal{I}_{r}$, providing a bijective correspondence between the highest weight $\Lambda$ and $\{N_{i}\}$
which determine a sector with fixed values of $\U(1)$ charges.
Creating a quasi-particle of type $a$ therefore reduces the maximal weight $L\,\Lambda_{\rm p}$ of the
reference ferromagnetic vacuum by the corresponding simple root $\alpha_{i}$.

Irreducible representations $\mathcal{V}_{\Lambda}$ which appear in decomposition \eqref{eqn:W_decomposition} occur with
multiplicities $Z_{\Lambda_{\rm p}}(\{N_{i}\})$. These can be computed with aid of combinatorial formulae,
first conjectured by Kirillov and Reshetikhin \cite{KR86,KR90} (afterwards extensively investigated in a series
of papers \cite{KN92,KNS94,Hernandez09}), reading
\begin{equation}
Z_{\Lambda_{\rm p}}(\{N_{i}\}) = \sum_{\{n\}}\prod_{a=1}^{r}\prod_{s}\binom{\mathscr{P}_{a,s}+n_{a,s}}{n_{a,s}},
\label{eqn:multiplicity_factor}
\end{equation}
with
\begin{align}
\mathscr{P}_{a,s} &= L\,{\rm min}(s,s_{\rm p})\delta_{a,a_{\rm p}} -2\sum_{s'\geq 1}{\rm min}(s,s')n_{a,s'} \nonumber \\
&+ \sum_{s'\geq 1}\sum_{b\neq a}^{r}{\rm min}(-\mathcal{K}_{ab}s,-\mathcal{K}_{ba}s')n_{b,s'}.
\end{align}
Here the sum runs over all integer sets (partitions) $\{n_{a,s}\}$ such that
\begin{equation}
N_{a} = \sum_{s\geq 1}s\,n_{a,s},\qquad n_{a,s}\geq 0,\quad 1\leq a \leq r,
\end{equation}
and $\mathscr{P}_{a,s}\geq 0$ for all $a$ and $s$.

\paragraph*{Remark.}
While the described partitioning scheme assigns a unique pattern of quantum numbers to each eigenstate of the highest-weight type,
it is remarkable that the is no bijective mapping with bound-state excitations (Bethe string). The correspondence is
spoiled by exceptional solutions which involve so-called collapsed strings. The precise correspondence can nonetheless be
established within the Wronskian formulation of Bethe Ansatz equations, as recently formulated in \cite{Chernyak20}.

\subsection{Multiplicities}
\label{app:multiplicities}

The aim of this section is to illustrate the `multiplicity formula' on two simple examples.
We begin with $A_{1}\equiv \mathfrak{su}(2)$ of rank $r=1$, with a single type of magnons in the spectrum.
Within each fixed $N$-particle sector, each highest-weight eigenstate is characterized by a pattern of $s$-strings $\{n_{s}\}$,
such that $N=\sum_{s\geq 1}s\,n_{s}$. The multiplicity factor is
\begin{equation}
Z_{\omega_{1}}(N) = \sum_{\{n_{s}\}}\prod_{s\geq 1}\binom{\mathscr{P}_{s}+n_{s}}{n_{s}},
\end{equation}
where $\mathscr{P}_{s}=L-2\sum_{s'\geq 1}{\rm min}(s,s')n_{s'}$.

For instance, in the fundamental $\SU(2)$ chain (the Heisenberg XXX model), $Z_{\omega_{1}}(N)$ for $0\leq N\leq L/2$ gives 
the numbers of distinct highest-weight Bethe eigenstates in the charge sector with $N$ magnons
that carries weight $\Lambda=(L-2N)\omega_{1}$. The latter coincides with the branching coefficients in the decomposition of 
$\mathcal{V}^{\otimes L}_{\omega_{1}}$, that is $b_{N}=\binom{L}{N}-\binom{L}{N-1}$.

As a concrete example we consider for the homogeneous spin-$1/2$ chain with $L=5$ sites,
whose Hilbert space decomposes into the following multiplets
\begin{equation}
\mathcal{H}\cong \mathcal{V}^{\otimes 5}_{\omega_{1}} = 5({\bf 2}) + 4({\bf 4}) + ({\bf 6}).
\end{equation}
Here $({\bf n})$ denote the $\SU(2)$ irreducible representation of dimension $n$, and the integer number in front is their
respective degeneracies. There are all together $5+4+1=10$ distinct highest-weight Bethe eigenstates divided into
three particles sectors with $N\in \{0,1,2\}$. Notice first that the empty state $Z(0)=1$ is the highest-weight ferromagnetic 
vacuum which resides in the largest multiplet $({\bf 6})$. The other two multiplicities are
\begin{equation}
Z_{\omega_{1}}(1) = \underbrace{\binom{3+1}{1}}_{(n_{1},n_{2})=(1,0)} = 4,
\end{equation}
yielding $4$ one-magnon eigenstates belonging to inequivalent multiplets $({\bf 4})$, and
\begin{equation}
Z_{\omega_{1}}(2) = \underbrace{\binom{1+2}{2}}_{(n_{1},n_{2})=(2,0)}
+ \underbrace{\binom{1+1}{1}}_{(n_{1},n_{2})=(0,1)} = 5,
\end{equation}
yielding $5$ distinct doublets $({\bf 2})$, among which $3$ contain two unbound magnons and $2$ states a bound pair ($2$-string).

As a basic example of a rank-$2$ algebra, we inspect as our second example the $B_{2}$ case.
Consider now the $L=3$ site fundamental $\SO(5)$ chain, in which case the Hilbert decomposes as
\begin{equation}
\mathcal{H}\cong \mathcal{V}^{\otimes 3}_{\omega_{1}} = 3({\bf 5}) + ({\bf 10}) + ({\bf 30}) + 2({\bf 35}).
\end{equation}
There are $3+1+1+2=7$ highest-weight Bethe eigenstates in the spectrum, whose weights $\Lambda=(\lambda_{1},\lambda_{2})$
are given by $\lambda_{1}=L-2N_{1}+N_{2}$ and $\lambda_{2}=2(N_{1}-N_{2})$.
Apart from the ferromagnetic vacuum, $\Lambda=[3,0]=({\bf 30})$, the are the following particle sectors:

\begin{center}
\centering
\renewcommand{\arraystretch}{1.2}
\begin{tabular}{|c|c|}
\hline
$(N_{1},N_{2})$ & $\Lambda$ \\
\hline
$(1,0)$ & $[1,2]$ \\
$(1,1)$ & $[2,0]$ \\
$(2,1)$ & $[0,2]$ \\
$(2,2)$ & $[1,2]$ \\
\hline
\end{tabular}
\end{center}

To compute multiplicities, we need
\begin{align}
\mathscr{P}_{1,s} &= L - 2\sum_{s'\geq 1}{\rm min}(s,s')n_{1,s'} \nonumber \\
&+ \sum_{s'\geq 1}{\rm min}(-\mathcal{K}_{12}s,-\mathcal{K}_{21}s')n_{2,s'},\\
\mathscr{P}_{2,s} &=  - 2\sum_{s'\geq 1}{\rm min}(s,m)n_{2,m} \nonumber \\
&+ \sum_{s'\geq 1}{\rm min}(-\mathcal{K}_{21}s,-\mathcal{K}_{12}s')n_{1,s'},
\end{align}
where $\mathcal{K}_{12}=-2$, $\mathcal{K}_{21}=-1$. Here $n_{1,s'}$ and $n_{2,s'}$ denote numbers of $s'$-strings of
flavor $1$ and $2$, respectively. We find
\begin{align}
Z_{\omega_{1}}(1,0) &= \binom{1+1}{1} = 2,\\
Z_{\omega_{1}}(2,1) &= \underbrace{\binom{0+2}{2}\binom{0+1}{1}}_{n_{1,1}=2,\,n_{2,1}=1} = 1,\\
Z_{\omega_{1}}(2,2) &= \underbrace{\binom{1+2}{2}}_{n_{1,1}=2,\,n_{2,2}=1} = 3.
\end{align}
Notice that the $(1,1)$-sector admits no solutions, $Z_{\omega_{1}}(1,1) = 0$, since not all $\mathcal{P}_{a,s}$ can be simultaneously
non-negative. For the exact same reason $Z_{\omega_{1}}(2,1)$ has no contribution from partition $n_{1,2}=n_{2,1}=1$, that is
$2$-strings of flavor $a=1$ are not allowed.
Lastly, in the $(N_{1},N_{2})=(2,2)$ sector, we find for all $n_{1,2}=n_{2,1}=1$, $n_{1,2}=n_{2,2}=1$
and $n_{1,1}=n_{2,1}=2$ partitions $\mathscr{P}_{2,1}<0$.

\subsection{Classical characters}
\label{app:characters}

For the purpose of the next section, we collect the Weyl determinant formulae for characters of the classical simple Lie algebras 
$\mathfrak{g}$ which can be found in standard text such as \cite{Weyl_book} or \cite{FultonHarris_book}.
Knowing closed-form expressions for the classical $\mathfrak{g}$-characters proves helpful in finding explicit solutions to
the constant $\scrT$-system relations (are provided blow in section \ref{app:Tsystem}) using the `children expansion'
to decompose the Kirillov--Resthetikhin modules over $\mathfrak{g}$-modules.

\paragraph*{$A$-series.}
We begin with the unitary series $\SU(n)$. A generic element $g_{0}\in T\subset G$ from torus subgroup $T$,
can be parametrized as
\begin{equation}
g_{0}={\rm diag}(x_{1},x_{2},\ldots,x_{n}),\qquad \prod_{i=1}^{n}x_{i}=1.
\end{equation}
Every finite dimension irreducible representation  $\mathcal{V}_{\Lambda}$ of $\mathfrak{g}$ can be associated
a character $\chi_{\Lambda}$, which by definition is the trace of the corresponding group element.
The first Weyl formula for $\SU(n)$ characters of rectangular representations
$\Lambda=s\,\omega_{a}=[0,\ldots,s,0,\ldots,0]$ (rectangular Young diagrams of dimension $a\times s$) is a ratio of two determinants
\begin{equation}
\chi_{a,s}(\{x_{i}\}) = \frac{\|x^{n-i-\sigma_{a,i}}_{j}\|}{\|x^{n-i}_{j}\|},
\end{equation}
where $\sigma_{a,i}=1$ if $a\geq i$. The totally symmetric characters can be computed with aid of the generating function
\begin{equation}
w(z) = \prod_{i=1}^{n}\frac{1}{1-z\,x_{i}} = \sum_{s=0}^{\infty}\chi_{1,s}(\{x_{i}\})z^{s}.
\end{equation}
For example, in the $\SU(2)$ case we can put $x_{1}=1/x_{2}=e^{h}$, and we have
\begin{align}
\chi_{1,s}(x_{1},x_{2}) &= (x^{s+1}_{1}-x^{s+1}_{2})/(x_{1}-x_{2}) \nonumber \\
&=\frac{\sinh{((s+1)h)}}{\sinh{(h)}}.
\end{align}

\paragraph*{$B$-series and $C$-series.}
Classical characters $\chi_{\Lambda}=\chi_{\Lambda}(x_{1},\ldots,x_{r})$ for Lie algebras $C_{r}$ corresponding
to unitary irreducible representations $\mathcal{V}_{\Lambda}$ of highest weight
$\Lambda=(\lambda_{1},\lambda_{2},\ldots,\lambda_{r})$ take the following form
\begin{equation}
\chi_{\Lambda} = \frac{\|x^{\lambda_{j}+r+1-j}_{i}-x^{-(\lambda_{j}+r+1-j)}_{i}\|}
{\|x^{r+1-j}_{i}-x^{-(r+1-j)}_{i}\|}.
\end{equation}
Similarly, classical characters for Lie algebra $B_{r}$ depend on parameters $\{x_{i}\}_{i=1}^{r}$,
$\chi(\{x_{i}\}) \in \mathbb{Z}[x^{\pm 1}_{1},x^{\pm 1}_{2},\ldots,x^{\pm 1}_{r}]$. They are given by ratios of
two $r$-dimensional determinants
\begin{equation}
\chi_{\Lambda} = \frac{\|x^{\lambda_{i}+\sigma_{i}}_{j}-x^{-(\lambda_{i}+\sigma_{i})}_{j}
+x^{\lambda_{i}+\sigma_{i}-1}_{j}-x^{-(\lambda_{i}+\sigma_{i}-1)}_{j}\|}
{\|x^{\sigma_{i}}_{j}-x^{-\sigma_{i}}_{j}\|},
\end{equation}
where $\sigma_{i}\equiv r+1-i$. They are manifestly invariant under $x_{i}\to 1/x_{i}$.

\paragraph*{$D$-series.}

Consider a unitary finite-dimensional irreducible representations of $D_{r}=\mathfrak{so}(2r)$ (with $r\geq 3$) characterized by
Dynkin labels $\Lambda=(\lambda_{1},\lambda_{2},\ldots,\lambda_{r})$. Using the assignment
\begin{align}
\ell_{i} &= \sum_{k=i}^{r-2}\lambda_{k} + \tfrac{1}{2}(\lambda_{r-1} + \lambda_{r}),\quad i=1,\ldots,r-1\\
\ell_{r} &= \tfrac{1}{2}(\lambda_{r-1} - \lambda_{r}),
\end{align}
one obtains the corresponding Young tableaux with row lengths $\ell_{i}$. We have $\ell_{i}\in \mathbb{N}$ for non-spinor
representations and $\ell_{i}=\tfrac{1}{2}\mathbb{N}$ for spinor representations.
Writing $\sigma_{i}\equiv \ell_{i}-i+r$, the characters reads
\begin{equation}
\chi_{\Lambda} = \frac{\|\lambda^{\sigma_{j}}_{i}+\lambda^{-\sigma_{j}}_{i}\|+\|\lambda^{\sigma_{j}}_{i}-\lambda^{-\sigma_{j}}_{i}\|}
{\|\lambda^{r-j}_{i}+\lambda^{-(r-j)}_{i}\|}.
\end{equation}

\subsection{$\calT$-system functional relations}
\label{app:Tsystem}

We spell out the $\calT$-system functional relations \cite{KP92,KLV16} for the thermodynamic $\calT$-functions $\calT_{a,s}(\theta)$.
Here we only consider $\calT$-functions in the large-$\theta$ limit, $\scrT_{a,s}\equiv \lim_{|\theta|\to \infty}\calT_{a,s}(\theta)$
which formally correspond to quantum characters associated to rectangular Kirillov--Reshetikhin modules of Yangians $Y(\mathfrak{g})$, 
\cite{KR86,Kirillov89,KR90}. General formulae which include dependence on $\theta$ are collected in e.g. \cite{KNS11}.

Constant $\scrT$-functions encode grand-canonical Gibbs equilibrium state in the $\beta \to 0$ limit.
They thus depend solely on the $\U(1)$ chemical potentials $x_{i}$ associated with $r$ Cartan charges of $\mathfrak{g}$.
For simply-laced algebras, that is $A_{r}$ and $D_{r}$ and $E_{r=6\sim 8}$, the constant $\calT$-system assumes the form
\begin{equation}
\scrT^{2}_{a,s} = \scrT_{a,s-1}\scrT_{a,s+1} + \prod_{b\sim a}\scrT_{b,s},
\end{equation}
where $\sim$ stands for the neighboring nodes in the Dynkin diagram, that is the sum run over all $b$ for which $\mathcal{K}_{ab}=-1$.
The $A$-series is special in that $\mathcal{W}$-modules are also irreducible as $\mathfrak{g}$-modules
and therefore the $\scrT$-functions are given by classical $\mathfrak{g}$-characters $\chi_{a,s}$ themselves.

\begin{widetext}

For non-simply-laced simple Lie algebras, that is for $B_{r}\equiv \mathfrak{so}(2r+1)$ and $C_{r}\equiv \mathfrak{sp}(2r)$,
$\scrT$-functions decompose non-trivially in terms of $\chi_{a,s}$.
The constant $\scrT$-systems read
\begin{align}
\scrT^{2}_{a,s} &= \scrT_{a,s-1}\scrT_{a,s+1} + \scrT_{a-1,s}\scrT_{a+1,s},\qquad 1\leq a \leq r-2, \nonumber \\
\scrT^{2}_{r-1,s} &= \scrT_{r-1,s-1}\scrT_{r-1,s+1} + \scrT_{r-2,s}\scrT_{r,2s},\nonumber \\
\scrT^{2}_{r,2s} &=  \scrT_{r,2s-1}\scrT_{r,2s+1} + \scrT^{2}_{r-1,s},\\
\scrT^{2}_{r,2s+1} &= \scrT_{r,2s}\scrT_{r,2s+2} + \scrT_{r-1,s}\scrT_{r-1,s+1}, \nonumber
\end{align}
for the $B_{r}$-series and
\begin{align}
\scrT^{2}_{a,s} &= \scrT_{a,s-1}\scrT_{a,s+1} + \scrT_{a-1,s}\scrT_{a+1,s},\qquad 1\leq a \leq r-2,\nonumber \\
\scrT^{2}_{r-1,2s} &= \scrT_{r-1,2s-1}\scrT_{r-1,2s+1} + \scrT_{r-2,2s}\scrT^{2}_{r,s},\nonumber \\
\scrT^{2}_{r-1,2s+1} &= \scrT_{r-1,2s}\scrT_{r-1,2s+1} + \scrT_{r-2,2s+1}\scrT_{r,s}\scrT_{r,s+1}, \nonumber \\
\scrT^{2}_{r,s} &= \scrT_{r,s-1}\scrT_{r,s+1} + \scrT_{r-1,2s},
\end{align}
for the $C_{r}$-series. Although $B_{r}$ and $C_{r}$ are non-isomorphic algebras in general, there is an exceptional isomorphism
$\mathfrak{so}(5)\cong \mathfrak{sp}(4)$ for $r=2$, in which case the equations maps to each other under permutation of flavors 
$1\leftrightarrow 2$.

There are analogous formuale for the exceptional algebras which can be found in e.g. \cite{KNS11}.

\end{widetext}

\subsection{Character expansions}
\label{app:character_expansions}

The $\scrT$-functions are completely determined by the structure of the underlying classical 
Lie algebra $\mathfrak{g}$. Formally, the thermodynamic $\scrT$-functions are identified as `quantum characters' of
the Kirillov--Resthetikhin modules $\mathcal{W}_{a,s}$ associated to rectangular type representations of Yangians $Y(\mathfrak{g})$.

When viewed as $\mathfrak{g}$-modules, $\mathcal{W}_{a,s}$ are generically not irreducible.
The case of unitary algebras $\mathfrak{g}=A_{n}$ are exceptional since the 
$\mathcal{W}$-modules $\mathcal{W}_{a,s}$ are also irreducible as $\mathfrak{g}$-modules, implying that
$\scrT$-functions are precisely the $\mathfrak{su}(2)$ characters $\chi_{a,s}$ of rectangular irreducible 
representations $\mathcal{V}_{s\,\omega_{a}}\equiv \mathcal{V}_{a,s}$, cf. Eq.~\eqref{eqn:character_Hirota}.
For general $\mathfrak{g}$ such a correspondence no longer holds as $\mathcal{W}_{a,s}$ decompose non-trivially in terms
of $\mathfrak{g}$-modules $\mathcal{V}_{a,s}$. It is known that such $\mathcal{W}$-modules
admit an expansion over $\mathfrak{g}$-characters
$\chi(\mathcal{V}_{\lambda})\equiv \chi_{\lambda}$ of the form
\begin{equation}
\scrT_{a,s} = \underbrace{\chi(\mathcal{V}_{s\,\omega_{a}})}_{\rm parent}
+ \underbrace{\sum_{\lambda < s\,\omega_{a}}b_{\lambda}\chi(\mathcal{V}_{\lambda})}_{\rm children},
\label{eqn:children_expansion}
\end{equation}
for some branching coefficients $b_{\lambda}$. Decomposition \eqref{eqn:children_expansion} is sometimes suggestively referred to
as the `children expansion'. The general structure of such decomposition has been first conjectured in Refs.~\cite{Kirillov89,KR90}
and subsequently proved in a number of follow-up papers such as e.g. \cite{HTOKY99,Nakajima03,Hernandez09}.

Using that functions $\scrT_{a,s}$ satisfy the Hirota functional relation \eqref{eqn:general_Hirota},
the higher $\scrT$-functions can be recursively constructed given the `initial' ones $\scrT_{a,1}$ associated with characters of
$a$th fundamental $\mathcal{W}$-module $\mathcal{W}_{a,1}$. For $A_{r}$ and $C_{r\geq 2}$, we have simply
\begin{equation}
\scrT_{a,1}=\chi(\omega_{a}).
\end{equation}
For $B_{r\geq 2}$ and $D_{r\geq 3}$ on the other have, we have
\begin{align}
\scrT_{a\leq r-1,1} &= \chi(\omega_{a}) + \chi(\omega_{a-2}) + \ldots,\\
\calT_{r,1} &= \chi(\omega_{r}),
\end{align}
and
\begin{align}
\scrT_{a\leq r-2,1} &= \chi(\omega_{a}) + \chi(\omega_{a-2}) + \ldots,\nonumber \\
\scrT_{r-1,1} &= \chi(\omega_{r-1}), \quad \scrT_{r,1} = \chi(\omega_{r}),
\end{align}
respectively.

There is a nifty graphical representation behind expansions of the form \eqref{eqn:children_expansion} which can be carried out
at the level of Young diagrams: for the $C_{r}\equiv \mathfrak{sp}(2r)$ case, the children diagrams are obtained from the
parent rectangular diagram $a\times s$ by successively removing $1\times 2$ `dominos', whereas for $B_{r}\equiv \mathfrak{so}(2r+1)$ 
one likewise removes horizontal $2\times 1$ dominos. One has to pay special attention to the presence of `half-partitions' in the 
case of $B$-series, as we clarify below on an explicit example. Explicit combinatorial expressions for $\scrT_{a,s}$ in terms 
of appropriately restricted sum can be found in Ref.~\cite{KNS11}.

\begin{figure}[htb]
\centering
\begin{tikzpicture}
\begin{scope}
\node at (0,0) {$A_{r}:$};
\node at (2,0) {$\mathcal{W}_{2,3} = $};
\node at (3.4,0) {$\yng(3,3)$};
\end{scope}

\begin{scope}[yshift=-50pt]
\node at (0,0) {$B_{r},D_{r}:$};
\node at (2,0) {$\mathcal{W}_{2,3} = $};
\node at (3.4,0) {$\yng(3,3)$};
\node at (4.4,0) {$\oplus$};
\node at (5.2,0) {$\yng(2,2)$};
\node at (6.2,0) {$\oplus$};
\node at (6.8,0) {$\yng(1,1)$};
\node at (7.4,0) {$\oplus$};
\node at (7.8,0) {$\cdot$};
\end{scope}

\begin{scope}[yshift=-100pt]
\node at (0,0) {$C_{r}:$};
\node at (2,0) {$\mathcal{W}_{2,3} = $};
\node at (3.4,0) {$\yng(3,3)$};
\node at (4.4,0) {$\oplus$};
\node at (5.4,0) {$\yng(3,1)$};
\node at (6.4,0) {$\oplus$};
\node at (7,0) {$\yng(1,1)$};
\end{scope}
\end{tikzpicture}
\caption{Decomposition of irreducible Kirillov--Reshetikhin modules $\mathcal{W}_{a,s}$ over $\mathfrak{g}$-modules $\chi_{a,s}$
(depicted by Young diagrams) for the classical series, with $\mathcal{W}_{2,3}$ shown as an example. In the unitary case $A_{r}$,
the $\mathcal{W}$-modules are already indecomposable. In the case of $\{B_{r}$ and $D_{r}\}$, they simply decompose in terms of 
$\mathfrak{g}$-modules with summands obtained by removal of $2\times 1$ dominos (provided $a<r$ in the case of $B_{r}$, cf. remarks in 
appendix \ref{app:character_expansions}). Similarly, for the $C$-series, one successively removes $1\times 2$ dominos.} 
\end{figure}

\begin{widetext}

\subsubsection*{Examples}

We conclude by providing a few explicit examples of the `children expansion' of Yangian characters $\scrT_{a,s}$
in terms of $\mathfrak{g}$-characters $\chi_{a,s}$.

As our first example we consider algebra $B_{2}=\mathfrak{so}(5)$ of rank $2$. The initial (i.e. fundamental) $\scrT$-functions read
\begin{equation}
\scrT_{1,1}=\chi(\omega_{1}),\qquad \scrT_{2,1}=\chi(\omega_{2}),
\end{equation}
whereas the higher $\scrT$-functions are given by the following restricted sums
\begin{align}
\scrT_{1,s} &= \sum_{k_{1}\in \mathbb{Z}_{\geq 0}} \chi(k_{1}\omega_{1}),&\quad k_{1} &= s,\\
\scrT_{2,s} &= \sum_{k_{0},k_{2}\in \mathbb{Z}_{\geq 0}} \chi(k_{0}\omega_{1}+k_{2}\omega_{2}),&\quad k_{0}+k_{2} &= s.
\end{align}
The first sum indeed involves a single term, yielding $\scrT_{1,s}=\chi(s\,\omega_{1})$.
The situation with the $2\times s$ quantum characters $\calT_{2,s}$ is more subtle since
the corresponding highest weights $\Lambda=[0,s]$ are associated with so-called half-partitions
$\Lambda=(\tfrac{s}{2},\tfrac{s}{2})$ made of basic $1\times \tfrac{1}{2}$ rectangles.
The children characters, which are obtained by successive removal of $2\times 1$ dominos, therefore involve four basic rectangles.
The consequence of this is a staggered odd-even structure.

The constant $\scrT$-systems and $\scrY$-system functional relations have the form
\begin{equation}
\scrT^{2}_{1,s} = \scrT_{1,s-1}\scrT_{1,s+1} + \scrT_{2,2s},\qquad
\scrT^{2}_{2,s} = \scrT_{1,s-1}\scrT_{1,s+1} + \scrT_{1,\lfloor s/2\rfloor}\scrT_{1,\lfloor (s+1)/2\rfloor},
\end{equation}
and
\begin{align}
\scrY^{2}_{1,s} &= \frac{(1+\scrY_{1,s-1})(1+\scrY_{1,s+1})}{(1+1/\scrY_{2,2s-1})(1+1/\scrY_{2,2s+1})(1+1/\scrY_{2,2s})^{2}}
\nonumber \\
\scrY^{2}_{2,2s} &= \frac{(1+\scrY_{2,2s-1})(1+\scrY_{2,2s+1})}{1+1/\scrY_{1,s}},\\
\scrY^{2}_{2,2s+1} &= (1+\scrY_{2,2s})(1+\scrY_{2,2s+2}), \nonumber
\end{align}
respectively. In the `classical limit' $x_{i}\to 1$, we find explicitly
\begin{equation}
\scrT_{1,s} = d_{1,s} = \frac{1}{6}(s+1)(s+2)(2s+3),\qquad
\scrT_{2,2s} = \begin{cases}
\tfrac{1}{48}(s+1)(s+3)^{2}(s+5) & {\rm mod}(s,2)=1,\\
\tfrac{1}{48}(s+2)(s+4)(s^{2}+6s+6) & {\rm mod}(s,2)=0.
\end{cases}
\end{equation}
The dimensions $d_{\Lambda}\equiv {\rm dim}(\mathcal{V}_{\Lambda})$ in terms of Dynkin labels $\Lambda=[m_{1},m_{2}]$ read explicitly
\begin{equation}
d_{[m_{1},m_{2}]} = \frac{1}{6}(m_{1}+1)(m_{2}+1)(m_{1}+m_{2}+2)(2m_{1}+m_{2}+3).
\end{equation}

As the second example we consider $C_{3}=\mathfrak{sp}(6)$ or rank $3$.
The $\scrT$-system has the form
\begin{align}
\scrT^{2}_{1,s} &= \scrT_{1,s-1}\scrT_{1,s+1} + \scrT_{2,s},\\
\scrT^{2}_{2,2s} &= \scrT_{2,2s-1}\scrT_{2,2s+1} + \scrT_{1,2s}\scrT^{2}_{3,s},\\
\scrT^{2}_{2,2s+1} &= \scrT_{2,2s}\scrT_{2,2s+2} + \scrT_{1,2s+1}\scrT_{3,s}\scrT_{3,s+1},\\
\scrT^{2}_{3,s} &= \scrT_{3,s-1}\scrT_{3,s+1} + \scrT_{2,2s},
\end{align}
for $s\in \mathbb{N}$, with boundary conditions $\scrT_{0,s}\equiv 1$ and $\scrT_{0,s}\equiv 0$.
The equation for the middle row $a=2$ can be also written uniformly as
$\scrT^{2}_{2,s}=\scrT_{2,s-1}\scrT_{2,s+1}+\scrT_{1,s}\scrT_{3,\lfloor s/2 \rfloor}\scrT_{3,\lfloor (s+1)/2\rfloor}$.
The corresponding $\scrY$-system algebraic relations read
\begin{align}
\scrY^{2}_{1,s} &= \frac{(1+\scrY_{1,s-1})(1+\scrY_{1,s+1})}{1+1/\scrY_{2,s}},\\
\scrY^{2}_{2,2s} &= \frac{(1+\scrY_{2,2s-1})(1+\scrY_{2,2s+1})}{(1+1/\scrY_{1,2s})(1+1/\scrY_{3,s})},\\
\scrY^{2}_{2,2s+1} &= \frac{(1+\scrY_{2,2s})(1+\scrY_{2,2s+2})}{1+1/\scrY_{1,2s+1}},\\
\scrY^{2}_{3,s} &= \frac{(1+\scrY_{3,s-1})(1+\scrY_{3,s+1})}{(1+1/\scrY_{2,2s-1})(1+1/\scrY_{2,2s+1})(1+1/\scrY_{2,2s})^{2}}.
\end{align}
The fundamental quantum characters coincide with the classical ones, $\scrT_{a,1}=\chi_{a,1}$. Functions
$\scrT_{3,s}$ at $r=3$ coincide with classical rectangular $\mathfrak{sp}(6)$ characters $\scrT_{3,s}=\chi_{3,s}$,
whereas for $a\in \{1,2\}$ we have non-trivial sum over children characters obtain by iteratively removing $1\times 2$ dominos
from the parently $a\times s$ Young diagram:
\begin{equation}
\scrT_{1,2s} = \sum_{k=0}^{s}\chi_{1,2k},\qquad
\scrT_{1,2s+1} = \sum_{k=0}^{s}\chi_{1,2k+1},\qquad
\label{eqn:T1_C3}
\end{equation}
and for the two-row diagrams
\begin{equation}
\scrT_{2,s}(\{x_{i}\}) = \sum_{0\leq f_{1}\geq f_{2}\leq s;\,{\rm mod}(f_{i})=s}
\chi_{\Lambda=m_{1}\omega_{1}+m_{2}\omega_{2}}(\{x_{i}\}),
\qquad m_{i} = \ell_{i}-\ell_{i+1}
\label{eqn:T2_C3}
\end{equation}
with $\ell_{4}\equiv 0$. Associating to finite-dimension $\mathcal{V}$-module $\mathcal{V}_{\Lambda}$ of
$\mathfrak{g}=\mathfrak{sp}(6)$ Dynkin labels $[m_{1},m_{2},m_{3}]$ the partition (Young diagram) $(\ell_{1},\ell_{2},\ell_{3})$
with $\ell_{i}$ boxes in the $\ell$th row ($\ell_{i}\geq \ell_{i+1}$) we have for example
\begin{align}
\mathcal{W}_{2,2} &= (2,2,0) \oplus (2,0,0) \oplus (),\\
\mathcal{W}_{2,3} &= (3,3,0) \oplus (3,1,0) \oplus (1,1,0),\\
\mathcal{W}_{2,4} &= (4,4,0) \oplus (4,2,0) \oplus (4,0,0) \oplus (2,2,0) \oplus (2,0,0) \oplus (),
\end{align}
and so forth. Notice that $\scrT_{2,2s}$ is in fact fully determined by functions $\scrT_{3,s}$ owing to
\begin{equation}
\scrT_{2,2s} = \scrT^{2}_{3,s} - \scrT_{3,s-1}\scrT_{3,s+1}.
\end{equation}
and consequently together with $\scrT_{0,s}=\scrT_{4,s}=1$ and fundamental $\scrT$-functions $\scrT_{a,1}$, the remaining
unknown $\scrT$-functions $\scrT_{1,s}$ and $\scrT_{2,2s+1}$ are implicitly fixed by virtue of the Hirota equation.

In the limit $x_{i}\to 1$, $\chi_{a,s}$ reduce to dimensions $d_{a,s}=\lim_{\{x_{i}\}\to 1}\chi_{a,s}(\{x_{i}\})$
of rectangular representations $\mathcal{V}_{a,s}$, reading explicitly
\begin{align}
d_{1,s} &= \frac{1}{120}(s+1)(s+2)(s+3)(s+4)(s+5),\\
d_{2,s} &= \frac{1}{720}(s+1)(s+2)^{2}(s+3)^{2}(s+4)(2s+5),\\
d_{3,s} &= \frac{1}{360}(s+1)(s+2)(s+3)(2s+3)(2s+4)(2s+5),
\end{align}
Plugging these back into Eqs.~\eqref{eqn:T1_C3} and \eqref{eqn:T2_C3}, the obtain the following rational expressions:
\begin{align}
\scrT_{1,2s-1} &= \frac{2s^{6}}{45} + \frac{2s^{5}}{5} + \frac{49s^{4}}{36} + \frac{13s^{3}}{6} + \frac{287s^{2}}{180}
+\frac{13s}{30},\\
\scrT_{1,2s} &= \frac{2s^{6}}{45} + \frac{8s^{5}}{15} + \frac{91s^{4}}{36} + 6s^{3} + \frac{1337s^{2}}{180} + \frac{67s}{15} + 1,\\
\scrT_{2,2s-1} &= \frac{2s^{10}}{675} + \frac{2s^{9}}{45} + \frac{13s^{8}}{45} + \frac{16s^{7}}{15} + \frac{14771s^{6}}{600}
+\frac{147s^{5}}{40} + \frac{3823s^{4}}{1080} \nonumber \\
&+ \frac{761s^{3}}{360} + \frac{53s^{2}}{75} + \frac{s}{10},\\
\scrT_{2,2s} &= \frac{2s^{10}}{675} + \frac{8s^{9}}{135} + \frac{47s^{8}}{90} + \frac{8s^{7}}{3} + \frac{7853s^{6}}{900}+\frac{859s^{5}}{45} + \frac{3812s^{4}}{135} \nonumber \\
&+ \frac{752s^{3}}{27} + \frac{15761s^{2}}{900} + \frac{19s}{3} + 1.
\end{align}

The Fermi functions,
\begin{equation}
n_{a,s} \equiv 1 - \ol{n}_{a,s} = \frac{1}{1+\scrY_{a,s}} = 1 - \frac{\scrT_{a,s-1}\scrT_{a,s+1}}{\scrT^{2}_{a,s}}
\end{equation}
become algebraic (rational functions) of $s$. Now one can explicitly verify that the Fermi functions $n_{a,s}$ decay
in the $\sim s^{-2}$ manner at large $s$. For example,
\begin{equation}
\ol{n}_{3,s} = \frac{s(s+4)(2s+1)(2s+7)}{(s+2)^{2}(2s+3)(2s+5)} \simeq 1 - \frac{6}{s^{2}} + \mathcal{O}(s^{-3}).
\end{equation}

\end{widetext}

\pagebreak
\section{Coset spaces for exceptional Lie groups}
\label{app:exceptional}

\begin{center}
\centering
\begin{table}[b]
\renewcommand{\arraystretch}{1.2}
\setlength{\tabcolsep}{0pt}
\begin{tabular}{|c|c|c|c|}
\hline
\rowcolor{pink}
$\mathfrak{g}$ & onsite irrep & Stabilizer $H$ & $n_{\rm G}$ \\
\hline \hline
$\mathfrak{g}_{2}$ & $({\bf 7}),\,({\bf 14})$ & $G_{2}/\U(1)\times \SU(2)$ & $5$  \\
\hline
$\mathfrak{f}_{4}$ & $({\bf 52})$ & $\U(1)\times \USp(6)$ & $15$  \\
 & $({\bf 26})$ & $\U(1)\times \SO(7)$ & $15$  \\
 & $({\bf 1274}),\,({\bf 73})$ & $\U(1)\times \SU(2)\times \SU(3)$ & $20$  \\
\hline
$\mathfrak{e}_{6}$ & $({\bf 27}),\,(\overline{\bf 27})$ & $\U(1)\times \SO(10)$ & $16$ \\
 & $({\bf 78})$ & $\U(1)\times \SU(6)$ & $21$ \\
 & $({\bf 351}),\,(\overline{\bf 351})$ & $\U(1)\times \SU(2)\times \SU(5)$ & $25$ \\
 & $({\bf 2925})$ & $\U(1)\times \SU(2)\times \SU(2)\times \SU(3)$ & $29$ \\
\hline
$\mathfrak{e}_{7}$ & $({\bf 133})$ & $\U(1)\times \SO(12)$ & $33$ \\
 & $({\bf 8645})$ & $\U(1)\times \SU(2)\times \SU(6)$ & $47$ \\
 & $({\bf 365750})$ & $\U(1)\times \SU(2)\times \SU(3)\times \SU(4)$ & $53$ \\
 & $({\bf 27664})$ & $\U(1)\times \SU(3)\times \SU(5)$ & $50$ \\
 & $({\bf 1539})$ & $\U(1)\times \SU(2)\times \SO(10)$ & $42$ \\
 & $({\bf 56})$ & $\U(1)\times E(6)$ & $27$ \\
 & $({\bf 912})$ & $\U(1)\times \times \SU(7)$ & $42$ \\
\hline
$\quad \mathfrak{e}_{8} \quad$ & $({\bf 3875})$ & $\U(1)\times \SO(14)$ & $78$ \\
 & $({\bf 6696000})$ & $\U(1)\times \SU(2)\times \SU(7)$ & $98$ \\
 & $\,({\bf 6899079264})\,$ & $\,\U(1)\times \SU(2)\times \SU(3)\times \SU(5)\,$ & $\quad 106 \quad$ \\
 & $({\bf 146325270})$ & $\U(1)\times \SU(4)\times \SU(5)$ & $104$ \\
 & $({\bf 2450240})$ & $\U(1)\times \SU(3)\times \SO(10)$ & $97$ \\
 & $({\bf 30380})$ & $\U(1)\times \SU(2)\times E_{6}$ & $83$ \\
 & $({\bf 248})$ & $\U(1)\times E_{7}$ & $57$ \\
 & $({\bf 147250})$ & $\U(1)\times \SU(8)$ & $92$ \\
\hline
\end{tabular}
\caption{Complete list of coset spaces for the family of fundamental ferromagnets with symmetry of exceptional Lie algebras.}
\end{table}
\end{center}

\end{document}